\documentclass[12pt]{article}
\pdfoutput=1

\usepackage[usenames,dvipsnames,svgnames,table]{xcolor} 
\usepackage[obeyspaces,hyphens,spaces]{url}
\usepackage{jcapmod}
\usepackage{verbatim}
\usepackage{graphics}
\usepackage{epsfig}
\usepackage{booktabs}
\usepackage{booktabs}
\usepackage[english]{babel}
\usepackage{amsmath, amssymb, amsbsy, amstext, amsthm, simplewick}
\usepackage{hyperref}
\usepackage{graphicx}
\usepackage{amsfonts}
\usepackage{upgreek}
\usepackage{framed}
\usepackage{tensor}
\usepackage{pifont}
\usepackage{latexsym, mathrsfs}
\usepackage{array}
\usepackage{hyperref}
\usepackage{xspace}
\usepackage{longtable}
\usepackage{multirow}
\usepackage{cases}
\usepackage{empheq}

\usepackage{bm}
\usepackage{bbm}
\usepackage{float}
\usepackage{afterpage}
\usepackage{setspace}
\usepackage{caption}
\usepackage[load-configurations=astronomy, range-units=brackets, range-phrase=-, per-mode=reciprocal, mode=math]{siunitx}
\usepackage{threeparttable}
\usepackage{subfigure}
\usepackage{tcolorbox}
\allowdisplaybreaks 

\usepackage{braket}

\newcommand{\be}{\begin{equation}}
\newcommand{\ee}{\end{equation}}
\newcommand{\bea}{\begin{eqnarray}}
\newcommand{\eea}{\end{eqnarray}}
\newcommand{\bml}{\begin{subequations}}
\newcommand{\eml}{\end{subequations}}
\newcommand{\bfig}{\begin{figure}}
\newcommand{\efig}{\end{figure}}

\newcommand{\bmat}{\begin{pmatrix}}
\newcommand{\emat}{\end{pmatrix}}

\usepackage{ragged2e}

\usepackage{hhline}
\usepackage{array}
\newcolumntype{P}[1]{>{\centering\arraybackslash}p{#1}}
\usepackage{booktabs}
\usepackage{tikz}
\usetikzlibrary{calc,shadings,patterns,tikzmark,fadings}
\usepackage{cleveref} 
\Crefname{equation}{Eq.}{Eqs.}
\Crefname{section}{Sec.}{Secs.}
\Crefname{figure}{Fig.}{Figs.}
\Crefname{table}{Table}{Tables}

\definecolor{Blue}{rgb}{0.25, 0.41, 0.88}
\definecolor{Red}{rgb}{0.92,0.,0.}
\definecolor{darkorange}{rgb}{1.0,0.549,0.}
\definecolor{cobalt}{RGB}{44, 98, 120}
\definecolor{Mathematica1}{rgb}{0.368417, 0.506779, 0.709798}
\definecolor{Mathematica2}{rgb}{0.880722, 0.611041, 0.142051}
\definecolor{Mathematica3}{rgb}{0.560181, 0.691569, 0.194885}
\definecolor{Mathematica4}{rgb}{0.922526, 0.385626, 0.209179}
\definecolor{Mathematica5}{rgb}{0.528488, 0.470624, 0.701351}
\definecolor{Mathematica6}{rgb}{0.772079, 0.431554, 0.102387}
\definecolor{Mathematica7}{rgb}{0.363898, 0.618501, 0.782349}
\definecolor{Mathematica8}{rgb}{1, 0.75, 0}
\definecolor{Mathematica9}{rgb}{0.647624, 0.37816, 0.614037}
\definecolor{plotBlue}{RGB}{94, 130, 181}
\definecolor{plotRed}{RGB}{233, 85, 54}
\definecolor{plotGreen}{RGB}{142, 176, 50}
\definecolor{plotPurple}{RGB}{135, 120, 178}

\newcolumntype{C}[1]{>{\centering\let\newline\\\arraybackslash\hspace{0pt}}m{#1}}

\def\x{{\bf x}}

\makeatletter
\newlength{\apb@width}
\newcommand{\autoparbox}[2][c]{\settowidth{\apb@width}{#2}\parbox[#1]{\apb@width}{#2}}

\makeatother

\makeatletter
\newsavebox\myboxA
\newsavebox\myboxB
\newlength\mylenA

\newcommand*\xoverline[2][0.75]{
	\sbox{\myboxA}{$\m@th#2$}%
	\setbox\myboxB\null
	\ht\myboxB=\ht\myboxA%
	\dp\myboxB=\dp\myboxA%
	\wd\myboxB=#1\wd\myboxA
	\sbox\myboxB{$\m@th\overline{\copy\myboxB}$}
	\setlength\mylenA{\the\wd\myboxA}
	\addtolength\mylenA{-\the\wd\myboxB}%
	ifdim\wd\myboxB<\wd\myboxA%
	\rlap{\hskip 0.5\mylenA\usebox\myboxB}{\usebox\myboxA}%
	\else
	\hskip -0.5\mylenA\rlap{\usebox\myboxA}{\hskip 0.5\mylenA\usebox\myboxB}%
	\fi}
\makeatother

\numberwithin{equation}{section}
\numberwithin{figure}{section}
\numberwithin{table}{section}

\def\beq{\begin{equation}}
\def\eeq{\end{equation}}

\def\bea{\begin{eqnarray}}
\def\eea{\end{eqnarray}}

\def\beq{\begin{equation}}
\def\eeq{\end{equation}}
\def\bea{\begin{eqnarray}}
\def\eea{\end{eqnarray}}

\numberwithin{equation}{section}
\def\beq{\begin{equation}}
\def\eeq{\end{equation}}

\def\bea{\begin{eqnarray}}
\def\eea{\end{eqnarray}}

\DeclareRobustCommand{\SkipTocEntry}[4]{}

\usepackage{colortbl}
\definecolor{blue2}{cmyk}{1, 0.1, 0.1, 0.1}

\definecolor{pyBlue}{RGB}{31, 119, 180}
\definecolor{pyRed}{RGB}{214, 39, 40}
\definecolor{pyGreen}{RGB}{44, 160, 44}
\definecolor{pyBlue2}{RGB}{0, 111, 237}
\definecolor{pyRed2}{RGB}{224, 52, 36}

\newcolumntype{P}[1]{>{\centering\arraybackslash}p{#1}}
\newcolumntype{M}[1]{>{\centering\arraybackslash}m{#1}}

\usepackage{tikz,xcolor,hyperref}

\usepackage{slashed}

\definecolor{lime}{HTML}{A6CE39}
\DeclareRobustCommand{\orcidicon}{
	\begin{tikzpicture}
	\draw[lime, fill=lime] (0,0) 
	circle [radius=0.2] 
	node[white] {{\fontfamily{qag}\selectfont \tiny ID}};
	\draw[white, fill=white] (-0.0625,0.095) 
	circle [radius=0.007];
	\end{tikzpicture}
	\hspace{-2mm}
}

\foreach \x in {A, ..., Z}{\expandafter\xdef\csname orcid\x\endcsname{\noexpand\href{https://orcid.org/\csname orcidauthor\x\endcsname}
			{\noexpand\orcidicon}}
}

\begin{document}

\tcbset{colframe=black,arc=0mm,box align=center,halign=left,valign=center,top=-10pt}

\renewcommand{\thefootnote}{\fnsymbol{footnote}}

\pagenumbering{roman}
\begin{titlepage}
	\baselineskip=2.5pt \thispagestyle{empty}
	
	
\begin{center}
 ~{\Huge 
	 {\fontsize{50}{35} \textcolor{Sepia}{\bf\sffamily ${\cal Q}$uantum ${\cal D}$iscord in de-${\cal S}$itter ${\cal A}$xiverse}}}\\  \vspace{0.25cm}
\end{center}

		\begin{center}
	
	{\fontsize{14}{18}\selectfont Sayantan Choudhury\orcidA{}${}^{\textcolor{Sepia}{1}}$\footnote{{\sffamily \textit{ Corresponding author, E-mail}} : {\ttfamily sayantan\_ccsp@sgtuniversity.org,  sayanphysicsisi@gmail.com}}}${{}^{,}}$
	\footnote{{\sffamily \textit{ NOTE: This project is the part of the non-profit virtual international research consortium ``Quantum Aspects of Space-Time \& Matter" (QASTM)} }. }${{}^{,}}$.~

\end{center}


\begin{center}
	
{ 
	
	\textit{${}^{1}$Centre For Cosmology and Science Popularization (CCSP),
SGT University, Gurugram, Delhi-NCR, Haryana- 122505, India.}
	}
\end{center}

\vspace{1.2cm}
\hrule \vspace{0.3cm}
\begin{center}
\noindent {\bf Abstract}\\
\end{center}

In this work, we compute quantum discord between two causally independent areas in $3+1$ dimensions global de Sitter Axiverse to investigate the signs of quantum entanglement. For this goal, we study a bipartite quantum field theoretic setting driven by an Axiverse that arises from the compactification of Type IIB strings on a Calabi-Yau three fold. We consider a spherical surface that separates the interior and exterior causally unconnected subregions of the spatial slice of the global de Sitter space.  The Bunch-Davies state is the most straightforward initial quantum vacuum that may be used for computing purposes. Two observers are introduced, one in an open chart of de Sitter space and the other in a global chart. The observers calculate the quantum discord generated by each detecting a mode. The relationship between an observer in one of the two Rindler charts in flat space and another in a Minkowski chart is comparable to this circumstance.
We see that when the curvature of the open chart increases, the state becomes less entangled. Nevertheless, we see that even in the limit when entanglement vanishes, the quantum discord never goes away.

\vskip10pt
\hrule
\vskip10pt

\text{Keywords:~~De-Sitter vacua,  Quantum Correlation, Cosmology of Theories beyond}\\
 \text{ the SM, Quantum Information Theory aspects of Gravity,  String Cosmology.}

\end{titlepage}

\newpage
\setcounter{tocdepth}{2}

\tableofcontents

\newpage
	
	\clearpage
	\pagenumbering{arabic}
	\setcounter{page}{1}
	
	\renewcommand{\thefootnote}{\arabic{footnote}}

\clearpage

\section{Introduction}

    The very counterintuitive qualities of quantum entanglement, which was predicted by Einstein-Podolsky-Rosen (EPR) \cite{Einstein:1935rr,Bell:1964kc,Aspect:1981zz,Aspect:1982fx,Horodecki:2009zz}, have captivated many scientists. It was long thought to be an untestable philosophical dilemma. The authors were able to demonstrate experimental proof of quantum entanglement later in references \cite{Aspect:1981zz,Aspect:1982fx}. Since then, the applications of quantum entanglement of EPR pairs in quantum teleportation, quantum information, and quantum cryptography have drawn more attention \cite{Horodecki:2009zz}.

    In non-relativistic regimes, quantum entanglement resulting from pair production has been thoroughly examined.
Observer dependency is an intriguing aspect of pair production in relativistic quantum field theory \cite{Garriga:2012qp,Garriga:2013pga,Frob:2014zka}. For example, if observers are substantially accelerated when detecting each of the two free modes of a scalar field, the quantum entanglement between them decreases. They examined two scalar field free modes in flat space in references \cite{Fuentes-Schuller:2004iaz,Richter:2015wha}. A uniformly accelerated observer detects one, whereas an observer in an inertial frame detects the other. In order to describe the quantum entanglement, they calculated the entanglement negativity—a measure of entanglement for mixed states—between the two free modes, which began in a maximally entangled state. They discovered that the entanglement vanished for the observer in the limit of infinite acceleration.

An expanding universe may also be used to illustrate observer-dependent entanglement. Pair production results from the universe's expansion. Different quantum information theoretic measurements of quantum entanglement are an amazing theoretical physics probe that aids in differentiating between different kinds of long-range correlated quantum mechanical states. In this regard, it is crucial to investigate the explicit function of long-range quantum correlations within the context of quantum field theory, which is an intriguing area of study in and of itself.  See refs \cite{Maldacena:2012xp,Casini:2009sr,Amico:2007ag,Laflorencie:2015eck,Plenio:2007zz,Cerf:1996nb,Calabrese:2004eu,Martin-Martinez:2012chf,Nambu:2008my,Nambu:2011ae,VerSteeg:2007xs,Fischler:2013fba,Iizuka:2014rua,Choudhury:2017bou,Choudhury:2017qyl,Choudhury:2018ppd,Kanno:2014bma,Kanno:2014ifa,Kanno:2014lma,Albrecht:2018prr,Kanno:2017wpw,Kanno:2016qcc,Kanno:2015ewa,Colas:2022kfu,Burgess:2022nwu,Martin:2021znx,Martin:2021xml,Grain:2020wro,Martin:2018zbe,Martin:2015qta,Horodecki:2009zz} for further reference. The initial quantum mechanical vacuum states—Chernikov-Tagirov, Bunch-Davies, Hartle-Hawking, $\alpha$, and Motta-Allen vacua \cite{Allen:1985ux,Adhikari:2021ked,Banerjee:2021lqu,Choudhury:2021tuu,Choudhury:2020yaa,Choudhury:2018ppd,Choudhury:2017qyl,Choudhury:2017glj,Choudhury:2017bou,Kanno:2014lma}—are the essential component of this investigation. One of the amazing results of the fundamental theoretical elements of quantum mechanics is discussed as quantum entanglement.  This idea is primarily motivated by the possibility that a local measurement in quantum mechanics might instantly have a substantial influence on the measurement's result outside of the physical light cone.   

 Recent research has demonstrated that quantum entanglement is only one type of conceivable quantum correlation and that it is only one way to describe quantumness. It is now recognised that quantum discord is a measure of all quantum correlations, including entanglement, and that more quantum correlations have been discovered experimentally \cite{Richter:2015wha,Ball:2005xa}. Even when there is no entanglement, this metric may be non-zero. The performance of quantum computers may be discussed using quantum discord, which has led to several studies on the subject \cite{Fuentes:2010dt}. The quantum discord between two free modes of a scalar field in flat space, which are perceived by two observers in inertial and non-inertial frames, respectively, has also been studied in references \cite{Nambu:2011ae,Maldacena:2012xp} in order to see the observer dependency of all quantum correlations. They discovered that even at the limit of infinite acceleration, the quantum discord never vanishes, in contrast to the entanglement. One of the most difficult problems in contemporary physics is developing a theory of gravity that is consistent with quantum field theory. Therefore, studying quantumness in the context of curved space is crucial to attempting to properly comprehend this challenge. Furthermore, one of the main tenets of inflationary cosmology is that quantum fluctuations during the early inflationary epoch are the source of both the large-scale structure of our universe and the temperature variations of the CMB. Thus, a deeper comprehension of the early phases of our universe and more accurate forecasts for cosmic discoveries may result from this investigation of quantumness in curved environments. 

 Ryu and Takayanagi in refs. \cite{Ryu:2006bv,Ryu:2006ef} first computed the theoretically consistent entanglement entropy for a strongly coupled quantum field theory with a gravitational dual counterpart \cite{Maldacena:1997re}. Further, Maldacena and Pimentel proposed a very effective computing technique in \cite{Maldacena:2012xp} using Bunch Davies initial vacuum. In refs \cite{Iizuka:2014rua,Kanno:2014lma}, the suggested approach was generalised for the identical problem with non-standard $\alpha$ vacua.  In references \cite{Choudhury:2017bou,Choudhury:2017qyl,Choudhury:2018ppd}, these notions were utilised within the framework of Axiverse, described from Type II string theory compactification \cite{Panda:2010uq,Svrcek:2006yi,Beasley:2005iu} in the presence of Bunch-Davies and $\alpha$ quantum vacua. See refs.  \cite{Kanno:2014ifa,Choudhury:2018ppd,Kanno:2022kve,Kanno:2017wpw,Soda:2017yzu,Kanno:2015ewa,Adil:2022rgt,Bolis:2019fmq,Holman:2019spa,Albrecht:2014aga,Lello:2013qza,Lello:2013bva} where various related concepts were studied in cosmological context.

In this work,  we compute the quantum discord from an Axiverse which is obtained from Type IIB string theory compactification on a Calabi-Yau three fold in presence of NS5 brane.  This Axiverse model was studied before in the context of inflationary model building purpose \cite{McAllister:2014mpa, McAllister:2008hb,Silverstein:2008sg,Flauger:2014ana}.A study examined how a non-inertial frame affects quantum correlations in Rindler space. The curvature of an open chart in de Sitter space affects the quantum discord between two free modes of a scalar field controlled by the Axiverse. This is because the non-inertial observer in Rindler space corresponds to the observer in the open chart.

The organization of this paper is as follows:
 In \underline{\textcolor{purple}{\bf Section \ref{ka1}}}, we briefly review the physical mechanism of quantum discord, which is the central idea of the study of this paper.
In \underline{\textcolor{purple}{\bf Section \ref{ka2}}}, we  discuss the geometry and wave function of the global to open charts in Axiverse. In \underline{\textcolor{purple}{\bf Section \ref{ka3}}}, we compute entanglement negativity in Axiverse. 
In \underline{\textcolor{purple}{\bf Section \ref{ka4}}}, we discuss the preparation of ground, excited and maximal entangled state in Axiverse.  Further in \underline{\textcolor{purple}{\bf Section \ref{ka6}}},  we discuss the construction of reduced density matrix.  Next in \underline{\textcolor{purple}{\bf Section \ref{ka7}}},  we discuss the partial transposition operation in detail. Further in \underline{\textcolor{purple}{\bf Section \ref{ka8}}},  we discuss the detailed computation and the physical impacts of the logarithmic negativity in Axiverse. Next in \underline{\textcolor{purple}{\bf Section \ref{ka9}}},  we discuss the detailed computations and implications of the quantum discord in Axiverse, which is one of the key findings of this paper. Finally in \underline{\textcolor{purple}{\bf Section \ref{ka10}}},  we conclude with the future prospects.

\section{Quantum discord: The basic overview}\label{ka1}
		
All quantum correlations, including entanglement for two subsystems, are measured via quantum discord \cite{Richter:2015wha,Ball:2005xa}. Even in the case of an unentangled mixed state, this metric may not be zero. It is calculated by optimising over all feasible measurements that may be made on one of the subsystems, and it is specified by quantum mutual information. Classical information theory defines mutual information between two random variables ($A$ and $B$) as:
\be \label{mutt} {\cal I}(A,B):={\cal H}(A)+{\cal H}(B)-{\cal H}(A,B),\ee
where the classicalized Shannon entropy is defined by the following expressions:
\bea {\cal H}(A)&=&-\sum_{A}{\cal P}(A)\log_2{\cal P}(A),\\
{\cal H}(B)&=&-\sum_{B}{\cal P}(B)\log_2{\cal P}(B).\eea
Here ${\cal H}(A)$ and ${\cal H}(B)$ are describing the ignorance of the classical information regarding the variables $A$ and $B$ with probability ${\cal P}(A)$ and ${\cal P}(B)$ respectively. Also, the joint entropy ${\cal H}(A,B)$ in the present context is defined as:
\bea \label{joi} {\cal H}(A,B)=-\sum_A \sum_B {\cal P}(A,B) \log_2 {\cal P}(A,B),\eea
where the joint version of the probability ${\cal P}(A,B)$ of the two random variables $A$ and $B$. Specifically, the expression for the mutual information stated in equation (\ref{mutt}), measures the amount of classical informmation encoded in $A$ and $B$ in common.

Now further using the underlying concept of the well-known {\it Bayes theorem}, the previously mentioned oint probability ${\cal P}(A,B)$ of the two random variables $A$ and $B$ can be written in terms of the conditional probability y the following  expression:
\bea {\cal P}(A,B)={\cal P}(B){\cal P}(A|B),\eea
where ${\cal P}(A|B)$ representing the probability of 
having the random varibale $A$ for given $B$. Consequently, the previously stated joint entropy in equation (\ref{joi}), can be further rewritten as:
\bea \label{cond} {\cal H}(A,B)&=&-\sum_A \sum_B {\cal P}(A,B) \Bigg(\log_2{\cal P}(B)+\log_2 {\cal P}(A|B)\Bigg)\nonumber\\
&=&-\sum_A \sum_B {\cal P}(B){\cal P}(A|B) \Bigg(\log_2{\cal P}(B)+\log_2 {\cal P}(A|B)\Bigg)\nonumber\\
&=&-\sum_A \sum_B {\cal P}(B){\cal P}(A|B) \log_2{\cal P}(B)+{\cal H}(A|B)\nonumber\\
&=&{\cal H}(B)+{\cal H}(A|B),\eea
where we have explicitly used the following crucial fact:
\bea {\cal P}(B)=\sum_A {\cal P}(A,B).\eea
Also, in the present context of discussion, the conditional entropy is defined by the following expression:
\bea {\cal H}(A|B)=-\sum_A \sum_B {\cal P}(B){\cal P}(A|B)\log_2 {\cal P}(A|B).\eea
The above expression physically represents that we need to take the average over $B$ of Shannon entropy of $A$, for a given $B$.

Further using equation (\ref{cond}) in equation (\ref{mutt}), we get the following simplified expression for the mutual information:
\bea \label{muttb} {\cal I}(A,B)={\cal H}(A)-{\cal H}(A|B).\eea
Here it is important to note that equations (\ref{mutt}) and (\ref{muttb}) represent the classically equivalent results to describe the form of mutual information.	

However, things are not going to be the same when we introduce the underlying physical concept of mutual information in the context of quantum mechanical system.  Specifically in the case of a quantum system, the previously mentioned  two expressions for the mutual information content (see equations (\ref{mutt}) and (\ref{muttb})) do not correspond to the identical results. The prime reason for such a crucial difference is measurements  on the subsystem $B$ disturb/perturb the subsystem $A$.  It is a very well known fact that in the context of quantum mechanical system,  the classical Shannon entropy is described by the following expression for the von Neumann entropy:
\bea S(\rho):=-{\rm Tr}(\rho\log_2\rho)\eea
where $\rho$ represents the density matrix in the present context of discussion. Additionally, it is important to note that the expression for the probabilities i.e. ${\cal P}(A)$, ${\cal P}(B)$ and, ${\cal P}(A,B)$ are respectively  replaced by  the reduced density matrix of the subsystems $A$ and $B$, which are: \bea \rho_{A}&=&{\rm Tr}_B \rho_{A,B},\\
\rho_{B}&=&{\rm Tr}_A \rho_{A,B},\eea
and density matrix of the total system $\rho_{A,B}$.

Further, in the context of quantum mechanical systems, the underlying concept of conditional probability ${\cal P}(A|B)$ is described by projective measurements of $B$, which is further described by the following complete set of projectors:
\bea \left\{\Pi_i\right\}=\left\{|\psi_i\rangle\langle\psi_i\right\} \forall i,\eea
where the index $i$ represents various distinctive outcomes of a specific measurement on subsystem $B$. However, this is not at all unique. Here we can make many different sets of measurements instead of a single one. In this connection, the state of $A$ after the  measurement on $B$  is described  by the following expression:
\bea \rho_{A|i}=\frac{1}{\omega_i}{\rm Tr}_B \bigg(\Pi_i\rho_{A,B}\Pi_i\bigg),\eea
where the factor $\omega_i$ is defined as:
\bea \omega_i={\rm Tr}_{A,B}\bigg(\Pi_i\rho_{A,B}\Pi_i\bigg).\eea
Now, the quantum mechanical version of the conditional entropy is described by the following expression:
\bea S(A|B):\equiv {\rm min}_{\left\{\Pi_i\right\}}\sum_i \omega_i  S(\rho_{A|i}).\eea
To avoid relying on projectors, choose the measurement that causes the least disturbance to the entire quantum state.

Then the quantum mechanical analogue of mutual information corresponding to the previously mentioned classical expressions stated in equation (\ref{mutt}) and equation (\ref{muttb}) are defined by the following expressions:
\bea {\cal I}_{Q}(A,B):&=&S(\rho_A)+S(\rho_B)-S(\rho_{A,B}),\\ 
{\cal E}_{Q}(A,B):&=&S(\rho_A)-S(A|B).\eea
Finally, the quantum discord is defined as:
\bea {\cal D}_{Q}(A,B):&=&{\cal I}_{Q}(A,B)-{\cal E}_{Q}(A,B)\nonumber\\
&=&S(\rho_B)-S(\rho_{A,B})+S(A|B).\eea
Thus, in classical mechanics, the quantum discord disappears, but in certain quantum systems, it doesn't seem to.

	\section{Global to open charts in Axiverse}
	\label{ka2} 
	Our main goal in this part is to provide a general overview of how the reduced density matrix for the open chart of global de Sitter space is calculated inside the Axiverse framework. See refs.  \cite{Maldacena:2012xp,Iizuka:2014rua,Kanno:2014lma,Choudhury:2017bou,Choudhury:2017qyl,Choudhury:2018ppd} for more details.

    \subsection{Geometry of open chart}
    \label{ka2b}
    \begin{figure*}[htb]
    	\centering
    	{
    		\includegraphics[width=15.5cm,height=13.5cm] {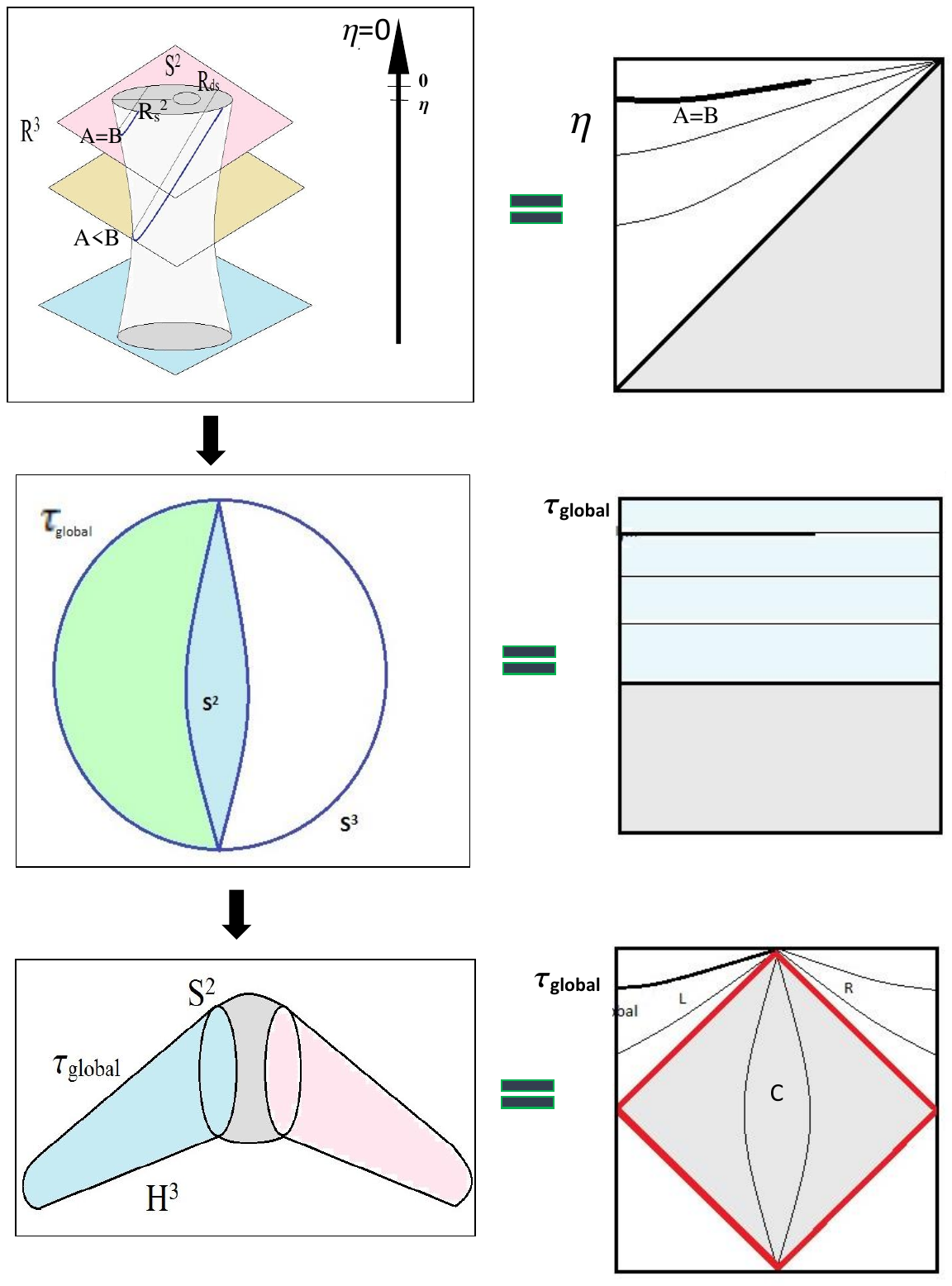}
    	}
    	\caption[Optional caption for list of figures]{Geometry of open chart. } 
    	\label{penrose}
    \end{figure*}
     Let us consider two sphere (${ \bf S}^2$) geometry of which is described as:
    \bea \sum^{3}_{i=1}x^2_{i}=R^2\quad\quad x_{i}\forall i=1,2,3,\eea
    where $R$ is the radius of the two sphere.  Here we considered that the radius of the two sphere is sufficiently large enough compared to the horizon size.   Since the conformal time scale describes the horizon size, we require $R\gg \tau$, where $\tau$ stands for the conformal time.   This may be accomplished practically by taking the late time limit $\tau\rightarrow 0$.   It is crucial to notice that the surface of the two spheres discussed in this section is invariant under the transformation ${\rm \bf SO(1,3)}$.  In this construction, the symmetry group is ${\rm \bf SO(1,4)}$, also known as the isometry group of de Sitter space. ${\rm \bf SO(1,3)}$ is a subgroup of this larger group.  This particular geometry is appearing in the left region of figure~(\ref{penrose}).  In our algorithm, we select equal conformal time slices on three spheres ${\bf S}^3$ that appear in the context of the global coordinates of de Sitter space.  Most critically, it is recognised that the surface of two spheres is considered as the equator of three spheres in this geometrical structure.  This conformal map on the border of the $3+1$-dimensional global de Sitter space produces divergent contributions, which may be readily controlled by applying another conformal reverse map from three spheres to two spheres via a global time surface.  
    
    Now,  we consider a $1+4$ dimensional Euclidean global hyperbolic geometry:
    \bea \sum^{5}_{j=1}Y^2_{j}= H^{-2},\eea
  where the coordinates are described as:  
    \bea
           \label{hypg}
  \displaystyle  Y_{j}&=&\displaystyle H^{-1}\times\left\{\begin{array}{ll}
 \displaystyle  \cos\tau_{\bf E}~\sin\sigma_{\bf E}~\hat{\bf n}_{j}~~~~~~~~~~~~ &
                                                           \mbox{\small {\textcolor{black}{\bf  for $j=1,2,3$}}}  
                                                          \\ 
          \displaystyle \sin\tau_{\bf E} & \mbox{\small { \textcolor{black}{\bf for $j=4$}}}\\ 
                    \displaystyle \cos\tau_{\bf E}~\cos\sigma_{\bf E} & \mbox{\small { \textcolor{black}{\bf for $j=5$}}}.~~~~~~~~
                                                                    \end{array}
                                                          \right.
                                                          \eea 
     Here $\hat{\bf n}_{j}\forall j=1,2,3$
    are in ${\bf R}^{3}$.  Here, the Euclidean metric is given by:
    \bea  ds^2_{\bf E}= H^{-2}\left\{d\tau^2_{\bf E}+\cos^2\tau_{\bf E}\left(d\sigma^2_{\bf E}+\sin^2\sigma_{\bf E}~d\Omega^2_{\bf 2}\right)\right\},
    \eea
    where $d\Omega^2_{\bf 2}$ is defined in two sphere. Further taking analytic continuation in the fifth coordinate, we have:
    \bea Y_{5}=H^{-1}\cos\tau_{\bf E}~\cos\sigma_{\bf E}\xrightarrow[]{\bf Analytic~continuation} X_{0}=iY_{5}=iH^{-1}\cos\tau_{\bf E}~\cos\sigma_{\bf E}\eea
  and we consider the coordinate redefinition,
   $X_{k}=Y_{k}{ \forall k=1,2,3,4}$.
    Hence,  the corresponding Lorentzian geometry is described by:
    \bea  \sum^{4}_{\mu=0}X^2_{\mu}=\left(-X^2_0+\sum^{4}_{j=1}X^2_j\right)=H^{-2}.\eea
    Since we have changed the coordinates in Lorentzian signature, the appropriate Lorentzian geometry for the distinct areas is described as:
    \bea
               \label{penr2}
      \displaystyle \textcolor{black}{\bf Region-R}&:\Longrightarrow&\displaystyle\left\{\begin{array}{ll}
     \displaystyle \tau_{\bf E}=\frac{\pi}{2}-it_{\bf R}~~~~~~~~~~~~ &
                                                               \mbox{\small {\textcolor{black}{\bf for $t_{\bf R}\geq 0$}}}  
                                                              \\ 
              \displaystyle \sigma_{\bf E}=-ir_{\bf R} & \mbox{\small { \textcolor{black}{\bf for $r_{\bf R}\geq 0$}}}.~~~~~~~~
                                                                        \end{array}
                                                              \right.\\
\label{penr3}
      \displaystyle \textcolor{black}{\bf Region-C}&:\Longrightarrow&\displaystyle\left\{\begin{array}{ll}
     \displaystyle \tau_{\bf E}=t_{\bf C}~~~~~~~~~~~~~~~~~~ &
                                                               \mbox{\small {\textcolor{black}{\bf for $\displaystyle-\frac{\pi}{2}\leq t_{\bf C}\leq \frac{\pi}{2}$}}}  
                                                              \\ 
              \displaystyle \sigma_{\bf E}=\frac{\pi}{2}-ir_{\bf C} & \mbox{\small { \textcolor{black}{\bf for $-\infty<r_{\bf C}< \infty$}}}.~~~~~~~~~~~~~~
                                                                        \end{array}
                                                              \right. \\
\label{penr4}
      \displaystyle \textcolor{black}{\bf Region-L}&:\Longrightarrow&\displaystyle\left\{\begin{array}{ll}
     \displaystyle \tau_{\bf E}=-\frac{\pi}{2}+it_{\bf L}~~~~~~~~~~~~ &
                                                               \mbox{\small { \textcolor{black}{\bf for $t_{\bf L}\geq 0$}}}  
                                                              \\ 
              \displaystyle \sigma_{\bf E}=-ir_{\bf L} & \mbox{\small { \textcolor{black}{\bf for $r_{\bf L}\geq 0$}}}.~~~~~~~~
                                                                        \end{array}
                                                              \right.                                         \eea
 Finally, the Lorentzian metric is described by  \cite{Maldacena:2012xp,Iizuka:2014rua,Kanno:2014lma,Choudhury:2017bou,Choudhury:2017qyl,Choudhury:2018ppd}:
  \bea
                 \label{r2z}
        \displaystyle \textcolor{black}{\bf Region-R}&:\Longrightarrow&\displaystyle\left\{\begin{array}{ll}
       \displaystyle ds^2_{\bf R}=H^{-2}\left[-dt^2_{\bf R}+\sinh^2t_{\bf R}\left(dr^2_{\bf R}+\sinh^2r_{\bf R}~d\Omega^2_{\bf 2}\right)\right], 
                                                                          \end{array}
                                                                \right.\\
  \label{r3}
        \displaystyle \textcolor{black}{\bf Region-C}&:\Longrightarrow&\displaystyle\left\{\begin{array}{ll}
       \displaystyle  ds^2_{\bf C}=H^{-2}\left[dt^2_{\bf C}+\cos^2t_{\bf C}\left(-dr^2_{\bf C}+\cosh^2r_{\bf C}~d\Omega^2_{\bf 2}\right)\right], \end{array}
                                                                \right. \\
  \label{r4}
        \displaystyle \textcolor{black}{\bf Region-L}&:\Longrightarrow&\displaystyle\left\{\begin{array}{ll}
       \displaystyle  ds^2_{\bf L}=H^{-2}\left[-dt^2_{\bf L}+\sinh^2t_{\bf L}\left(dr^2_{\bf L}+\sinh^2r_{\bf L}~d\Omega^2_{\bf 2}\right)\right].  \end{array}
                                                                \right.                                         \eea

   In the next paragraph, we will compute the formula for the wave function in the presence of string theory generated axion effective interactions in the area {\bf L} or {\bf R} of the open chart of hyperbolic slices of global de Sitter space. The typical pensore diagram in figure~(\ref{penrose}) makes it clear that the regions {\bf L} and {\bf R} are identical replicas of one another based on our creation of the geometrical setup. Because of this, we calculate the wave function in both regions; however, while creating the reduced density matrix, we only take the effective contributions from the region {\bf L}, taking the partial trace over the contributions of the region {\bf R}.  The area {\bf R} can also benefit from the same method.

    \subsection{Wave function of Axiverse}
    \label{ka2c}

In this section, we calculate the hyperbolic open mode functions, related wave functions, and quantum number and momentum dependent mode functions for the global de Sitter space time hyperbolic open chart. To do this, we employ the axion effective potential developed from string theory, which appears as a scalar field in the matter sector. This result will be extremely useful in calculating the quantum discord and the decreased density matrix expression in the next paragraph. It is important to highlight that we are interested in the axion monodromy model in this discussion.  This result will be extremely useful in calculating the quantum discord and the reduced density matrix expression in the next paragraph. It is important to highlight that we are interested in the axion monodromy model in this discussion. The RR sector of the Type IIB string theory setup, where the effective potential and associated interaction occur as a result of the compactification on a Calabi-Yau three fold in the presence of an NS5 brane, generates the axion field in the current architecture \cite{Panda:2010uq,Svrcek:2006yi,Beasley:2005iu}.

    \begin{figure*}[htb]
    \centering
    \subfigure[For $b<0$]{
        \includegraphics[width=7.5cm,height=8cm] {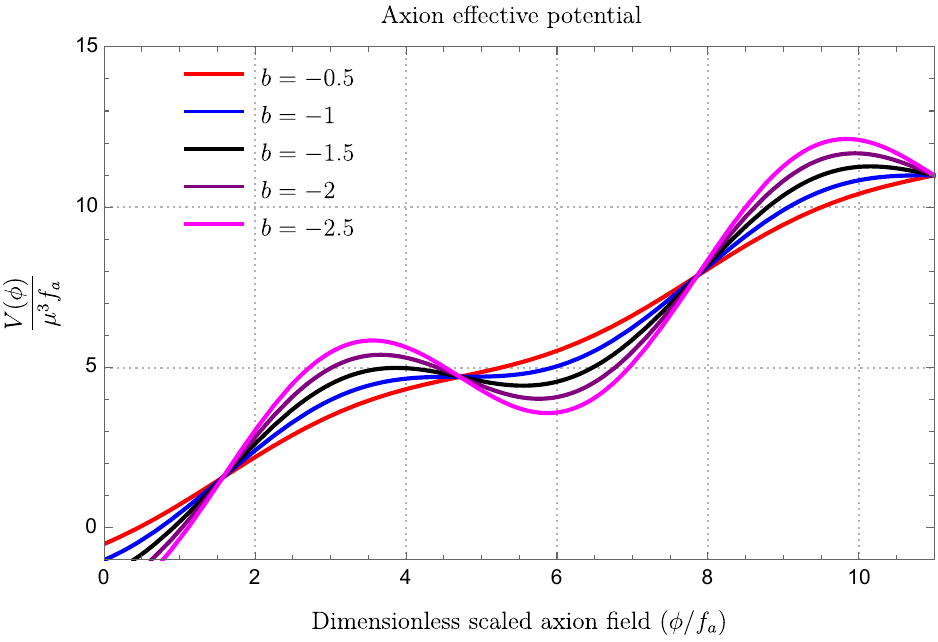}
        \label{fig1a}
    }
    \subfigure[For $b>0$]{
        \includegraphics[width=7.5cm,height=8cm] {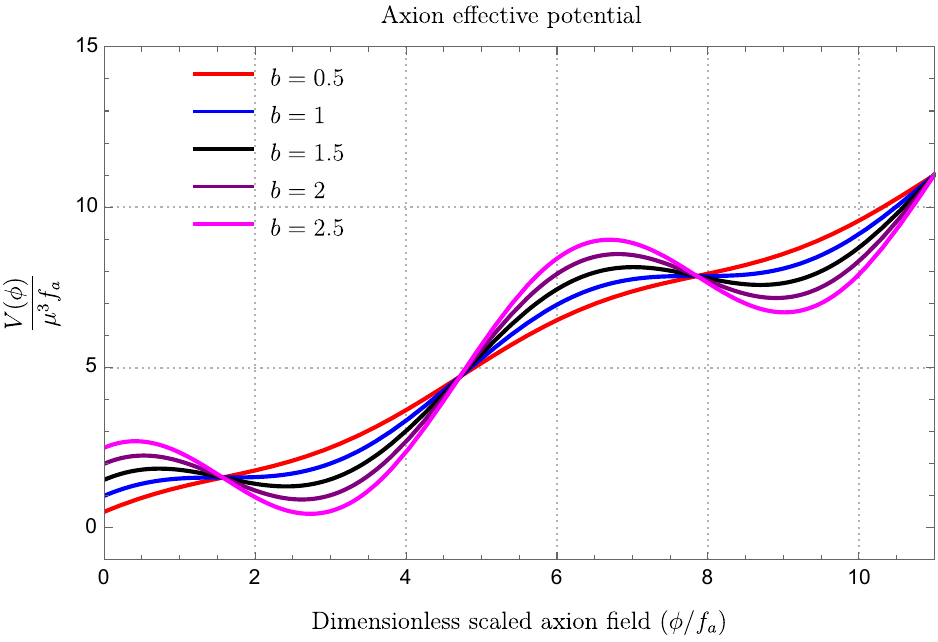}
        \label{fig2a}
       }
    \caption[Optional caption for list of figures]{Axion potential with respect to the field obtained from Type IIB String Theory compactification.} 
    \label{pote1}
    \end{figure*}  
      Let's begin with the standard effective action for an axion field in $1+3$ dimensions that is minimally connected to the gravitational sector via space-time metric:
         \be\label{axeff}   S= \int d^{4}x \sqrt{-g}\left[-\frac{1}{2}(\partial \phi)^2 -V(\phi)\right],\ee
        Here $\phi$ is the dimensionful axion field,  which is described by the potential \cite{Panda:2010uq,Svrcek:2006yi,Beasley:2005iu}:
    \bea\label{axion} ~V(\phi)&=&\mu^3\phi+\Lambda^4_{G}\cos\left(\frac{\phi}{f_{a}}\right)=\mu^3f_{a}\left[\left(\frac{\phi}{f_{a}}\right)+b\cos\left(\frac{\phi}{f_{a}}\right)\right].~~~~~~~~~~\eea        
    In figure (\ref{fig1a}) and figure (\ref{fig2a}) we have shown the behaviour of the axion potential $V(\phi)/\mu^3 f_a$ with $\phi/f_a$ for various signatures of $b= \Lambda^4_{G}/\mu^3 f_{a}$. The coupling parameter of linear interaction, denoted by $\mu^3$, is related to the underlying theoretical scale and may be written as:
  \bea \mu^3=\frac{1}{f_a \alpha^{'2}g_s}\exp(4A_0)+\frac{R^2 m^4_{SUSY}}{f_a \alpha^{'}L^4}\exp(2A_0),~~~~~\eea
 where $\exp(A_0)$ is the warp factor of the lower part of the Klebanov-Strassler throat geometry, $R$ is the radius that stabilised the $5$ brane and antibrane in the corresponding string theory construction, $m_{SUSY}$ is the only mass scale used in this construction, which actually represents the underlying supersymmetry breaking scale in this particular setup, $\alpha^{'}$ is the Regge slope, which is proportional to the inverse string tension, $g_s$ is the string coupling, and $L^6$ is the world volume.
     Here, the shift symmetry is broken by the first half of the effective potential, but the symmetry $\phi\rightarrow \phi+2\pi f_a$ is preserved by the remaining part.  The axionic decay parameter is quantified here by $f_a$, and we have selected the following helpful profile for it:
                     \bea f_a/H=\sqrt{100-\frac{80}{1+\left(\ln\left[\frac{ \ln\left[{\rm tanh}\left(\frac{t}{2}\right)\right]}{\ln\left[{\rm tanh}\left(\frac{t_c}{2}\right)\right]}\right]\right)^2}},\eea
                         which was used in refs.~\cite{Maldacena:2015bha,Choudhury:2016cso,Choudhury:2016pfr}.  Here $t_c$ is the characteristic time scale at which  $f_a/H=2\sqrt{5}$.

Here we introduce an energy scale,  $\Lambda_{G}$, which is defined as:
\bea  \Lambda_{G}=\sqrt{\frac{m_{SUSY} L^3}{ \sqrt{\alpha^{'}}g_{s}}}~\underbrace{\exp\left(-cS_{inst}\right)}_{\bf Instanton~decay }.\eea
Here, $S_{inst}$ stands for the instantonic action that ultimately gives birth to the current structure of the effective potential within the context of string theory. In this computation, the instanton coupling parameter $c\sim{\cal O}(1)$ is really considered as a constant term.   At last, the for of the warp factor and string scale may be fixed in terms of all the stringy parameters, which is provided by:
\bea  \exp(A_0)=\left(\frac{\Lambda_G}{m_{SUSY}}\right)^2\frac{L}{R}\sqrt{\alpha^{'}g_s},\quad\quad M_s=\frac{1}{\sqrt{\alpha^{'}}} \exp(A_0).\eea

Here we consider the following two cases:
    \begin{enumerate}
    \item \underline{\textcolor{black}{\bf Case~A:}} 
   In this particular scenario, we only take into account the portion of the effective potential that violates the shift symmetry $\phi\rightarrow \phi+2\pi f_a$, which is provided by:
        \bea\label{af1}  V(\phi)\approx \mu^3f_a\left(\frac{\phi}{f_{a}}\right).\eea
      The aforementioned potential contributes as a source term in terms of $\mu^3$ in the field equation for the current computational purpose, which essentially fixes the total energy scale in terms of the stringy model parameters.

    \item \underline{\textcolor{black}{\bf Case~B:}}
   We examine the small field limiting approximation in this particular instance, where the dimensionless field variable $\phi/f_a\ll 1$.   Because of this, the shift symmetry $\phi\rightarrow \phi+2\pi f_a$ may be approximated while maintaining non-perturbative contribution as,
    $\cos\left(\frac{\phi}{f_{a}}\right)\approx 1-\frac{1}{2}\left(\frac{\phi}{f_{a}}\right)^2.$ The previously indicated shift symmetry is violated because the quadratic order term is truncated at the end.  The entire non-perturbative term may be taken in Prince, but dealing with such terms at the level of equations of motion and later is quite difficult in the field theory language. Higher order terms can be ignored in the current computational purpose due to the small field limit without sacrificing generality, at the expense of violating the shift symmetry.  Hence the effective potential can be written as:
        \bea\label{af2} V(\phi)&\approx&\mu^3f_{a}\left(b+\left(\frac{\phi}{f_{a}}\right)-\frac{b}{2}\left(\frac{\phi}{f_{a}}\right)^2\right)\quad{\rm where}\quad  m^2_{eff}=\mu^3 b f_{a}=\Lambda^4_{G}.\quad\eea

        \end{enumerate} 

    The field equations of axion may be found by altering the effective action indicated in equation (\ref{axeff}) with regard to the axion field itself, giving rise to the following formulations for the above-mentioned two situations:
    \bea \underline{\textcolor{black}{\bf For ~Case~A}:}~~~~~~~~~~~~~~~~~~~~~~~~~~~~~~\left[\frac{1}{a^3(t)}\partial_{t}\left(a^3(t)\partial_{t}\right)-\frac{1}{H^2a^2(t)}\hat{\bf L}^2_{\bf H^3}\right]\phi&=&\mu^3,\\
    \underline{\textcolor{black}{\bf For ~Case~B}:}~~~~~~~~~~~~~~~\left(\left[\frac{1}{a^3(t)}\partial_{t}\left(a^3(t)\partial_{t}\right)-\frac{1}{H^2a^2(t)}\hat{\bf L}^2_{\bf H^3}\right]+m^2_{eff}\right)\phi&=&\mu^3,\quad
    \eea
 where the following equation yields the scale factor $a(t)$ for global de Sitter space:
 \bea
                   \label{r2zzaa}
          \displaystyle a(t)&=&\displaystyle\frac{1}{H}\sinh t~~~~{\rm where}~~~t=\Bigg(t_{\bf R}({\rm in}~ {\bf R}),t_{\bf L}({\rm in}~ {\bf L})\Bigg).\eea
 A Laplacian operator $\hat{\bf L}^2_{\bf H^3}$ is introduced in ${\bf H^3}$ which is expressed as:
          \bea \hat{\bf L}^2_{\bf H^3}&=&\frac{1}{\sinh^2r}\left[\partial_{r}\left(\sinh^2r~\partial_{r}\right)+\frac{1}{\sin\theta}\partial_{\theta}\left(\sin\theta~\partial_{\theta}\right)+\frac{1}{\sin^2\theta}\partial^2_{\Phi}\right].\eea Here the Laplacian operator $\hat{\bf L}^2_{\bf H^3}$ satisfies the following eigenvalue equation:
              \bea\hat{\bf L}^2_{\bf H^3}{\rm\cal Y}_{plm}(r,\theta,\Phi)&=&\lambda_p{\rm\cal Y}_{plm}(r,\theta,\Phi)\quad {\rm  where}\quad \lambda_p=-(1+p^2).    \eea
             Also the eigenfunction of this operator is defined as:
             \bea   {\cal Y}_{plm}(r,\theta,\Phi)&=&\frac{\Gamma\left(ip+l+1\right)}{\Gamma\left(ip+1\right)}~\frac{p}{\sqrt{\sinh r}}~{\cal P}^{-\left(l+\frac{1}{2}\right)}_{\left(ip-\frac{1}{2}\right)}\left(\cosh r\right)Y_{lm}(\theta,\Phi),\eea         
             where $p$, $l$ and $m$ are three quantum numbers.  Last but not least, the radial solution is characterised by the function, ${\cal P}^{-\left(l+\frac{1}{2}\right)}_{\left(ip-\frac{1}{2}\right)}\left(\cosh r\right)$, which is the well-known associated Legendre polynomial in this context. Here, $Y_{lm}(\theta,\Phi)$ is the well-known spherical harmonics, which is dependent on two quantum numbers $l$ and $m$ and on two angular coordinates as defined in ${\bf S}^2$.
         
              Following quantisation, the field equation's classical solution is promoted in terms of the quantum operator. Using the well-known canonical quantisation technique, the corresponding quantum operator can be expressed in terms of the creation and annihilation operators along with the basis Bunch Davies mode function, which is simply the field equation's classical counterpart.  The following compact form may be used to express the entire quantum solution for the axion field operator for both the {\bf Case A} and {\bf Case B}:         
    \bea\label{total} \widehat{ \phi}(t,r,\theta,\Phi)&=&\int^{\infty}_{0} dp \sum_{\sigma=\pm 1}\sum^{p-1}_{l=0}\sum^{+l}_{m=-l}\left[a_{\sigma plm}{\cal U}_{\sigma plm}(t,r,\theta,\Phi)+a^{\dagger}_{\sigma plm}{\cal U}^{*}_{\sigma plm}(t,r,\theta,\Phi)\right],\quad\quad\eea   where, $t=(t_{\bf R},t_{\bf L})$. Here the Bunch-Davies vacuum is defined as:
\bea  ~a_{\sigma p l m}|{\bf BD}\rangle&=&0~~~~~~~~~~ \forall \sigma=(+1,-1);0<p<\infty;\nonumber\\
&&~~~~~~~~~~~~l=0,\cdots,p-1,m=-l,\cdots,+l.~~~~~~~~~~~ \eea
The classical solution of the field equation for the axion for both the {\bf Case A} and {\bf Case B}, which constitute the whole basis, is represented here by ${\cal U}_{\sigma plm}(t,r,\theta,\Phi)$.  The three quantum numbers, $p$, $l$, and $m$, which emerge as a result of the canonical quantisation of the modes in the current context of discussion, are used to tag these basic functions, which are often referred to as the mode functions, after quantisation.   For the {\bf Case A} and {\bf Case B}, the mode functions may be solved by solving the relevant axion field equations, which are essentially partial differential equations solved using the well-known approach of separation of variables. This gives:
  \bea {\cal U}_{\sigma plm}(t,r,\theta,\Phi)&=&\frac{1}{a(t)}\chi_{p,\sigma}(t){\cal Y}_{plm}(r,\theta,\Phi)=\frac{H}{\sinh t}\chi_{p,\sigma}(t){\cal Y}_{plm}(r,\theta,\Phi).\eea 
  It's crucial to note that the time-dependent part of the mode function $\chi_{p,\sigma}(t)$ only works for positive frequencies and hence forms a complete set under the current theoretical setup.  This element of the answer is reliant on the momentum $p$, which is the wave number in the quantum mechanical description, as previously stated.  We are interested in the dynamical behaviour of the mode function in the {\bf R} and {\bf L} regions of the open chart of global de Sitter space time, therefore the time-dependent part of the wave function is very important for our discussion.

  Half of the computational work is completed if we can extract the hidden features from the time-dependent portion of the field equations from the {\bf Case A} and {\bf Case B}.  The entire solution may be expressed as the sum of the complementary part ($\chi^{(c)}_{p,\sigma}(t)$) and the particular integral part ($\chi^{(p)}_{p,\sigma}(t)$), as in both situations we are working with inhomogeneous second order differential equations. 
          \bea \chi_{p,\sigma}(t)=\underbrace{\chi^{(C)}_{p,\sigma}(t)}_{\bf Complementary~part}+\underbrace{\chi^{(P)}_{p,\sigma}(t)}_{\bf Particular~integral~part}.\eea
         The complementary part ($\chi(c)_{p,\sigma}(t)$) of the time-dependent solution of the mode function satisfies the homogeneous part of the field equation may be stated as:
              \bea
                                 \label{com}
                        \displaystyle 0&=&\displaystyle\left\{\begin{array}{ll}
                       \displaystyle \left[\partial^2_t +3\coth t~ \partial_t+\frac{(1+p^2)}{\sinh^2t}\right]\chi^{(C)}_{p,\sigma}(t)~~~~~~~~~~~~~~~~~ &
                                                                                 \mbox{\small {\textcolor{black}{\bf for Case A}}}  
                                                                                \\ \\
                                \displaystyle \left[\partial^2_t +3\coth t~ \partial_t+\frac{(1+p^2)}{\sinh^2t}+\frac{m^2_{eff}}{H^2}\right]\chi^{(C)}_{p,\sigma}(t) & \mbox{\small {\textcolor{black}{\bf for Case B}}}.~~~~~~~~
                                                                                          \end{array}
                                                                                \right.\eea
                                                                                The solution of the above equationcan be written as:
 \bea
\label{sol}
                         \displaystyle \chi^{(c)}_{p,\sigma}(t)&=&\displaystyle\left\{\begin{array}{ll}
                        \displaystyle \left\{\frac{1}{2\sinh\pi p}\left[\frac{\left(e^{\pi p}-i\sigma~e^{-i\pi\nu}\right)}{\Gamma\left(\nu+\frac{1}{2}+ip\right)}{\cal P}^{ip}_{\left(\nu-\frac{1}{2}\right)}(\cosh t_{\bf R})\displaystyle\right.\right.\\
                        \left.\left.  \displaystyle~~~ ~~~~~~~~~~~-\frac{\left(e^{-\pi p}-i\sigma~e^{-i\pi\nu}\right)}{\Gamma\left(\nu+\frac{1}{2}-ip\right)}{\cal P}^{-ip}_{\left(\nu-\frac{1}{2}\right)}(\cosh t_{\bf R})\right]\right\}_{\sigma=\pm 1}~~ &
                                                                          ~~~ ~~~~~~~~~~~~~~~~~~~~~ \mbox{\small {\textcolor{black}{\bf for R}}}  
                                                                                 \\ 
                                 \displaystyle \left\{\frac{\sigma}{2\sinh\pi p}\left[\frac{\left(e^{\pi p}-i\sigma~e^{-i\pi\nu}\right)}{\Gamma\left(\nu+\frac{1}{2}+ip\right)}{\cal P}^{ip}_{\left(\nu-\frac{1}{2}\right)}(\cosh t_{\bf L})\displaystyle\displaystyle\right.\right.\\
                        \left.\left.  \displaystyle~~~ ~~~~~~~~~~~-\frac{\left(e^{-\pi p}-i\sigma~e^{-i\pi\nu}\right)}{\Gamma\left(\nu+\frac{1}{2}-ip\right)}{\cal P}^{-ip}_{\left(\nu-\frac{1}{2}\right)}(\cosh t_{\bf L})\right]\right\}_{\sigma=\pm 1} & ~~~ ~~~~~~~~~~~~~~~~~~~~~~ \mbox{\small {\textcolor{black}{\bf for L}}},~~
                                                                                           \end{array}                          \right.                                                                                           \eea 
where 
we introduce a new parameter $\nu$:
   \bea
                                 \label{nu}
                        \displaystyle \nu&=&\displaystyle\left\{\begin{array}{ll}
                       \displaystyle \frac{3}{2}~~~~~~~~~~~~~~~~~ &
                                                                                 \mbox{\small {\textcolor{black}{\bf for Case A}}}  
                                                                                \\ 
                                \displaystyle \sqrt{\frac{9}{4}-\frac{m^2_{eff}}{H^2}} ~~~~~~~~~& \mbox{\small {\textcolor{black}{\bf for Case B}}}.~~~~~~~~
                                                                                          \end{array}
                                                                                \right.\eea
 Here $\sigma=\pm 1$ for {\bf R} and {\bf L} regions. 
 Also,  $\chi^{(C)}_{p,\sigma}(t)=\chi^{(C)}_{-p,\sigma}(t)$. Now, we can define Klien-Gordon inner product :
\bea \left(\left(\chi^{(C)}_{p,\sigma}(t),\chi^{(C)}_{p,\sigma^{'}}(t)\right)\right)_{\bf KG}={\cal N}_{p\sigma}\delta_{\sigma\sigma^{'}}, \eea
where ${\cal N}_{p\sigma}$ is the normalization constant,  which is described by:
\bea
                                 \label{norm}
                        \displaystyle {\cal N}_{p\sigma}&=&\displaystyle
                                \displaystyle \frac{4}{\pi}\frac{\left[\cosh\pi p-\sigma\cos\left(\nu-\frac{1}{2}\right)\right]}{|\Gamma\left(\nu+\frac{1}{2}+ip\right)|^2}\forall \sigma=\pm 1.\eea

Also for the particular integral part we have:
   \bea
                                 \label{com2}
                       &&\displaystyle\left\{\begin{array}{ll}
                       \displaystyle \left[\partial^2_t +3\coth t~ \partial_t+\frac{(1+p^2)}{\sinh^2t}\right]\chi^{(P)}_{p,\sigma}(t)=\mu^3~~~~~~~~~~~~~~~~~ &
                                                                                 \mbox{\small {\textcolor{black}{\bf for Case A}}}  
                                                                                \\ 
                                \displaystyle \left[\partial^2_t +3\coth t~ \partial_t+\frac{(1+p^2)}{\sinh^2t}+\frac{m^2_{eff}}{H^2}\right]\chi^{(P)}_{p,\sigma}(t)=\frac{m^2_{eff}f_a}{b}~~~~~~~~ & \mbox{\small {\textcolor{black}{\bf for Case B}}}.~~~~~~~~
                                                                                          \end{array}
                                                                                \right.\eea

Using the Green's function method the solution of the particular integral part can be written as:
\bea
                                 \label{com3} \textcolor{black}
                        \displaystyle  \chi^{(P)}_{p,\sigma}(t)&=&\displaystyle\left\{\begin{array}{ll}
                       \displaystyle\int dt^{'}~G_{\sigma}(t,t^{'})~\mu^3~~~~~~~~~~~~~~~~~ &
                                                                                 \mbox{\small {\textcolor{black}{\bf for Case A}}}  
                                                                                \\ 
                                \displaystyle\int dt^{'}~G_{\sigma}(t,t^{'})~\frac{m^2_{eff}f_a(t^{'})}{b}~~~~~~~~ & \mbox{\small {\textcolor{black}{\bf for Case B}}}.~~~~~~~~
                                                                                          \end{array}
                                                                                \right.\eea
   where $G_{\sigma}(t,t^{'})$ is the Green's function for axion field, defined as:
   \bea\label{green} G_{\sigma}(t,t^{'})&=&\sinh^2 t\sum^{\infty}_{n=0}\frac{1}{\left(p^2-p^2_{n}\right)}\chi^{(C)}_{p_{n},\sigma}(t)\chi^{(C)}_{p_{n},\sigma}(t^{'})\quad\quad{\rm where}\quad\sigma=\pm 1.\eea 
   We also use the following shorthand notation:
\bea 
                                   \label{bbvx}
                          \displaystyle {\cal P}^{q}&=& {\cal P}^{ip}_{\left(\nu-\frac{1}{2}\right)}(\cosh t_{q}),\quad{\cal P}^{{q},n}=
                                  \displaystyle {\cal P}^{ip_n}_{\left(\nu-\frac{1}{2}\right)}(\cosh t_{q}) 
                           \eea                                                  where $q=\left({\bf R},{\bf L}\right)$.  Hence the total solution is given by:                                               
   \bea
                                   \label{cxcxx}  
                          \displaystyle \boxed{\boxed{\chi_{p,\sigma}(t)=\sum_{q={\bf R},{\bf L}}\left\{\underbrace{\frac{1}{{\cal N}_{p}}\left[\alpha^{\sigma}_{q}~{\cal P}^{q}+\beta^{\sigma}_{q}~{\cal P}^{q*}\right]}_{\textcolor{black}{\bf Complementary}}+\underbrace{\sum^{\infty}_{n=0}\frac{1}{{\cal N}_{p_n}\left(p^2-p^2_n\right)}\left[\bar{\alpha}^{\sigma}_{q,n}~\bar{\cal P}^{q,n}+\bar{\beta}^{\sigma}_{q,n}~\bar{\cal P}^{*q,n}\right]}_{\textcolor{black}{\bf Particular}}\right\}}},\quad\quad\eea 
                          where we use the following notations:
                                                   \bea
                                 \label{com5}                              
                        \displaystyle \overline{\cal P}^{q,n}&=& \sinh^2t~  {\cal P}^{q,n}\times\displaystyle\left\{\begin{array}{ll}
                       \displaystyle\int dt^{'}~\chi^{(C)}_{p_n,\sigma,q}(t^{'})~\mu^3~~~~~~~~~~~~~~~~~ &
                                                                                 \mbox{\small {\textcolor{black}{\bf for Case A}}}  
                                                                                \\ 
                                \displaystyle\int dt^{'}~\chi^{(C)}_{p_n,\sigma,q}(t^{'})~\frac{m^2_{eff}f_a(t^{'})}{b}~~~~~~~~ & \mbox{\small {\textcolor{black}{\bf for Case B}}}.~~~~~~~~
                                                                                          \end{array}
                                                                                \right.\\
         {\cal N}_{p}&=&2\sinh \pi p ~\sqrt{{\cal N}_{p\sigma}}=4\sinh \pi p ~\sqrt{ \frac{\left[\cosh\pi p-\sigma\cos\left(\nu-\frac{1}{2}\right)\right]}{\pi|\Gamma\left(\nu+\frac{1}{2}+ip\right)|^2}}~~~~~~~~~~~~\\
         {\cal N}_{p_n}&=&2\sinh \pi p_n ~\sqrt{{\cal N}_{p_n\sigma}}=4\sinh \pi p_n ~\sqrt{ \frac{\left[\cosh\pi p_n-\sigma\cos\left(\nu-\frac{1}{2}\right)\right]}{\pi|\Gamma\left(\nu+\frac{1}{2}+ip_n\right)|^2}}.~~~~~~~~~~~~\eea                                                                      
  In equation~(\ref{cxcxx}),  we introduce few coefficients,  which are defined as:
   \bea 
   \label{dex1}
   \alpha^{\sigma}_{\bf R}&=&\frac{1}{\sigma}\alpha^{\sigma}_{\bf L}=\frac{\left(e^{\pi p}-i\sigma e^{-i\pi\nu}\right)}{\Gamma\left(\nu+\frac{1}{2}+ip\right)},\alpha^{\sigma}_{{\bf R},n}=\frac{1}{\sigma}\alpha^{\sigma}_{{\bf L},n}=\frac{\left(e^{\pi p_n}-i\sigma e^{-i\pi\nu}\right)}{\Gamma\left(\nu+\frac{1}{2}+ip_n\right)}\\
                  \beta^{\sigma}_{\bf R}&=& \frac{1}{\sigma}\beta^{\sigma}_{\bf L}= -\frac{\left(e^{-\pi p}-i\sigma e^{-i\pi\nu}\right)}{\Gamma\left(\nu+\frac{1}{2}-ip\right)}., \beta^{\sigma}_{{\bf R},n}= \frac{1}{\sigma}\beta^{\sigma}_{{\bf L},n}= -\frac{\left(e^{-\pi p_n}-i\sigma e^{-i\pi\nu}\right)}{\Gamma\left(\nu+\frac{1}{2}-ip_n\right)}.   ~~~~~~~~
 \eea 
                                                                                                                                      
 Further equation~(\ref{cxcxx}) can be recast in matrix for as:
 \bea\label{xmatrix} \boxed{\boxed{{\bf \chi}^{I}=\underbrace{\frac{1}{{\cal N}_p}{\cal  M}^{I}_{J}{\cal P}^{J}}_{\bf Complementary}+\underbrace{\sum^{\infty}_{n=0}\frac{1}{{\cal N}_{p,(n)}}\left({\cal  M}_{(n)}\right)^{I}_{J}{\cal P}^{J}_{(n)}}_{\bf Particular}}}\eea 
 where we define two square matrices for the complementary and particular part as:
  \bea {\cal M}^{I}_{J}&=&\left(\begin{array}{ccc} \alpha^{\sigma}_{q} &~~~ \beta^{\sigma}_{q} \\ \beta^{\sigma^{*}}_{q} &~~~ \alpha^{\sigma^{*}}_{q}  \end{array}\right),~~
  \left({\cal M}_{(n)}\right)^{I}_{J}=\left(\begin{array}{ccc} \bar{\alpha}^{\sigma}_{q,n} &~~~ \bar{\beta}^{\sigma}_{q,n} \\ \bar{\beta}^{\sigma^{*}}_{q,n} &~~~ \bar{\alpha}^{\sigma^{*}}_{q,n}  \end{array}\right),\\
  {\cal P}^{J}_{(n)}&=&\left(\begin{array}{ccc} {\cal P}^{q,n} \\ {\cal P}^{{q^*},n}\\
          \end{array}\right),~~
  \chi^{I}=\left(\begin{array}{ccc} \chi_{\sigma}(t) \\ \chi^{*}_{\sigma}(t),
   \end{array}\right),~~
   {\cal P}^{J}=\left(\begin{array}{ccc} {\cal P}^{q} \\ {\cal P}^{{q^*}},\\
      \end{array}\right).~~~~~~~~~~\eea
      Here $(I,J)=1,2,3,4.$  Also we use ${\cal N}_{p,(n)}$, which is defined by:
\bea {\cal N}_{p,(n)}&=&2\sinh \pi p_n ~\sqrt{{\cal N}_{p_n\sigma}} ~\left(p^2-p^2_n\right).~~~~~\eea
Hence the Bunch-Davies mode function can be expressed as:  
          \bea   \frac{H}{\sinh t}a_{I}\chi^{I}=\frac{H}{\sinh t}a_{I}\left[\frac{1}{{\cal N}_p}{\cal  M}^{I}_{J}{\cal P}^{J}+\sum^{\infty}_{n=0}\frac{1}{{\cal N}_{p,(n)}}\left({\cal  M}_{(n)}\right)^{I}_{J}{\cal P}^{J}_{(n)} \right],~~{\rm where}~~ a_{I}=(a_{\sigma},
           a^{\dagger}_{\sigma}).~~~~~~~~
           \eea                                  
Here we define:
           \bea\label{def1} b_{J} &=& a^{(C)}_{I}{\cal M}^{I}_{J},~~~ b_{J(n)} = a^{(P)}_{I(n)}\left({{\cal M}_{(n)}}\right)^{I}_{J},~~{\rm where}~~ a^{(C)}_{I}=(a^{(C)}_{\sigma},
                      a^{(C)\dagger}_{\sigma}), a^{(P)}_{I(n)}=(a^{(P)}_{\sigma,n},a^{(P)\dagger}_{\sigma,n}).~~~~~~~~~~~
           \eea
We use the operator ansatz:
 \bea a_{I}&=& \left[a^{(c)}_{I}+\sum^{\infty}_{n=0}a^{(p)}_{I(n)}\right], a^{(c)}_{I} = b_{J}\left({\cal M}^{-1}\right)^{I}_{J},~~~ a^{(p)}_{I(n)} = b_{J(n)}\left({\cal M}^{-1}_{(n)}\right)^{I}_{J},\eea 
  where inverse matrices are defined as:
   \bea \left({\cal M}^{-1}\right)^{I}_{J}&=&\left(\begin{array}{ccc} \gamma_{\sigma q} &~~~ \delta_{\sigma q} \\ \delta^{*}_{\sigma q} &~~~ \gamma^{*}_{\sigma q}  \end{array}\right),
        ~~~~\left({\cal M}^{-1}_{(n)}\right)^{I}_{J}=\left(\begin{array}{ccc}\overline{\gamma}_{\sigma q,n} &~~~ \overline{\delta}_{\sigma q,n} \\ \overline{\delta}^{*}_{\sigma q,n} &~~~\overline{\gamma}^{*}_{\sigma q,n}  \end{array}\right),\eea
  where,
           \bea 
           \label{r1}\gamma_{j\sigma}&=& \displaystyle     \displaystyle \frac{\Gamma\left(\nu+\frac{1}{2}+ip\right)~e^{\pi p+i\pi\left(\nu+\frac{1}{2}\right)}}{4\sinh\pi p}\left(\begin{array}{ccc} \frac{1}{\displaystyle e^{\pi p+i\pi \left(\nu+\frac{1}{2}\right)}+1} &~~~ \frac{1}{\displaystyle e^{\pi p+i\pi \left(\nu+\frac{1}{2}\right)}-1} \\ \frac{1}{\displaystyle e^{\pi p+i\pi \left(\nu+\frac{1}{2}\right)}+1} &~~~ -\frac{1}{\displaystyle e^{\pi p+i\pi \left(\nu+\frac{1}{2}\right)}-1}  \end{array}\right)~~~~~~~ \\
         \label{r2}\delta^{*}_{j\sigma}&=&
                                                                               \displaystyle \frac{\Gamma\left(\nu+\frac{1}{2}-ip\right)~e^{i\pi\left(\nu+\frac{1}{2}\right)}}{4\sinh\pi p}\left(\begin{array}{ccc} \frac{1}{\displaystyle e^{\pi p}+e^{i\pi \left(\nu+\frac{1}{2}\right)}} &~~~ -\frac{1}{\displaystyle e^{\pi p}-e^{i\pi \left(\nu+\frac{1}{2}\right)}} \\  \frac{1}{\displaystyle e^{\pi p}+e^{i\pi \left(\nu+\frac{1}{2}\right)}} &~~~ \frac{1}{\displaystyle e^{\pi p}-e^{\pi p+i\pi \left(\nu+\frac{1}{2}\right)}}  \end{array}\right)~~~~~~~     \\
              \label{r3}  \overline{\gamma}_{j\sigma,n}&=& \displaystyle \frac{\Gamma\left(\nu+\frac{1}{2}+ip_n\right)~e^{\pi p_n+i\pi\left(\nu+\frac{1}{2}\right)}}{4\sinh\pi p_n}\left(\begin{array}{ccc} \frac{1}{\displaystyle e^{\pi p_n+i\pi \left(\nu+\frac{1}{2}\right)}+1} &~~~ \frac{1}{\displaystyle e^{\pi p_n+i\pi \left(\nu+\frac{1}{2}\right)}-1} \\ \frac{1}{\displaystyle e^{\pi p_n+i\pi \left(\nu+\frac{1}{2}\right)}+1} &~~~ -\frac{1}{\displaystyle e^{\pi p_n+i\pi \left(\nu+\frac{1}{2}\right)}-1}  \end{array}\right)~~ \\
                   \label{r4}       \overline{\delta}^{*}_{j\sigma,n}&=& \displaystyle \frac{\Gamma\left(\nu+\frac{1}{2}-ip_n\right)~e^{i\pi\left(\nu+\frac{1}{2}\right)}}{4\sinh\pi p_n}\left(\begin{array}{ccc} \frac{1}{\displaystyle e^{\pi p_n}+e^{i\pi \left(\nu+\frac{1}{2}\right)}} &~~~ -\frac{1}{\displaystyle e^{\pi p_n}-e^{i\pi \left(\nu+\frac{1}{2}\right)}} \\  \frac{1}{\displaystyle e^{\pi p_n}+e^{i\pi \left(\nu+\frac{1}{2}\right)}} &~~~ \frac{1}{\displaystyle e^{\pi p_n}-e^{\pi p_n+i\pi \left(\nu+\frac{1}{2}\right)}}  \end{array}\right)~~~~~~~   \eea 
                   We also use the following constraints:
 \bea a^{(C)}_{I}\underbrace{\left[\sum^{\infty}_{n=0}\frac{1}{{\cal N}_{p,(n)}}\left({\cal  M}_{(n)}\right)^{I}_{J}{\cal P}^{J}_{(n)}\right]}_{\bf Particular}= 0,~~~~~~~ 
 a^{(P)}_{I(n)}\underbrace{\left[\frac{1}{{\cal N}_p}{\cal  M}^{I}_{J}{\cal P}^{J}\right]}_{\bf Complementary}=0.\eea
Here the annihilation and creation operators are defined as:
  \bea\label{vb1} a_{\sigma}&=&\sum_{q={\bf R},{\bf L}}\left\{\left[\gamma_{q\sigma}b_{q}+\delta^{*}_{q\sigma}b^{\dagger}_{q}\right]+\sum^{\infty}_{n=0}\left[\overline{\gamma}_{q\sigma,n}\overline{b}_{q,n}+\overline{\delta}^{*}_{q\sigma,n}\overline{b}^{\dagger}_{q,n}\right]\right\}\forall \sigma=\pm 1,\\ \label{vb2}  a^{\dagger}_{\sigma}&=&\sum_{q={\bf R},{\bf L}}\left\{\left[\gamma^{*}_{q\sigma}b^{\dagger}_{q}+\delta_{q\sigma}b_{q}\right]+\sum^{\infty}_{n=0}\left[\overline{\gamma}^{*}_{q\sigma,n}\overline{b}^{\dagger}_{q,n}+\overline{\delta}_{q\sigma,n}\overline{b}_{q,n}\right]\right\}\forall \sigma=\pm 1.~~~~~~~~~~\eea   
 Now using Bogoliubov transformation the Bunch-Davies quantum vacuum can be written in terms of the direct product of ${\bf R}$ and ${\bf L}$ vacua can be written as:
  \bea|{\bf BD}\rangle &=&\exp\left(\widehat{\cal K}\right)~\bigg(|{\bf R}\rangle \otimes|{\bf L}\rangle\bigg),\eea 
  where Bogoliubov operator $\hat{\cal K}$ is expressed as:
  \bea \widehat{\cal K}&=&\Bigg(\underbrace{\frac{1}{2}\sum_{i,j={\bf R},{\bf L}}m_{ij}~b^{\dagger}_{i}~b^{\dagger}_{j}}_{\bf Complementary}+\underbrace{\frac{1}{2}\sum_{i,j={\bf R},{\bf L}}\sum^{\infty}_{n=0}\overline{m}_{ij,n}~\overline{b}^{\dagger}_{i,n}~\overline{b}^{\dagger}_{j,n}}_{\bf Particular~integral}\Bigg),~~~~~~~\eea
  where we determine the coefficients $m_{ij}$ and $\bar{m}_{ij,n}$. Also we define:
  \bea   |{\bf R}\rangle&=&\bigg(|{\bf R}\rangle_{(C)}+\sum^{\infty}_{n=0}|{\bf R}\rangle_{(P),n}\bigg),~~~  |{\bf L}\rangle=\bigg( |{\bf L}\rangle_{(C)}+\sum^{\infty}_{n=0}|{\bf L}\rangle_{(P),n}\bigg),
  \eea 
where,
      \bea b_{{\bf L}}|{\bf L}\rangle_{(C)}&=&0,b_{\bf R}|{\bf R}\rangle_{(C)}=0,
      \overline{b}_{{\bf L},n}|{\bf L}\rangle_{(P)}= 0, \overline{b}_{{\bf R},n}|{\bf R}\rangle_{(P)}=0.\eea
     Also we have:
      \bea 
      \left[ b_i,b^{\dagger}_j\right]&=&\delta_{ij},~~~~ \left[ b_i,b_j\right]=0=
      \left[ b^{\dagger}_i,b^{\dagger}_j\right].~~~~~~~~~~~~~~~\\ 
            \left[ \overline{b}_{i,n},\overline{b}^{\dagger}_{j,m}\right]&=&\delta_{ij}{\delta}_{nm},~~~~\left[ \overline{b}_{i,n},\overline{b}_{j,m}\right]= 0=
            \left[ \overline{b}^{\dagger}_{i,m},\overline{b}^{\dagger}_{j,m}\right].~~~~~~~~~~~
            \eea
            This implies:
   \bea \Bigg(\underbrace{\left(m_{ij}\gamma_{j\sigma}+\delta^{*}_{i\sigma}\right)b^{\dagger}_{i}}_{\bf Complementary}+\underbrace{\sum^{\infty}_{n=0}\left(\overline{m}_{ij,n}\overline{\gamma}_{j\sigma,n}+\overline{\delta}^{*}_{i\sigma,n}\right)\overline{b}^{\dagger}_{i,n}}_{\bf Particular~integral}\Bigg)\left(|{\bf R}\rangle\otimes |{\bf L}\rangle\right)&=& 0,\eea
    which implies:
         \bea \left(m_{ij}\gamma_{j\sigma}+\delta^{*}_{i\sigma}\right)&=& 0, \left(\bar{m}_{ij,n}\bar{\gamma}_{j\sigma,n}+\bar{\delta}^{*}_{i\sigma,n}\right)= 0~~~~\forall n,\eea
         using which we define the following mass matrices,  which are given by: 
           \bea  m_{ij}&=& -\delta^{*}_{i\sigma}\left(\gamma^{-1}\right)_{\sigma j}
\displaystyle\approx \frac{e^{i\theta}\sqrt{2}~e^{-p\pi}}{\sqrt{\cosh 2\pi p+\cosh 2\pi \nu}}\left(\begin{array}{ccc} \cos\pi\nu &~~~ i\sinh p\pi \\ i\sinh p\pi &~~~ \cos\pi\nu  \end{array}\right), \\
   \bar{m}_{ij,n}&=&-\overline{\delta}^{*}_{i\sigma,n}\left(\overline{\gamma}^{-1}\right)_{\sigma j,n}\displaystyle\approx \frac{e^{i\theta}\sqrt{2}~e^{-p_n\pi}}{\sqrt{\cosh 2\pi p_n+\cosh 2\pi \nu}}\left(\begin{array}{ccc} \cos\pi\nu &~~~ i\sinh p_n\pi \\ i\sinh p_n\pi &~~~ \cos\pi\nu  \end{array}\right).   
       \eea 
  The eigenvalues of the the matrices are given by:
\bea
  \lambda_{\pm}&=&e^{i\theta}\frac{\sqrt{2}~e^{-p\pi}\left(\cos\pi\nu \pm i\sinh p\pi\right)}{\sqrt{\cosh 2\pi p+\cos 2\pi \nu}},
  \lambda_{\pm,n}= e^{i\theta}\frac{\sqrt{2}~e^{-p_n\pi}\left(\cos\pi\nu \pm i\sinh p_n\pi\right)}{\sqrt{\cosh 2\pi p_n+\cos 2\pi \nu}}.\quad\eea
  However, the ${\bf R}$ and ${\bf L}$ basis is unsuitable for this present calculation. To find a suitable basis another Bogoliubov transformation needs to be performed:
 \bea 
   c_{\bf R}&=& \bigg(u~b_{\bf R}+v~b^{\dagger}_{\bf R}\bigg),~~~C_{{\bf R},n}= \bigg(U_n~b_{{\bf R},n}+V_n~b^{\dagger}_{{\bf R},n}\bigg).\\
c_{\bf L}&=&\bigg( \overline{u}~b_{\bf L}+\overline{v}~b^{\dagger}_{\bf L}\bigg),~~~C_{{\bf L},n}= \bigg(\overline{U}_n~b_{{\bf L},n}+\overline{V}_n~b^{\dagger}_{{\bf L},n}\bigg),\eea   
   which satisfy:
        \bea 
          \bigg( |u|^2-|v|^2\bigg)&=& 1,~~~\bigg(|U_n|^2-|V_n|^2\bigg)= 1.\\  
                     \bigg(|\bar{u}|^2-|\bar{v}|^2\bigg)&=& 1 ,~~~\bigg(|\bar{U}_n|^2-|\bar{V}_n|^2\bigg)= 1.\eea  
                      In this new  basis Bunch-Davies vacuum state can rewritten as:
 \bea |{\bf BD}\rangle &=&\sqrt{\left[1-\left(|\gamma_p|^2+\sum^{\infty}_{n=0}|\Gamma_{p,n}|^2\right)\right]}\exp\left(\widehat{\cal W}\right)\left(|{\bf R}^{'}\rangle\otimes |{\bf L}^{'}\rangle\right),\eea  
 where $|{\bf R}^{'}\rangle$ and $ |{\bf L}^{'}\rangle$ are operators.  Now, we introduce a new operator:
   \bea  \widehat{\cal W}&=&\bigg(\underbrace{\gamma_{p}~c^{\dagger}_{\bf R}~c^{\dagger}_{\bf L}}_{\bf Complementary}+\underbrace{\sum^{\infty}_{n=0}\Gamma_{p,n}~C^{\dagger}_{{\bf R},n}~C^{\dagger}_{{\bf L},n}}_{\bf Particular~integral}\bigg).\quad\eea
 Here we have new sets of algebra: 
           \bea 
           \left[ c_i,c^{\dagger}_j\right]&=&\delta_{ij},~~~~ \left[ c_i,c_j\right]=0=
           \left[ c^{\dagger}_i,c^{\dagger}_j\right].~~~~~~~~~~~~~~~\\
                 \left[ C_{i,n},C^{\dagger}_{j,m}\right]&=&\delta_{ij}{\delta}_{nm},~~~~\left[ C_{i,n},C_{j,m}\right]= 0=
                 \left[ C^{\dagger}_{i,m},C^{\dagger}_{j,m}\right].~~~~~~~~~~~~~~~
                 \eea
   Also, these operators are described as:
      \bea  c_{\bf R}|{\bf BD}\rangle &=&\gamma_{p}~c^{\dagger}_{\bf L}|{\bf BD}\rangle,~~~
     c_{\bf L}|{\bf BD}\rangle =\gamma_{p}~c^{\dagger}_{\bf R}|{\bf BD}\rangle,\\ 
   C_{{\bf R},n}|{\bf BD}\rangle &=&\Gamma_{p,n}~C^{\dagger}_{{\bf L},n}|{\bf BD}\rangle,~~~
          C_{{\bf L},n}|{\bf BD}\rangle =\Gamma_{p,n}~C^{\dagger}_{{\bf R},n}|{\bf BD}\rangle.\eea
   Here we have:
            \bea c_{J}&=& b_{I}{\cal G}^{I}_{J},~
            C_{J(n)}=\bar{b}_{J(n)}\left({\cal G}_{(n)}\right)^{I}_{J}~{\rm where}~~{\cal G}^{I}_{J}=\left(\begin{array}{ccc} U_q &~~~ V^{*}_q \\ V_q &~~~ U^{*}_q  \end{array}\right)
                    ,~
                    \left({\cal G}_{(n)}\right)^{I}_{J}=\left(\begin{array}{ccc} \overline{U}_{ q,n} &~~~ \overline{V}^{*}_{\sigma q,n} \\ \overline{V}_{ q,n} &~~~ \overline{U}^{*}_{ q,n}  \end{array}\right),\quad\quad\quad\quad\eea
where,
                    \bea U_q &\equiv& {\rm \bf diag}\left(u,\overline{u}\right),~~V_q \equiv {\rm \bf diag}\left(v,\overline{v}\right),~~ \overline{U}_{q,n} \equiv {\rm \bf diag}\left(U_n,\overline{U}_n\right),~~\overline{V}_{q,n} \equiv {\rm \bf diag}\left(V_n,\overline{V}_n\right).~~\quad\quad\quad\eea
           Finally, we derive: 
  \bea
  m_{\bf RR}u+v-\gamma_{p} m_{\bf RL}\overline{v}^{*}&=& 0,\\
   m_{\bf RR}\overline{u}+\overline{v}-\gamma_{p} m_{\bf RL}v^{*}&=& 0,\\ 
   m_{\bf RL}u-\gamma_{p} \overline{u}^{*}-\gamma_{p}m_{\bf RR}\overline{v}^{*}&=& 0,\\
   m_{\bf RL}\overline{u}-\gamma_{p} u^{*}-\gamma_{p}m_{\bf RR}v^{*}&=& 0,\\
    \bar{m}_{{\bf RR},n}U_n+V_n-\Gamma_{p,n}\overline{m}_{{\bf RL},n}\overline{V}^{*}_n&=& 0,\\
      \overline{m}_{{\bf RR},n}\overline{U}_n+\overline{V}_n-\Gamma_{p,n} \overline{m}_{{\bf RL},n}V^{*}_n&=& 0,\\ 
       \overline{m}_{{\bf RL},n}U_n-\Gamma_{p,n} \overline{U}^{*}_n-\Gamma_{p,n} \overline{m}_{{\bf RR},n}\overline{V}^{*}_n&=& 0,\\
       \overline{m}_{{\bf RL},n}\overline{U}_n-\Gamma_{p,n}  U^{*}_n-\Gamma_{p,n} \overline{m}_{{\bf RR},n}V^{*}_n&=& 0,~~~~~~~~~~\eea
       Here we have:
         \bea m_{\bf RR}&=& m_{\bf LL}=m^{*}_{\bf RR}=\omega=\frac{\sqrt{2}~e^{-p\pi}\cos\pi\nu}{\sqrt{\cosh 2\pi p+\cos 2\pi \nu}},\\
       m_{\bf RL}&=& m_{\bf LR}=-m^{*}_{\bf RL}=\zeta=e^{i\frac{\pi}{2}}\frac{\sqrt{2}~e^{-p\pi}\sinh p\pi}{\sqrt{\cosh 2\pi p+\cos 2\pi \nu}},\\  \bar{m}_{{\bf RR},n}&=& \bar{m}_{{\bf LL},n}=\bar{m}^{*}_{{\bf RR},n}=\omega_n=\frac{\sqrt{2}~e^{-p_n\pi}\cos\pi\nu}{\sqrt{\cosh 2\pi p_n+\cos 2\pi \nu}},\\
                     \bar{m}_{{\bf RL},n}&=& \bar{m}_{{\bf LR},n}=-\bar{m}^{*}_{{\bf RL},n}=\zeta_n=e^{i\frac{\pi}{2}}\frac{\sqrt{2}~e^{-p_n\pi}\sinh p_n\pi}{\sqrt{\cosh 2\pi p_n+\cos 2\pi \nu}}.\eea
 Having the constraints, $\gamma^{*}_{p}=-\gamma_{p},
   \Gamma^{*}_{p,n}=-\Gamma_{p,n}$,
we have:
\bea v^{*}=\overline{v},~~
   u^{*}=\overline{u},~~ V^{*}_n=\overline{V}_n,~~ 
      U^{*}_n=\overline{U}_n. \eea
      This further satisfy:  
\bea  \bigg(|u|^2-|v|^2\bigg)=1, \quad\quad \bigg(|U_n|^2-|V_n|^2\bigg)=1.\eea 
 Finally, we have:
     \bea
         \gamma_{p}&=&i\frac{\sqrt{2}}{\sqrt{\cosh 2\pi p +\cos 2\pi \nu}+\sqrt{\cosh 2\pi p +\cos 2\pi \nu+2}}.~~~~~~~~~~~\\
          \Gamma_{p,n}&=&i\frac{\sqrt{2}}{\sqrt{\cosh 2\pi p_n +\cos 2\pi \nu}+ \sqrt{\cosh 2\pi p_n +\cos 2\pi \nu+2}}.~~~~~~~~~~\eea   
          Here we have,  $|\gamma_p|<1$ and  $|\Gamma_{p,n}|<1$.  Also we have:
 \bea  \bigg(|\overline{u}|^2-|\overline{v}|^2\bigg)=1, \quad\quad \bigg(|\overline{U}_n|^2-|\overline{V}_n|^2\bigg)=1.\eea 
 where the general solutions are given by:
\bea \overline{u}&=&\frac{1-\gamma_p \zeta}{\sqrt{|1-\gamma_p \zeta|^{2}-|\omega|^{2}}}=u^{*}=u,\quad\quad\overline{v}=\frac{\omega}{\sqrt{|1-\gamma_p \zeta|^{2}-|\omega|^{2}}}=v^{*}=v,\\
 \overline{U}_n&=&\frac{1-\Gamma_{p_n} \zeta_n}{\sqrt{|1-\Gamma_{p_n} \zeta_n|^{2}-|\omega|^{2}}}={U}^{*}_n={U}_n\quad\quad\overline{V}_n=\frac{\omega_n}{\sqrt{|1-\Gamma_{p_n} \zeta_n|^{2}-|\omega|^{2}}}={V}^{*}_n={V}_n.\quad\quad\quad\eea
  Here we have:
  \bea \omega^{*}=\omega,\quad\quad \zeta^{*}=-\zeta,\quad\quad \gamma^{*}_p=-\gamma_p,\quad\quad\Gamma^{*}_{p,n}=-\Gamma_{p,n}.\eea

     \section{Entanglement negativity in Axiverse}\label{ka3}

In terms of the quantum numbers $p$, $l$, and $m$, the complementary and specific integral sections of the solution may be used to factor the Bunch Davies quantum vacuum state as follows:
\bea |{\bf BD}\rangle &=&\Bigg\{\sqrt{\frac{\left(1-|\gamma_p|^2\right)}{\left(1+f_p\right)}}\sum^{\infty}_{k=0}|\gamma_{p}|^{k}\bigg(|k;p,l,m\rangle_{{\bf R}^{'}}\otimes|k;p,l,m\rangle_{{\bf L}^{'}}\bigg)\nonumber\\
        &&\quad\quad\quad\quad\quad\quad+\frac{f_{p}}{\sqrt{\left(1+f_p\right)}}\sum^{\infty}_{n=0}\sum^{\infty}_{r=0}|\Gamma_{p,n}|^{r}\bigg(|n,r;p,l,m\rangle_{{\bf R}^{'}}\otimes| n,r;p,l,m\rangle_{{\bf L}^{'}}\bigg)\Bigg\},~~~~~~~~~\quad\eea
        where, the factor $f_{p}$ is defined as:
        \bea f^{-1}_p&=&\left(\displaystyle \sum^{\infty}_{n=0}\frac{1}{1-|\Gamma_{p,n}|^2}\right).\eea  
     The expression for the eigenvalues, which is given by the following equation, may be immediately computed by applying the fundamental physical idea of Schmidt decomposition for a pure quantum state, which was covered in the preceding section of this study:
                    \bea \sqrt{\lambda_k}&=&\Bigg\{\sqrt{\frac{\left(1-|\gamma_p|^2\right)}{\left(1+f_p\right)}}|\gamma_{p}|^{k}+\frac{f_{p}}{\sqrt{\left(1+f_p\right)}}\sum^{\infty}_{n=0}|\Gamma_{p,n}|^{k}\Bigg\}\quad\quad\forall k=[0,\infty].\eea
 Then the logarithmic negativity can be expressed as:
 \bea {\cal L}{\cal N}(p,\nu)&=&2\ln\Bigg(\sum^{\infty}_{k=0}\lambda_k\Bigg)\nonumber\\
 &=&2\ln\Bigg(\sum^{\infty}_{k=0}\Bigg\{\sqrt{\frac{\left(1-|\gamma_p|^2\right)}{\left(1+f_p\right)}}|\gamma_{p}|^{k}+\frac{f_{p}}{\sqrt{\left(1+f_p\right)}}\sum^{\infty}_{n=0}|\Gamma_{p,n}|^{k}\Bigg\}\Bigg)\nonumber\\
 &=&\ln\Bigg(\frac{1}{\left(1+f_p\right)}\Bigg\{\sqrt{\frac{\left(1+|\gamma_p|\right)}{\left(1-|\gamma_p|\right)}}+\frac{f_p}{\overline{f}_p}\Bigg\}^{2}\Bigg),\eea                where $\overline{f}_p$ is defined as:
        \bea \overline{f}^{-1}_p&=&\displaystyle \sum^{\infty}_{n=0}\frac{1}{1-|\Gamma_{p,n}|}.\eea                          Here we use:
\bea \sum^{\infty}_{k=0}|\gamma_{p}|^{k}&=&\frac{1}{\left(1-|\gamma_p|\right)},\quad
\sum^{\infty}_{k=0}|\Gamma_{p,n}|^{k}=\frac{1}{\left(1-|\Gamma_{p,n}|\right)}.\eea
Now,  ultimately we have:
\bea {\cal N}(p,\nu)&=&\frac{1}{2}\Bigg(\exp\left({\cal L}{\cal N}(p,\nu)\right)-1\Bigg)=\frac{1}{2}\Bigg(\frac{1}{\left(1+f_p\right)}\Bigg\{\sqrt{\frac{\left(1+|\gamma_p|\right)}{\left(1-|\gamma_p|\right)}}+\frac{f_p}{\overline{f}_p}\Bigg\}^{2}-1\Bigg).\eea
  Under the current framework, the two causally unrelated areas, {\bf R} and {\bf L}, are expected to be quantum mechanically entangled with one another for any finite values of $p$. This is due to the non-vanishing contribution from both $|\gamma_{p}|$ and $|\Gamma_{p,n}|$. 
  
After integrating over $p$ and accounting for the density of quantum mechanical states under consideration in the open chart, we get the following formula for the logarithmic negativity in the volume of a hyperboloid:
  \bea\label{eeq1}  {\cal L}{\cal N}(\nu)&=&V^{\bf reg}_{\bf H^3}\int^{\infty}_{0}~dp~{\cal D}(p)~ {\cal L}{\cal N}(p,\nu),\eea   
  where ${\cal D}(p)=p^2/2\pi^2$ is the density of quantum states. Also, $V^{\bf reg}_{\bf H^3}=V_{\bf S^2}/2=4\pi/2=2\pi$.
    \begin{figure*}[htb]
    \centering
    \subfigure[For $f_p=0$.]{
        \includegraphics[width=14.2cm,height=8.5cm] {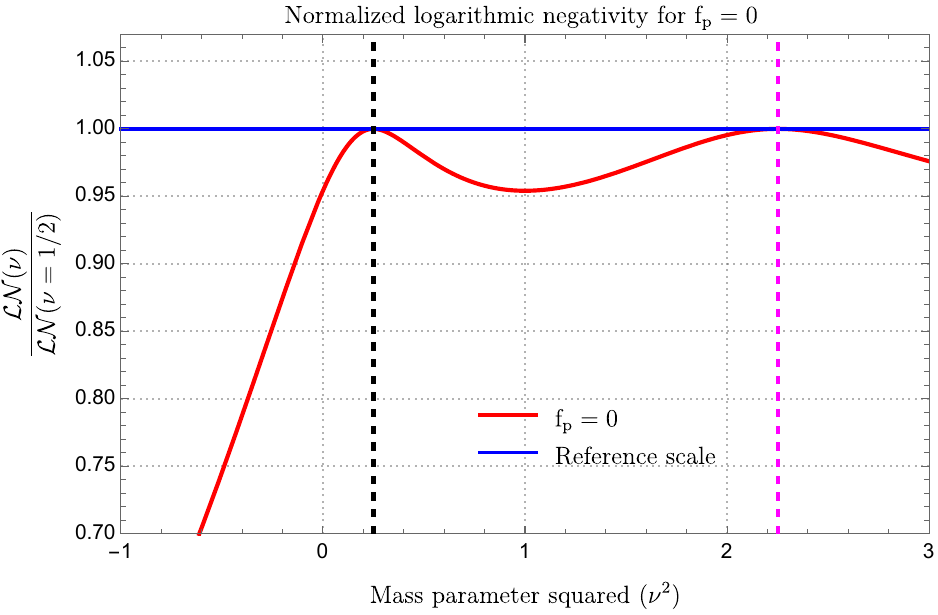}
        \label{L1}
    }
    \subfigure[For small $f_p\neq 0$.]{
        \includegraphics[width=14.2cm,height=8.5cm] {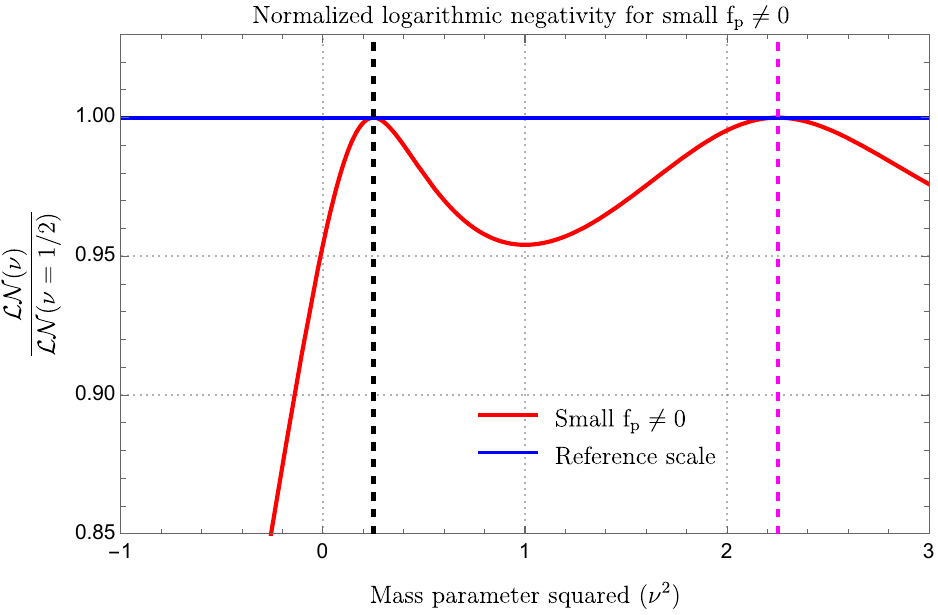}
        \label{L2}
       }
    \caption[Optional caption for list of figures]{ Logarithmic negativity (${\cal LN}(\nu)/{\cal LN}(\nu=1/2)$) with mass parameter squared ($\nu^2$) for both $f_p=0$ and small $f_p\neq 0$.  } 
    \label{LN}
    \end{figure*}  
    \begin{figure*}[htb]
    \centering
    \subfigure[For $f_p=0$.]{
        \includegraphics[width=14.2cm,height=8.5cm] {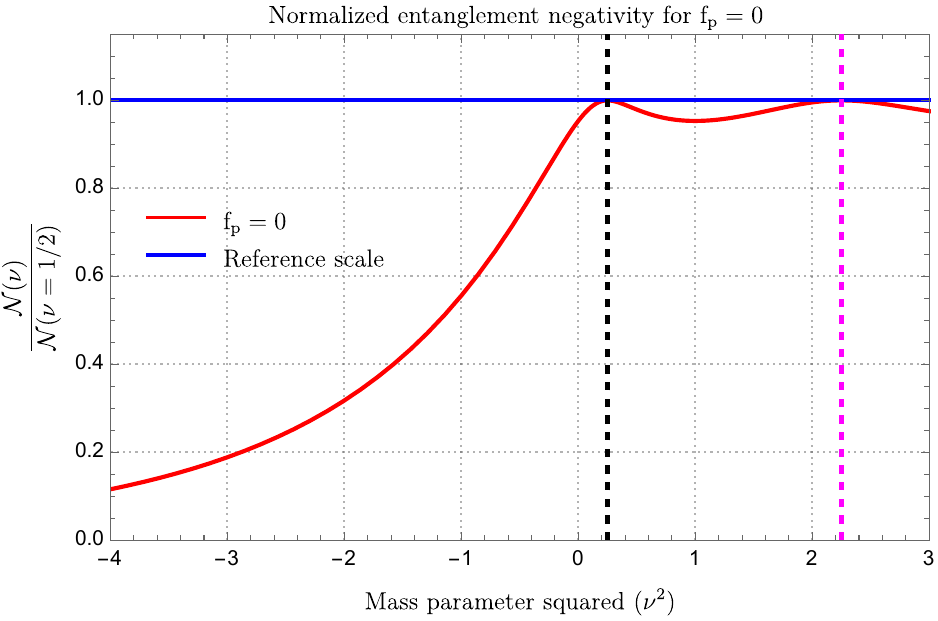}
        \label{N1}
    }
    \subfigure[For small $f_p\neq 0$.]{
        \includegraphics[width=14.2cm,height=8.5cm] {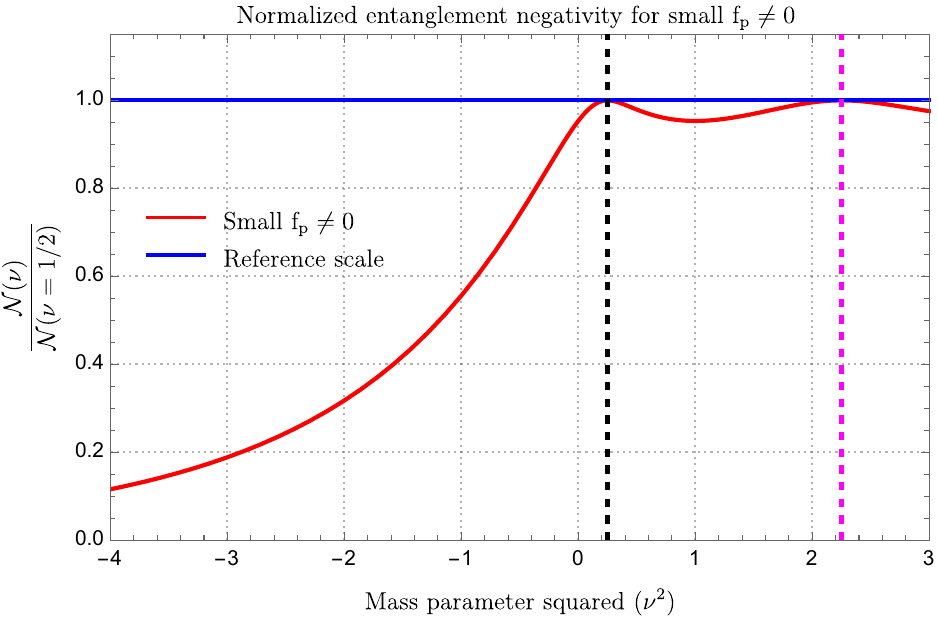}
        \label{N2}
       }
    \caption[Optional caption for list of figures]{Entanglement negativity (${\cal N}(\nu)/{\cal N}(\nu=1/2)$) with mass parameter squared ($\nu^2$) for both $f_p=0$ and small $f_p\neq 0$. } 
    \label{N}
    \end{figure*}  
  
  Consequently,  we have:     
  \bea\label{eexq1}  {\cal L}{\cal N}(\nu)&=&\frac{1}{\pi}\int^{\infty}_{0}~dp~p^2~ {\cal L}{\cal N}(p,\nu).\eea
  However, in most physical problems, the integrand diverges at the top limit of the aforementioned integration.  For this reason, one needs to incorporate a regulator $\Lambda$ as the upper limit of the integration instead of strictly declaring it to be infinity.  However, for computational purposes, we set the value of $\Lambda$ to be a huge integer.  In this case, the cut-off is physically regarded as the Ultra Violet (UV) cut-off. In quantum field theory, the UV cut-off is sometimes physically understood as a lattice regulator for the kind of computing performed in this research.  On the other hand, it is worth noting that in the majority of relevant physical situations, the integrand converges at the lower limit of integration. In technical terms, this lower limit corresponds to Infra Red (IR), which is safe for the specific situation we are addressing in this work. Then in the regulated version we have:
      \bea\label{eexq2}  {\cal L}{\cal N}(\nu)&=&\frac{1}{\pi}\int^{\Lambda}_{0}~dp~p^2~\ln\Bigg(\frac{1}{\left(1+f_p\right)}\Bigg\{\sqrt{\frac{\left(1+|\gamma_p|\right)}{\left(1-|\gamma_p|\right)}}+\frac{f_p}{\overline{f}_p}\Bigg\}^{2}\Bigg).\eea

In figure (\ref{LN}(a)) and (\ref{LN}(b)),   we have shown the normalized version of logarithmic negativity (${\cal LN}(\nu)/{\cal LN}(\nu=1/2)$) with mass parameter squared ($\nu^2$) for both $f_p=0$ and small $f_p\neq 0$.  Similarly,  in figure (\ref{N}(a)) and (\ref{N}(b)),   we have shown the normalized version of entanglement negativity (${\cal N}(\nu)/{\cal N}(\nu=1/2)$) with mass parameter squared ($\nu^2$) for both $f_p=0$ and small $f_p\neq 0$.  Vertical lines are shown for $\nu=1/2$ and $\nu=3/2$.

 \section{Preparing maximal entangled states in Axiverse}   
    \label{ka4}
  
   \subsection{Constructing ground state}
 Bogoliubov transformations connect the solutions in the Bunch-Davies vacuum to those in the {\bf R} and {\bf L} vacua. This transformation allows us to determine that a two-mode compressed state in the open charts corresponds to the ground state of a particular mode as observed by an observer in the global chart. The fields shown in the {\bf R} and {\bf L} charts match these two modes. We must trace over the inaccessible area as we cannot access the modes in the causally disconnected {\bf R} region if we merely explore one of the open charts, let's say {\bf L}.  In the sense that the global chart and Minkowski chart cover the entire spacetime geometry, whereas open charts and Rindler charts only cover a portion of the spacetime geometry and thus exist horizons, this situation is comparable to the relationship between an observer in a Minkowski chart and another in one of the two Rindler charts in flat space.

We start by examining the effects of entanglement negativity between two causally unrelated patches of open chart, which is actually represented by the product of the quantum vacuum states for each oscillator. For subsequent computations, it is important to keep in mind that each oscillator's quantum mechanical state is determined by one of the three quantum numbers $p$, $l$, or $m$. Because of this, the final expression of total quantum must take the product over $p$. In this setup, the ultimate Bunch Davies quantum vacuum state is expressed as:
\bea  |0\rangle_{\bf BD} &=&\prod_{p}|0_p\rangle_{\bf BD},\eea
where the contribution from the individual modes are given by \footnote{It is significant to notice that the tags of $l$ and $m$ on the individual direct product states specified in the regions {\bf R} and {\bf L} in the Bogoliubov transformed basis have been deleted from the above statement for writing purposes.  We will be better equipped to handle complex expressions thanks to this simplification.}:
    \bea |0_p\rangle_{\bf BD} &=&\Bigg\{\sqrt{\frac{\left(1-|\gamma_p|^2\right)}{\left(1+f_p\right)}}\sum^{\infty}_{k=0}|\gamma_{p}|^{k}\bigg(|k_p\rangle_{{\bf R}^{'}}\otimes|k_p\rangle_{{\bf L}^{'}}\bigg)\nonumber\\
        &&\quad\quad\quad\quad\quad\quad+\frac{f_{p}}{\sqrt{\left(1+f_p\right)}}\sum^{\infty}_{n=0}\sum^{\infty}_{r=0}|\Gamma_{p,n}|^{r}\bigg(|n,r_p\rangle_{{\bf R}^{'}}\otimes| n,r_p\rangle_{{\bf L}^{'}}\bigg)\Bigg\}.~~~~~~~~~\quad\eea

   Here we consider Axiverse, where many causally disconnected patches in the open chart of the de Sitter bubbles forms a maximally entangled state. Additionally, we have assumed that every causally unrelated patch corresponds to a Bunch Davies quantum vacuum state, which will ultimately create a maximally entangled state among all feasible Bunch Davies states. Now let us first consider two momentum modes having momenta $p=p_{\bf in}$ and $p=p_{\bf out}$ of the Axiverse.   The maximally entangled state in terms of the unique contributions of Bunch Davies vacuum can be written as:
   \bea |\Psi\rangle:&=&\frac{1}{\sqrt{2}}\sum_{i=0,1}\Bigg(|i_{p_{\bf out}}\rangle_{\bf BD}\otimes|i_{p_{\bf in}}\rangle_{\bf BD}\Bigg)\nonumber\\
   &=&\frac{1}{\sqrt{2}}\Bigg(|0_{p_{\bf out}}\rangle_{\bf BD}\otimes|0_{p_{\bf in}}\rangle_{\bf BD}+|1_{p_{\bf out}}\rangle_{\bf BD}\otimes|1_{p_{\bf in}}\rangle_{\bf BD}\Bigg),\eea   
  where,  $|0_{p_{\bf out}}\rangle_{\bf BD_1}$ and $|1_{p_{\bf out}}\rangle_{\bf BD_1}$ signify the ground and first single particle excited state with mode $p_{\bf out}$.  Also,       
   $|0_{p_{\bf in}}\rangle_{\bf BD_2}$ and $|1_{p_{\bf in}}\rangle_{\bf BD_2}$ signify the ground and first single particle excited quantum state with mode $p_{\bf in}$.

    \subsection{Constructing excited state}

   In this paragraph, our primary goal is to generate the excited quantum state for a single oscillator. This will be useful in generating the overall maximally entangled state needed for the particular operation. To illustrate this problem in more depth, let's begin with the characteristic matrix equation for the oscillators in the recently published Bogoliubov transformed basis:  
      \bea c_{J}&=& b_{I}{\cal G}^{I}_{J},~\quad
            C_{J(n)}=\bar{b}_{J(n)}\left({\cal G}_{(n)}\right)^{I}_{J}\quad\quad {\rm with}\quad\quad c_J=(c_q,c^{\dagger}_q), C_{J(n)}=(C_{q(n)},C^{\dagger}_{q(n)}),\quad\quad\quad\eea
  where we define:
         \bea {\cal G}^{I}_{J}=\left(\begin{array}{ccc} U_q &~~~ V^{*}_q \\ V_q &~~~ U^{*}_q  \end{array}\right)
                    ,~\quad\quad
                    \left({\cal G}_{(n)}\right)^{I}_{J}=\left(\begin{array}{ccc} \overline{U}_{ q,n} &~~~ \overline{V}^{*}_{\sigma q,n} \\ \overline{V}_{ q,n} &~~~ \overline{U}^{*}_{ q,n}  \end{array}\right),\quad\quad\quad\quad\eea
where:
                    \bea U_q &\equiv& {\rm \bf diag}\left(u,\overline{u}\right),
                    V_q \equiv {\rm \bf diag}\left(v,\overline{v}\right),
                     \overline{U}_{q,n} \equiv {\rm \bf diag}\left(U_n,\overline{U}_n\right),
                     \overline{V}_{q,n} \equiv {\rm \bf diag}\left(V_n,\overline{V}_n\right).\eea
The relationship between the $a$-type and $c$-type  oscillators are given by:

 \bea a^{(c)}_{J} &=& b_{J}\left({\cal M}^{-1}\right)^{I}_{J}=c_K\left({\cal G}^{-1}\right)^{K}_{I} \left({\cal M}^{-1}\right)^{I}_{J},\\
  a^{(p)}_{J(n)} &=& b_{J(n)}\left({\cal M}^{-1}_{(n)}\right)^{I}_{J}=C_{K(n)}\left({\cal G}^{-1}_{(n)}\right)^{K}_{I} \left({\cal M}^{-1}_{(n)}\right)^{I}_{J},\eea 
 Here we have:
 \bea a_{I}&=& \left[a^{(c)}_{I}+\sum^{\infty}_{n=0}a^{(p)}_{I(n)}\right]=\bigg[\underbrace{c_K\left({\cal G}^{-1}\right)^{K}_{I} \left({\cal M}^{-1}\right)^{I}_{J}}_{\bf Complementary}+\underbrace{\sum^{\infty}_{n=0}C_{K(n)}\left({\cal G}^{-1}_{(n)}\right)^{K}_{I} \left({\cal M}^{-1}_{(n)}\right)^{I}_{J}}_{\bf Particular~integral}\bigg].\quad\quad\eea 
The product of two inverse matrices can be written as:
   \bea \left({\cal G}^{-1}\right)^{K}_{I} \left({\cal M}^{-1}\right)^{I}_{J}&=&\left(\begin{array}{ccc} Q_{\sigma q} &~~~R^{*}_{\sigma q} \\ R_{\sigma q} &~~~ Q^{*}_{\sigma q}  \end{array}\right),
         \\
\left({\cal G}^{-1}_{(n)}\right)^{K}_{I} \left({\cal M}^{-1}_{(n)}\right)^{I}_{J}&=&\left(\begin{array}{ccc}\overline{Q}_{\sigma q,n} &~~~ \overline{R}^{*}_{\sigma q,n} \\ R_{\sigma q,n} &~~~\overline{Q}^{*}_{\sigma q,n}  \end{array}\right),\eea   
where:
\bea Q_{\sigma q}&=&\left(\begin{array}{ccc}  \widetilde{A}u~~~ &~~~-\widetilde{B}u+\widetilde{D}^{*}v\\ -\widetilde{B}u+\widetilde{D}^{*}v &~~~  \widetilde{A}u  \end{array}\right),\\
R_{\sigma q}&=&\left(\begin{array}{ccc} - \widetilde{A}v~~~ &\widetilde{B}v-\widetilde{D}^{*}u\\ \widetilde{B}v-\widetilde{D}^{*}u &~~~ - \widetilde{A}v  \end{array}\right),\\
Q_{\sigma q,n}&=&\left(\begin{array}{ccc} \widetilde{A}_nU_n~~~ &~~~-\widetilde{B}_nU_n+\widetilde{D}^{*}_nV_n\\ -\widetilde{B}_nU_n+\widetilde{D}^{*}_nV_n &~~~ \widetilde{A}_nU_n  \end{array}\right),\\
R_{\sigma q,n}&=&\left(\begin{array}{ccc} -\widetilde{A}_nV_n~~~ &\widetilde{B}_nV_n-\widetilde{D}^{*}_nU_n\\ \widetilde{B}_nV_n-\widetilde{D}^{*}_nU_n &~~~ -\widetilde{A}_nV_n  \end{array}\right).
         \eea
         The coefficients ($ \widetilde{A}$, $\widetilde{B}$,$\widetilde{D}$) and ($\widetilde{A}_n$, $\widetilde{B}_n$,$\widetilde{D}_n$) are given by:
         \bea  \widetilde{A}&=&\frac{\sqrt{\pi p}}{\left|\Gamma\left(\nu+\frac{1}{2}+ip\right)\right|}\frac{\exp\left(\frac{\pi p}{2}\right)}{\sqrt{\cosh 2\pi p+\cos 2\pi\nu}},\\ 
         \widetilde{B}&=& \widetilde{A}~\frac{\cos \pi\nu}{i \sinh \pi p}\nonumber\\
         &=&\frac{\sqrt{\pi p}}{\left|\Gamma\left(\nu+\frac{1}{2}+ip\right)\right|}\frac{\exp\left(\frac{\pi p}{2}\right)}{\sqrt{\cosh 2\pi p+\cos 2\pi\nu}}\frac{\cos \pi\nu}{i \sinh \pi p},\\
         \widetilde{D}&=&- \widetilde{A}~\frac{\cos (ip+\nu)\pi}{i \sinh \pi p}\exp\left(-\pi p\right)\frac{\Gamma\left(\nu+\frac{1}{2}+ip\right)}{\Gamma\left(\nu+\frac{1}{2}-ip\right)}\nonumber\\
         &=&-\frac{\sqrt{\pi p}}{\left|\Gamma\left(\nu+\frac{1}{2}+ip\right)\right|}\frac{\exp\left(-\frac{\pi p}{2}\right)}{\sqrt{\cosh 2\pi p+\cos 2\pi\nu}}~\frac{\cos (ip+\nu)\pi}{i \sinh \pi p}\frac{\Gamma\left(\nu+\frac{1}{2}+ip\right)}{\Gamma\left(\nu+\frac{1}{2}-ip\right)},\\
        \widetilde{A}_n &=&\frac{\sqrt{\pi p_n}}{\left|\Gamma\left(\nu+\frac{1}{2}+ip_n\right)\right|}\frac{\exp\left(\frac{\pi p_n}{2}\right)}{\sqrt{\cosh 2\pi p_n+\cos 2\pi\nu}},\\
       \widetilde{B}_n &=&\widetilde{A}_n~\frac{\cos \pi\nu}{i \sinh \pi p_n}\nonumber\\
         &=&\frac{\sqrt{\pi p_n}}{\left|\Gamma\left(\nu+\frac{1}{2}+ip_n\right)\right|}\frac{\exp\left(\frac{\pi p_n}{2}\right)}{\sqrt{\cosh 2\pi p_n+\cos 2\pi\nu}}\frac{\cos \pi\nu}{i \sinh \pi p_n},\\ 
       \widetilde{D}_n &=&-\widetilde{A}_n~\frac{\cos (ip_n+\nu)\pi}{i \sinh \pi p_n}\exp\left(-\pi p_n\right)\frac{\Gamma\left(\nu+\frac{1}{2}+ip_n\right)}{\Gamma\left(\nu+\frac{1}{2}-ip_n\right)}\nonumber\\
         &=&-\frac{\sqrt{\pi p_n}}{\left|\Gamma\left(\nu+\frac{1}{2}+ip_n\right)\right|}\frac{\exp\left(-\frac{\pi p_n}{2}\right)}{\sqrt{\cosh 2\pi p_n+\cos 2\pi\nu}}~\frac{\cos (ip_n+\nu)\pi}{i \sinh \pi p_n}\frac{\Gamma\left(\nu+\frac{1}{2}+ip_n\right)}{\Gamma\left(\nu+\frac{1}{2}-ip_n\right)}. \eea  
   Here we found:
   \bea &&Q_{q\sigma}=Q^{T}_{\sigma q}=Q_{\sigma q},\\
   &&R_{q\sigma}=R^{T}_{\sigma q,n}=R_{\sigma q,n},\\
   &&Q_{q\sigma,n}=Q^{T}_{\sigma q,n}=Q_{\sigma q,n},\\
   &&R_{q\sigma,n}=R^{T}_{\sigma q,n}=R_{\sigma q,n}.\eea
   We also use:
   \bea && \widetilde{A}^{*}= \widetilde{A},\quad\quad \widetilde{B}^{*}=-\widetilde{B},\quad\quad u^{*}=u=\overline{u},\quad\quad\quad\quad  v^{*}=v=\overline{v}, \\
   &&\widetilde{A}^{*}_n=\widetilde{A}_n,\quad\quad \widetilde{B}^{*}_n=-\widetilde{B}_n,\quad\quad U^{*}_n=U_n=\overline{U}_n,\quad\quad  V^{*}_n=V_n=\overline{V}_n.\eea  
   Setting $B=0$ and $\widetilde{B}=0$ in addition to $v=0$ and $V_n=0$ corresponds to the massless theory ($\nu=3/2$) and the conformal coupling ($\nu=1/2$).

Also, in the region {\bf L} we  provide the creation and annihilation operators:
   \bea a^{\dagger}_{\bf L} :&=&\underbrace{\Bigg( \widetilde{A}uc^{\dagger}_{\bf L}- \widetilde{A}vc_{\bf L}+\left(\widetilde{B}u+\widetilde{D}v\right)c^{\dagger}_{\bf R} -\left(\widetilde{B}v+\widetilde{D}u\right)c_{\bf R}   \Bigg)}_{\bf Complementary}\nonumber\\
   &&\quad\quad+\underbrace{\sum^{\infty}_{n=0}\Bigg(\widetilde{A}_nU_nC^{\dagger}_{{\bf L}(n)}-\widetilde{A}_nV_nC_{{\bf L}(n)}+\left(\widetilde{B}_nU_n+\widetilde{D}_nV_n\right)C^{\dagger}_{{\bf R}(n)} -\left(\widetilde{B}_nV_n+\widetilde{D}_nU_n\right)C_{{\bf R}(n)}   \Bigg)}_{\bf Particular~integral},\nonumber\\
   && \\
  a_{\bf L} :&=&\underbrace{\Bigg( \widetilde{A}uc_{\bf L}- \widetilde{A}vc^{\dagger}_{\bf L}+\left(-\widetilde{B}u+\widetilde{D}^{*}v\right)c_{\bf R} -\left(-\widetilde{B}v+\widetilde{D}^{*}u\right)c^{\dagger}_{\bf R}   \Bigg)}_{\bf Complementary}\nonumber\\
   &&\quad\quad+\underbrace{\sum^{\infty}_{n=0}\Bigg(\widetilde{A}_nU_nC_{{\bf L}(n)}-\widetilde{A}_nV_nC^{\dagger}_{{\bf L}(n)}+\left(-\widetilde{B}_nU_n+\widetilde{D}^{\dagger}_nV_n\right)C_{{\bf R}(n)} -\left(-\widetilde{B}_nV_n+\widetilde{D}^{\dagger}_nU_n\right)C^{\dagger}_{{\bf R}(n)}   \Bigg)}_{\bf Particular~integral},\nonumber\\
   && \eea
    Hence the excited state of the inside observer is given by:
   \bea |1_{p_{\bf in}}\rangle_{{\bf BD}}&=& a^{\dagger}_{\bf L} |0_{p_{\bf in}}\rangle_{{\bf BD_2}}\nonumber\\
   &=&\Bigg\{\underbrace{\Bigg( \widetilde{A}uc^{\dagger}_{\bf L}- \widetilde{A}vc_{\bf L}+\left(\widetilde{B}u+\widetilde{D}v\right)c^{\dagger}_{\bf R} -\left(\widetilde{B}v+\widetilde{D}u\right)c_{\bf R}   \Bigg)}_{\bf Complementary}\nonumber\\
   &&+\underbrace{\sum^{\infty}_{n=0}\Bigg(\widetilde{A}_nU_nC^{\dagger}_{{\bf L}(n)}-\widetilde{A}_nV_nC_{{\bf L}(n)}+\left(\widetilde{B}_nU_n+\widetilde{D}_nV_n\right)C^{\dagger}_{{\bf R}(n)} -\left(\widetilde{B}_nV_n+\widetilde{D}_nU_n\right)C_{{\bf R}(n)}   \Bigg)}_{\bf Particular~integral}\nonumber\\ 
   &&\quad\quad\quad\quad\quad\Bigg\}\Bigg\{\sqrt{\frac{\left(1-|\gamma_{p_{\bf in}}|^2\right)}{\left(1+f_{p_{\bf in}}\right)}}\sum^{\infty}_{k=0}|\gamma_{p_{\bf in}}|^{k}\bigg(|k_{p_{\bf in}}\rangle_{{\bf R}^{'}}\otimes|k_{p_{\bf in}}\rangle_{{\bf L}^{'}}\bigg)\nonumber\\
        &&\quad\quad\quad\quad\quad\quad+\frac{f_{{p_{\bf in}}}}{\sqrt{\left(1+f_{p_{\bf in}}\right)}}\sum^{\infty}_{n=0}\sum^{\infty}_{r=0}|\Gamma_{{p_{\bf in}},n}|^{r}\bigg(|n,r_{p_{\bf in}}\rangle_{{\bf R}^{'}}\otimes| n,r_{p_{\bf in}}\rangle_{{\bf L}^{'}}\bigg)\Bigg\}~~~~~~~~~\nonumber\\
   &=&\Bigg\{\sqrt{\frac{\left(1-|\gamma_{p_{\bf in}}|^2\right)}{\left(1+f_{p_{\bf in}}\right)}}\Bigg[\Delta_1\sum^{\infty}_{k=0}|\gamma_{p_{\bf in}}|^{k}\sqrt{k+1}\bigg(|k_{p_{\bf in}}\rangle_{{\bf R}^{'}}\otimes|(k+1)_{p_{\bf in}}\rangle_{{\bf L}^{'}}\bigg)\nonumber\\
   &&\quad\quad\quad\quad+\Delta_2\sum^{\infty}_{k=0}|\gamma_{p_{\bf in}}|^{k}\sqrt{k+1}\bigg(|(k+1)_{p_{\bf in}}\rangle_{{\bf R}^{'}}\otimes|k_{p_{\bf in}}\rangle_{{\bf L}^{'}}\bigg)\Bigg]\nonumber\\
        &&+\frac{f_{{p_{\bf in}}}}{\sqrt{\left(1+f_{p_{\bf in}}\right)}}\Bigg[\sum^{\infty}_{n=0} \Delta_{3,n}\sum^{\infty}_{r=0}|\Gamma_{{p_{\bf in}},n}|^{r}\sqrt{r+1}\bigg(|n,r_{p_{\bf in}}\rangle_{{\bf R}^{'}}\otimes| n,(r+1)_{p_{\bf in}}\rangle_{{\bf L}^{'}}\bigg)\nonumber\\
        &&\quad\quad\quad\quad+\sum^{\infty}_{n=0}\Delta_{4,n}\sum^{\infty}_{r=0} |\Gamma_{{p_{\bf in}},n}|^{r}\sqrt{r+1}\bigg(|n,(r+1)_{p_{\bf in}}\rangle_{{\bf R}^{'}}\otimes| n,r_{p_{\bf in}}\rangle_{{\bf L}^{'}}\bigg)\Bigg]\Bigg\}.\eea  
  Here the symbols $\Delta_1$,  $\Delta_2$,  $\Delta_{3,n}$  and  $\Delta_{4,n}$ are defined as:
   \bea \Delta_1 &=& \Bigg( \widetilde{A}u-(\widetilde{B}v+\widetilde{D}u)\gamma_{p_{\bf in}}\Bigg),
   \Delta_2 =\Bigg(- \widetilde{A}v\gamma_{p_{\bf in}}+(\widetilde{B}u+\widetilde{D}v)\Bigg),\\
   \Delta_{3,n} &=&\Bigg(\widetilde{A}_nU_n -\left(\widetilde{B}_nV_n+\widetilde{D}_nU_n\right)\Bigg),
   \Delta_{4,n} =\Bigg(-\widetilde{A}_nV_n \Gamma_{{p_{\bf in}},n}+\left(\widetilde{B}_nU_n+\widetilde{D}_nV_n\right)\Bigg).\quad\quad\eea   
   
    \subsection{Constructing maximally entangled state}

    Here the maximally entangled state is given by:
   \bea |\Psi\rangle:
   &=&\frac{1}{\sqrt{2}}\Bigg(|0_{p_{\bf out}}\rangle_{\bf BD}\otimes\Bigg\{\sqrt{\frac{\left(1-|\gamma_{p_{\bf in}}|^2\right)}{\left(1+f_{p_{\bf in}}\right)}}\sum^{\infty}_{k=0}|\gamma_{p_{\bf in}}|^{k}\bigg(|k_{p_{\bf in}}\rangle_{{\bf R}^{'}}\otimes|k_{p_{\bf in}}\rangle_{{\bf L}^{'}}\bigg)\nonumber\\
        &&\quad\quad\quad\quad\quad\quad+\frac{f_{{p_{\bf in}}}}{\sqrt{\left(1+f_{p_{\bf in}}\right)}}\sum^{\infty}_{n=0}\sum^{\infty}_{r=0}|\Gamma_{{p_{\bf in}},n}|^{r}\bigg(|n,r_{p_{\bf in}}\rangle_{{\bf R}^{'}}\otimes| n,r_{p_{\bf in}}\rangle_{{\bf L}^{'}}\bigg)\Bigg\}\nonumber\\
        &&+|1_{p_{\bf out}}\rangle_{\bf BD}\otimes\Bigg\{\sqrt{\frac{\left(1-|\gamma_{p_{\bf in}}|^2\right)}{\left(1+f_{p_{\bf in}}\right)}}\Bigg[\Delta_1\sum^{\infty}_{k=0}|\gamma_{p_{\bf in}}|^{k}\sqrt{k+1}\bigg(|k_{p_{\bf in}}\rangle_{{\bf R}^{'}}\otimes|(k+1)_{p_{\bf in}}\rangle_{{\bf L}^{'}}\bigg)\nonumber\\
   &&\quad\quad\quad\quad+\Delta_2\sum^{\infty}_{k=0}|\gamma_{p_{\bf in}}|^{k}\sqrt{k+1}\bigg(|(k+1)_{p_{\bf in}}\rangle_{{\bf R}^{'}}\otimes|k_{p_{\bf in}}\rangle_{{\bf L}^{'}}\bigg)\Bigg]\nonumber\\
        &&+\frac{f_{{p_{\bf in}}}}{\sqrt{\left(1+f_{p_{\bf in}}\right)}}\Bigg[\sum^{\infty}_{n=0} \Delta_{3,n}\sum^{\infty}_{r=0}|\Gamma_{{p_{\bf in}},n}|^{r}\sqrt{r+1}\bigg(|n,r_{p_{\bf in}}\rangle_{{\bf R}^{'}}\otimes| n,(r+1)_{p_{\bf in}}\rangle_{{\bf L}^{'}}\bigg)\nonumber\\
        &&\quad\quad\quad\quad+\sum^{\infty}_{n=0}\Delta_{4,n}\sum^{\infty}_{r=0} |\Gamma_{{p_{\bf in}},n}|^{r}\sqrt{r+1}\bigg(|n,(r+1)_{p_{\bf in}}\rangle_{{\bf R}^{'}}\otimes| n,r_{p_{\bf in}}\rangle_{{\bf L}^{'}}\bigg)\Bigg]\Bigg\}\Bigg).\quad\quad\quad\eea   
    The scale dependence in the maximal entangled state constructed in the present setup is observed through the quantities $\gamma_{\bf p_{in}}$, $\Gamma_{{p_{\bf in}},n}$, $\Delta_1$, $\Delta_2$, $\Delta_{3,n}$, and $\Delta_{4,n}$ appearing in the computation.  This finding is a direct result of the factorisation of the inside observer's subspace into two symmetric subspaces {\bf R} and {\bf L}.  Our next task is to investigate the fingerprints of this scale dependency on the physical outcomes of the systems in order to uncover the undiscovered truths from the theoretical framework under examination. 

       \section{Constructing reduced density matrix}\label{ka6}
    
   Since we now know that the inside observer's subspace receives no information content from region {\bf R}, we must now take a partial trace across its degrees of freedom. This will enable us to use this setup to construct the lower density matrix. We must be aware of the fact that the aforementioned freshly constructed maximally entangled quantum state—which is really a mixed state in the current prescription—will be considered throughout this computation. Consequently, the decreased density matrix may be written as follows: 
    \bea \label{wl}\rho_{\bf red}:&=&{\bf Tr}_{{\bf R}^{'}}\left[|\Psi\rangle_{\bf ME}\quad{}_{\bf ME}\langle\Psi|\right]\nonumber\\
    &=&\sum^{\infty}_{m_{\bf p_{in}}=0}{}_{{\bf R}^{'}}\langle m_{\bf p_{in}}|\Psi\rangle_{\bf ME}{}_{\bf ME}\langle\Psi|m\rangle_{{\bf R}^{'}}+\sum^{\infty}_{s=0}\sum^{\infty}_{m_{\bf p_{in}}=0}{}_{{\bf R}^{'}}\langle s,m_{\bf p_{in}}|\Psi\rangle_{\bf ME}{}_{\bf ME}\langle\Psi|s,m_{\bf p_{in}}\rangle_{{\bf R}^{'}}\nonumber\\
    &=& \Bigg\{\sum^{\infty}_{m_{\bf p_{in}}=0}\rho_{m_{\bf p_{in}}}+\sum^{\infty}_{m_{\bf p_{in}}=0}\sum^{\infty}_{s=0}\rho_{m_{\bf p_{in}},s}\Bigg\},\eea
   where we define:   
    \bea \rho_m &=& \frac{\left(1-|\gamma_{p_{\bf in}}|^2\right)}{2\left(1+f_{p_{\bf in}}\right)}|\gamma_{p_{\bf in}}|^{2m_{\bf p_{in}}}\Bigg\{|0_{\bf p_{out}}\rangle_{\bf BD}|m_{\bf p_{in}}\rangle_{{\bf L}^{'}}~{}_{\bf BD}\langle 0|{}_{{\bf L}^{'}}\langle m_{\bf p_{in}}|\nonumber\\
    &&+\Delta^{*}_2\gamma_{p_{\bf in}}\sqrt{m_{\bf p_{in}}+1}|0_{\bf p_{out}}\rangle_{\bf BD}|m_{\bf p_{in}}+1\rangle_{{\bf L}^{'}}~{}_{\bf BD}\langle 1|{}_{{\bf L}^{'}}\langle m_{\bf p_{in}}|\nonumber\\
    &&+\Delta_2\gamma^{*}_{p_{\bf in}}\sqrt{m_{\bf p_{in}}+1}|1_{\bf p_{out}}\rangle_{\bf BD}|m_{\bf p_{in}}\rangle_{{\bf L}^{'}}~{}_{\bf BD}\langle 0|{}_{{\bf L}^{'}}\langle m_{\bf p_{in}}+1|\nonumber\\
    &&+|\Delta_2|^{2}(m_{\bf p_{in}}+1)|1_{\bf p_{out}}\rangle_{\bf BD}|m_{\bf p_{in}}\rangle_{{\bf L}^{'}}~{}_{\bf BD}\langle 1|{}_{{\bf L}^{'}}\langle m_{\bf p_{in}}|\nonumber\\
    &&+\Delta^{*}_1 \sqrt{m_{\bf p_{in}}+1}|0_{\bf p_{out}}\rangle_{\bf BD}|m_{\bf p_{in}}\rangle_{{\bf L}^{'}}~{}_{\bf BD}\langle 1|{}_{{\bf L}^{'}}\langle m_{\bf p_{in}}+1|\nonumber\\
    &&+\Delta_1 \sqrt{m_{\bf p_{in}}+1}|1_{\bf p_{out}}\rangle_{\bf BD}|m_{\bf p_{in}}+1\rangle_{{\bf L}^{'}}~{}_{\bf BD}\langle 0|{}_{{\bf L}^{'}}\langle m_{\bf p_{in}}|\nonumber\\
    &&+\Delta^{*}_1 \Delta_2 \gamma^{*}_{p_{\bf in}}\sqrt{(m_{\bf p_{in}}+1)(m_{\bf p_{in}}+2)}|1_{\bf p_{out}}\rangle_{\bf BD}|m_{\bf p_{in}}\rangle_{{\bf L}^{'}}~{}_{\bf BD}\langle 1|{}_{{\bf L}^{'}}\langle m_{\bf p_{in}}+2|\nonumber\\
    &&+\Delta_1 \Delta^{*}_2 \gamma_{p_{\bf in}}\sqrt{(m_{\bf p_{in}}+1)(m_{\bf p_{in}}+2)}|1_{\bf p_{out}}\rangle_{\bf BD}|m_{\bf p_{in}}+2\rangle_{{\bf L}^{'}}~{}_{\bf BD}\langle 1|{}_{{\bf L}^{'}}\langle m_{\bf p_{in}}|\nonumber\\
    &&+|\Delta_1|^{2}(m_{\bf p_{in}}+1)|1_{\bf p_{out}}\rangle_{\bf BD}|m_{\bf p_{in}}+1\rangle_{{\bf L}^{'}}~{}_{\bf BD}\langle 1|{}_{{\bf L}^{'}}\langle m_{\bf p_{in}}+1|\Bigg\},\\
   \rho_{m,s} &=&\frac{f^{2}_{p_{\bf in}}}{2\left(1+f_{p_{\bf in}}\right)}|\Gamma_{{p_{\bf in}},s}|^{2m_{\bf p_{in}}}\Bigg\{|0_{\bf p_{out}}\rangle_{\bf BD}|s,m_{\bf p_{in}}\rangle_{{\bf L}^{'}}~{}_{\bf BD}\langle 0|{}_{{\bf L}^{'}}\langle s,m_{\bf p_{in}}|\nonumber\\
    &&+\Delta^{*}_{4,s}\Gamma_{{p_{\bf in}},s}\sqrt{m_{\bf p_{in}}+1}|0_{\bf p_{out}}\rangle_{\bf BD_1}|s,(m_{\bf p_{in}}+1)\rangle_{{\bf L}^{'}}~{}_{\bf BD}\langle 1|{}_{{\bf L}^{'}}\langle s,m_{\bf p_{in}}|\nonumber\\
    &&+\Delta_{4,s}\Gamma^{*}_{{p_{\bf in}},s}\sqrt{m_{\bf p_{in}}+1}|1_{\bf p_{out}}\rangle_{\bf BD}|s,m_{\bf p_{in}}\rangle_{{\bf L}^{'}}~{}_{\bf BD}\langle 0|{}_{{\bf L}^{'}}\langle s, (m_{\bf p_{in}}+1)|\nonumber\\
    &&+|\Delta_{4,s}|^{2}(m_{\bf p_{in}}+1)|1_{\bf p_{out}}\rangle_{\bf BD}|s,m_{\bf p_{in}}\rangle_{{\bf L}^{'}}~{}_{\bf BD}\langle 1|{}_{{\bf L}^{'}}\langle s,m_{\bf p_{in}}|\nonumber\\
    &&+\Delta^{*}_{3,s} \sqrt{m_{\bf p_{in}}+1}|0_{\bf p_{out}}\rangle_{\bf BD}|s,m_{\bf p_{in}}\rangle_{{\bf L}^{'}}~{}_{\bf BD}\langle 1|{}_{{\bf L}^{'}}\langle s,(m_{\bf p_{in}}+1)|\nonumber\\
    &&+\Delta_{3,s} \sqrt{m_{\bf p_{in}}+1}|1_{\bf p_{out}}\rangle_{\bf BD}|s,m_{\bf p_{in}}+1\rangle_{{\bf L}^{'}}~{}_{\bf BD}\langle 0|{}_{{\bf L}^{'}}\langle s,m_{\bf p_{in}}|\nonumber\\
    &&+\Delta^{*}_{3,s} \Delta_{4,s} \Gamma^{*}_{{p_{\bf in}},s}\sqrt{(m_{\bf p_{in}}+1)(m_{\bf p_{in}}+2)}|1_{\bf p_{out}}\rangle_{\bf BD}|s,m_{\bf p_{in}}\rangle_{{\bf L}^{'}}~{}_{\bf BD}\langle 1|{}_{{\bf L}^{'}}\langle s,(m_{\bf p_{in}}+2)|\nonumber\\
    &&+\Delta_{3,s} \Delta^{*}_{4,s} \Gamma_{{p_{\bf in}},s}\sqrt{(m_{\bf p_{in}}+1)(m_{\bf p_{in}}+2)}|1_{\bf p_{out}}\rangle_{\bf BD}|s,(m_{\bf p_{in}}+2)\rangle_{{\bf L}^{'}}~{}_{\bf BD}\langle 1|{}_{{\bf L}^{'}}\langle s,m_{\bf p_{in}}|\nonumber\\
    &&+|\Delta_{3,s}|^{2}(m_{\bf p_{in}}+1)|1_{\bf p_{out}}\rangle_{\bf BD}|s,(m_{\bf p_{in}}+1)\rangle_{{\bf L}^{'}}~{}_{\bf BD}\langle 1|{}_{{\bf L}^{'}}\langle s,(m_{\bf p_{in}}+1)|\Bigg\}.\eea
The quantum mechanical state that appears in this case for both the complementary and particular integral portions, when the corresponding observer is situated in one of the areas of the open chart of the global de Sitter space, effectively describes the internal observer. Additionally, it is important to keep in mind that all mode eigen values for the particular and complementary integral portions are the same and denoted by the notation $m_{p_{\bf in}}$.
This is owing to the fact that the index $s$ that appears in the particular integral portion as a result of placing the source term has no influence on the eigen values of the mode function at the end of the day.  The quantum states of the integral portion can be tagged with $s$ and $m_{p_{\bf in}}$ to distinguish them from the quantum modes of the complementary part, which are merely tagged with the quantum number $m_{p_{\bf in}}$.  This is a critical topic that must be addressed at this level of calculation in order to avoid additional misunderstanding.

      \section{Partial transposition operation} \label{ka7}
      
     The primary goal of this part is to identify the negative eigenvalues of the previously described formula for a decreased density matrix. To do this computation, we separate the contributions from the complimentary section and the specific integral component. Here is what we have:
         \bea \rho^{T,{\bf BD}}_m &=&\frac{\left(1-|\gamma_{p_{\bf in}}|^2\right)}{2\left(1+f_{p_{\bf in}}\right)}|\gamma_{p_{\bf in}}|^{2m_{\bf p_{in}}}\Bigg\{|0_{\bf p_{out}}\rangle_{\bf BD}|m_{\bf p_{in}}\rangle_{{\bf L}^{'}}~{}_{\bf BD}\langle 0|{}_{{\bf L}^{'}}\langle m_{\bf p_{in}}|\nonumber\\
    &&+\Delta^{*}_2\gamma_{p_{\bf in}}\sqrt{m_{\bf p_{in}}+1}|1_{\bf p_{out}}\rangle_{\bf BD}|m_{\bf p_{in}}+1\rangle_{{\bf L}^{'}}~{}_{\bf BD}\langle 0|{}_{{\bf L}^{'}}\langle m_{\bf p_{in}}|\nonumber\\
    &&+\Delta_2\gamma^{*}_{p_{\bf in}}\sqrt{m_{\bf p_{in}}+1}|0_{\bf p_{out}}\rangle_{\bf BD}|m_{\bf p_{in}}\rangle_{{\bf L}^{'}}~{}_{\bf BD}\langle 1|{}_{{\bf L}^{'}}\langle m_{\bf p_{in}}+1|\nonumber\\
    &&+|\Delta_2|^{2}(m_{\bf p_{in}}+1)|1_{\bf p_{out}}\rangle_{\bf BD}|m_{\bf p_{in}}\rangle_{{\bf L}^{'}}~{}_{\bf BD}\langle 1|{}_{{\bf L}^{'}}\langle m_{\bf p_{in}}|\nonumber\\
    &&+\Delta^{*}_1 \sqrt{m_{\bf p_{in}}+1}|1_{\bf p_{out}}\rangle_{\bf BD}|m_{\bf p_{in}}\rangle_{{\bf L}^{'}}~{}_{\bf BD}\langle 0|{}_{{\bf L}^{'}}\langle m_{\bf p_{in}}+1|\nonumber\\
    &&+\Delta_1 \sqrt{m_{\bf p_{in}}+1}|0_{\bf p_{out}}\rangle_{\bf BD}|m_{\bf p_{in}}+1\rangle_{{\bf L}^{'}}~{}_{\bf BD}\langle 1|{}_{{\bf L}^{'}}\langle m_{\bf p_{in}}|\nonumber\\
    &&+\Delta^{*}_1 \Delta_2 \gamma^{*}_{p_{\bf in}}\sqrt{(m_{\bf p_{in}}+1)(m_{\bf p_{in}}+2)}|1_{\bf p_{out}}\rangle_{\bf BD}|m_{\bf p_{in}}\rangle_{{\bf L}^{'}}~{}_{\bf BD}\langle 1|{}_{{\bf L}^{'}}\langle m_{\bf p_{in}}+2|\nonumber\\
    &&+\Delta_1 \Delta^{*}_2 \gamma_{p_{\bf in}}\sqrt{(m_{\bf p_{in}}+1)(m_{\bf p_{in}}+2)}|1_{\bf p_{out}}\rangle_{\bf BD}|m_{\bf p_{in}}+2\rangle_{{\bf L}^{'}}~{}_{\bf BD}\langle 1|{}_{{\bf L}^{'}}\langle m_{\bf p_{in}}|\nonumber\\
    &&+|\Delta_1|^{2}(m_{\bf p_{in}}+1)|1_{\bf p_{out}}\rangle_{\bf BD}|m_{\bf p_{in}}+1\rangle_{{\bf L}^{'}}~{}_{\bf BD}\langle 1|{}_{{\bf L}^{'}}\langle m_{\bf p_{in}}+1|\Bigg\},\\
   \rho^{T,{\bf BD}}_{m,s} &=&\frac{f^{2}_{p_{\bf in}}}{2\left(1+f_{p_{\bf in}}\right)}|\Gamma_{{p_{\bf in}},s}|^{2m_{\bf p_{in}}}\Bigg\{|0_{\bf p_{out}}\rangle_{\bf BD}|s,m_{\bf p_{in}}\rangle_{{\bf L}^{'}}~{}_{\bf BD}\langle 0|{}_{{\bf L}^{'}}\langle s,m_{\bf p_{in}}|\nonumber\\
    &&+\Delta^{*}_{4,s}\Gamma_{{p_{\bf in}},s}\sqrt{m_{\bf p_{in}}+1}|1_{\bf p_{out}}\rangle_{\bf BD}|s,(m_{\bf p_{in}}+1)\rangle_{{\bf L}^{'}}~{}_{\bf BD}\langle 0|{}_{{\bf L}^{'}}\langle s,m_{\bf p_{in}}|\nonumber\\
    &&+\Delta_{4,s}\Gamma^{*}_{{p_{\bf in}},s}\sqrt{m_{\bf p_{in}}+1}|0_{\bf p_{out}}\rangle_{\bf BD}|s,m_{\bf p_{in}}\rangle_{{\bf L}^{'}}~{}_{\bf BD}\langle 1|{}_{{\bf L}^{'}}\langle s, (m_{\bf p_{in}}+1)|\nonumber\\
    &&+|\Delta_{4,s}|^{2}(m_{\bf p_{in}}+1)|1_{\bf p_{out}}\rangle_{\bf BD}|s,m_{\bf p_{in}}\rangle_{{\bf L}^{'}}~{}_{\bf BD}\langle 1|{}_{{\bf L}^{'}}\langle s,m_{\bf p_{in}}|\nonumber\\
    &&+\Delta^{*}_{3,s} \sqrt{m_{\bf p_{in}}+1}|1_{\bf p_{out}}\rangle_{\bf BD}|s,m_{\bf p_{in}}\rangle_{{\bf L}^{'}}~{}_{\bf BD}\langle 0|{}_{{\bf L}^{'}}\langle s,(m_{\bf p_{in}}+1)|\nonumber\\
    &&+\Delta_{3,s} \sqrt{m_{\bf p_{in}}+1}|0_{\bf p_{out}}\rangle_{\bf BD}|s,m_{\bf p_{in}}+1\rangle_{{\bf L}^{'}}~{}_{\bf BD}\langle 1|{}_{{\bf L}^{'}}\langle s,m_{\bf p_{in}}|\nonumber\\
    &&+\Delta^{*}_{3,s} \Delta_{4,s} \Gamma^{*}_{{p_{\bf in}},s}\sqrt{(m_{\bf p_{in}}+1)(m_{\bf p_{in}}+2)}|1_{\bf p_{out}}\rangle_{\bf BD}|s,m_{\bf p_{in}}\rangle_{{\bf L}^{'}}~{}_{\bf BD}\langle 1|{}_{{\bf L}^{'}}\langle s,(m_{\bf p_{in}}+2)|\nonumber\\
    &&+\Delta_{3,s} \Delta^{*}_{4,s} \Gamma_{{p_{\bf in}},s}\sqrt{(m_{\bf p_{in}}+1)(m_{\bf p_{in}}+2)}|1_{\bf p_{out}}\rangle_{\bf BD}|s,(m_{\bf p_{in}}+2)\rangle_{{\bf L}^{'}}~{}_{\bf BD}\langle 1|{}_{{\bf L}^{'}}\langle s,m_{\bf p_{in}}|\nonumber\\
    &&+|\Delta_{3,s}|^{2}(m_{\bf p_{in}}+1)|1_{\bf p_{out}}\rangle_{\bf BD}|s,(m_{\bf p_{in}}+1)\rangle_{{\bf L}^{'}}~{}_{\bf BD}\langle 1|{}_{{\bf L}^{'}}\langle s,(m_{\bf p_{in}}+1)|\Bigg\}.\eea
     We may infer from our theoretical setup that the quantum mechanical states corresponding to the inside and outside observers are entangled if at least one eigenvalue is discovered to be negative.     
           
   \section{Logarithmic negativity in Axiverse}\label{ka8}
        Let us write:
     \bea \rho^{T,{\bf BD_1}}_{m}&=&\frac{\left(1-|\gamma_{p_{\bf in}}|^2\right)}{2\left(1+f_{p_{\bf in}}\right)}|\gamma_{p_{\bf in}}|^{2m_{\bf p_{in}}}\left(\begin{array}{ccc} A_{m_{\bf p_{in}}} &~~~B_{m_{\bf p_{in}}} &~~~C_{m_{\bf p_{in}}}\\ B^{*}_{m_{\bf p_{in}}} &~~~ D_{m_{\bf p_{in}}} &~~~ 0\\ C^{*}_{m_{\bf p_{in}}} &~~~ 0 &~~~ 0  \end{array}\right),\\\rho^{T,{\bf BD_1}}_{m_{\bf p_{in}},s}&=&\frac{f^{2}_{p_{\bf in}}}{2\left(1+f_{p_{\bf in}}\right)}|\Gamma_{{p_{\bf in}},s}|^{2m_{\bf p_{in}}}\left(\begin{array}{ccc} A_{m_{\bf p_{in}},s} &~~~B_{m_{\bf p_{in}},s} &~~~C_{m_{\bf p_{in}},s}\\ B^{*}_{m_{\bf p_{in}},s} &~~~ D_{m_{\bf p_{in}},s} &~~~ 0\\ C^{*}_{m_{\bf p_{in}},s} &~~~ 0 &~~~ 0  \end{array}\right),\eea
           where we define:
           \bea A_{m_{\bf p_{in}}} &=& 1+|\Delta_2|^{2}(m_{\bf p_{in}}+1),\\
           B_{m_{\bf p_{in}}} &=& \sqrt{m_{\bf p_{in}}+1}\left(\Delta_2 \gamma^{*}_{p_{\bf in}}+\Delta^{*}_1\right),\\
           C_{m_{\bf p_{in}}} &=&  \sqrt{(m_{\bf p_{in}}+1)(m_{\bf p_{in}}+2)}~\Delta^{*}_1\Delta_2 \gamma^{*}_{p_{\bf in}},\\
          D_{m_{\bf p_{in}}} &=& |\Delta_1|^{2}(m_{\bf p_{in}}+1),\\ A_{m_{\bf p_{in}},s} &=& 1+|\Delta_{4,s}|^{2}(m_{\bf p_{in}}+1),\\
           B_{m_{\bf p_{in}},s} &=& \sqrt{m_{\bf p_{in}}+1}\left(\Delta_{4,s} \Gamma^{*}_{{p_{\bf in}},s}+\Delta^{*}_{3,s}\right),\\
           C_{m_{\bf p_{in}},s} &=&  \sqrt{(m_{\bf p_{in}}+1)(m_{\bf p_{in}}+2)}~\Delta^{*}_{3,s}\Delta_{4,s} \Gamma^{*}_{{p_{\bf in}},s},\\
          D_{m_{\bf p_{in}},s} &=& |\Delta_{3,s}|^{2}(m_{\bf p_{in}}+1).\eea
          Here we have the following eigen value equation:
          \bea && \widetilde{\lambda}^{3}_{m_{\bf p_{in}}}-\overline{A}_{m_{\bf p_{in}}}\widetilde{\lambda}^{2}_{m_{\bf p_{in}}}+\overline{B}_{m_{\bf p_{in}}}\widetilde{\lambda}_{m_{\bf p_{in}}}+\overline{C}_{m_{\bf p_{in}}}=0.\eea
         where we define:
         \bea \overline{A}_{m_{\bf p_{in}}} &=&\frac{1}{2\left(1+f_{p_{\bf in}}\right)}\Bigg\{|\gamma_{p_{\bf in}}|^{2m_{\bf p_{in}}}\left(1-|\gamma_{p_{\bf in}}|^2\right) \left(A_{m_{\bf p_{in}}}+D_{m_{\bf p_{in}}}\right)\nonumber\\
         &&\quad\quad\quad\quad\quad\quad\quad\quad\quad\quad +f^{2}_{p_{\bf in}}\sum^{\infty}_{s=0}|\Gamma_{{p_{\bf in}},s}|^{2m}\left(A_{m_{\bf p_{in}},s}+D_{m_{\bf p_{in}},s}\right)\Bigg\},\quad\\
        \overline{B}_{m_{\bf p_{in}}} &=& \frac{1}{4\left(1+f_{p_{\bf in}}\right)^2}\Bigg\{|\gamma_{p_{\bf in}}|^{4m_{\bf p_{in}}}\left(1-|\gamma_{p_{\bf in}}|^2\right)^2\left(A_{m_{\bf p_{in}}} D_{m_{\bf p_{in}}}-\left(|B_{m_{\bf p_{in}}}|^{2}+|C_{m_{\bf p_{in}}}|^{2}\right)\right)\nonumber\\
        &&\quad\quad +f^{4}_{p_{\bf in}}\sum^{\infty}_{s=0}|\Gamma_{{p_{\bf in}},s}|^{4m_{\bf p_{in}}} \left(A_{m_{\bf p_{in}},s} D_{m_{\bf p_{in}},s}-\left(|B_{m_{\bf p_{in}},s}|^{2}+|C_{m_{\bf p_{in}},s}|^{2}\right)\right)\Bigg\},\\
        \overline{C}_{m_{\bf p_{in}}} &=&\frac{1}{8\left(1+f_{p_{\bf in}}\right)^3}\Bigg\{|\gamma_{p_{\bf in}}|^{6m_{\bf p_{in}}} \left(1-|\gamma_{p_{\bf in}}|^2\right)^3|C_{m_{\bf p_{in}}}|^{2}D_{m_{\bf p_{in}}}\nonumber\\
        &&\quad\quad\quad\quad\quad\quad\quad\quad\quad\quad+f^{6}_{p_{\bf in}}\sum^{\infty}_{s=0}|\Gamma_{{p_{\bf in}},s}|^{6m_{\bf p_{in}}}|C_{m_{\bf p_{in}},s}|^{2}D_{m_{\bf p_{in}},s}\Bigg\}.\eea

    \begin{figure*}[htb]
    \centering
    \subfigure[For $f_{p_{\bf in}}=0$.]{
        \includegraphics[width=14.2cm,height=8.9cm] {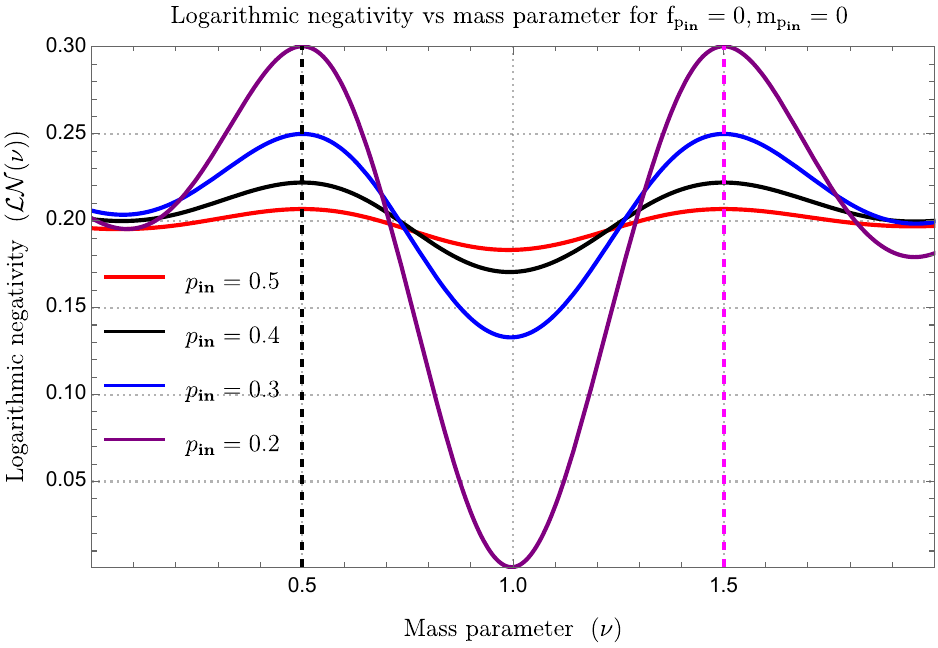}
        \label{LEN1}
    }
    \subfigure[For small $f_{p_{\bf in}}\neq 0$.]{
        \includegraphics[width=14.2cm,height=8.9cm] {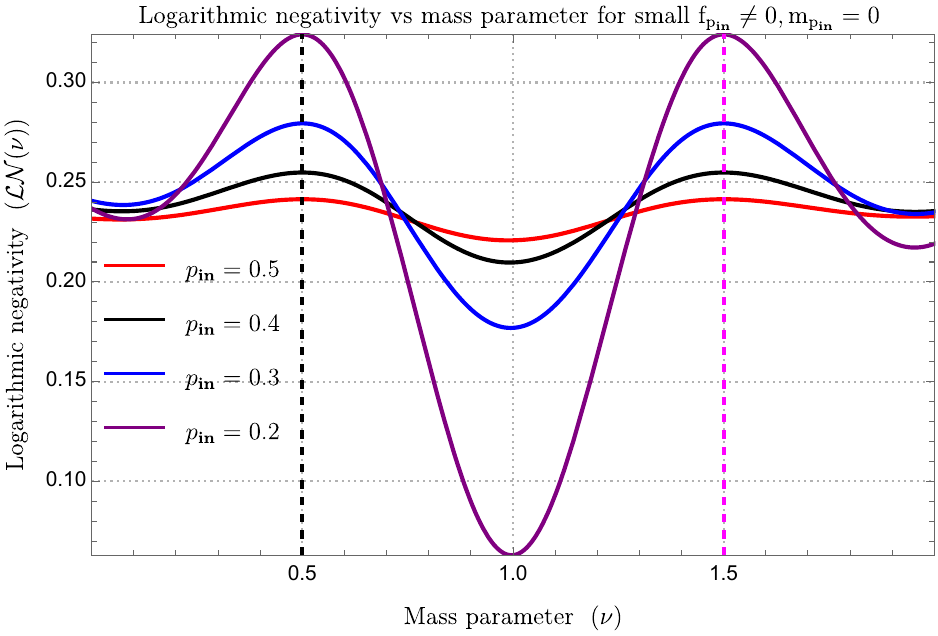}
        \label{LEN2}
       }
    \caption[Optional caption for list of figures]{Logarithmic negativity vs mass parameter for a specified momentum mode.   } 
    \label{LEN}
    \end{figure*}  
    \begin{figure*}[htb]
    \centering
    \subfigure[For $f_{p_{\bf in}}=0$.]{
        \includegraphics[width=14.2cm,height=8.9cm] {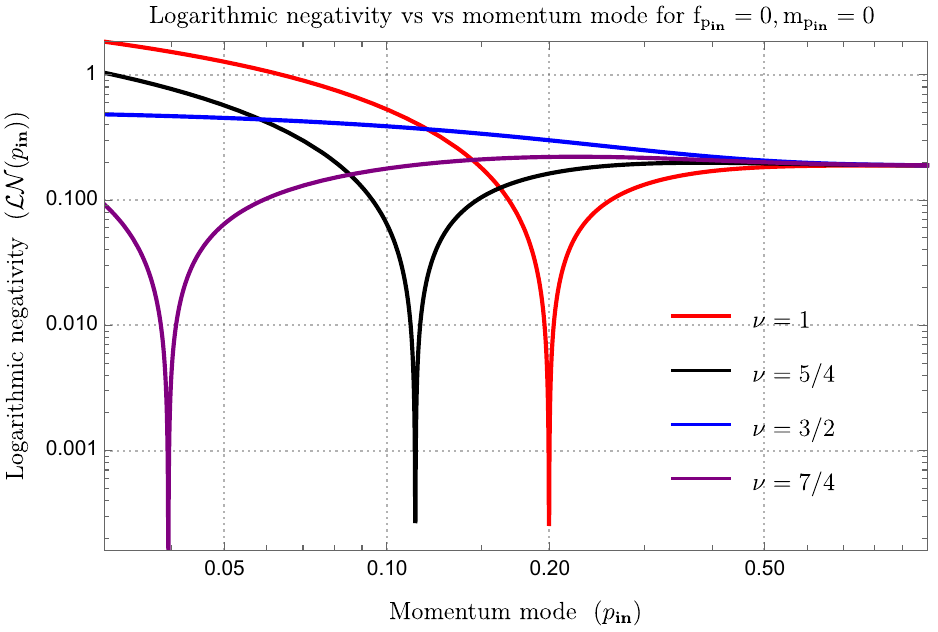}
        \label{LNp1}
    }
    \subfigure[For small $f_{p_{\bf in}}\neq 0$.]{
        \includegraphics[width=14.2cm,height=8.9cm] {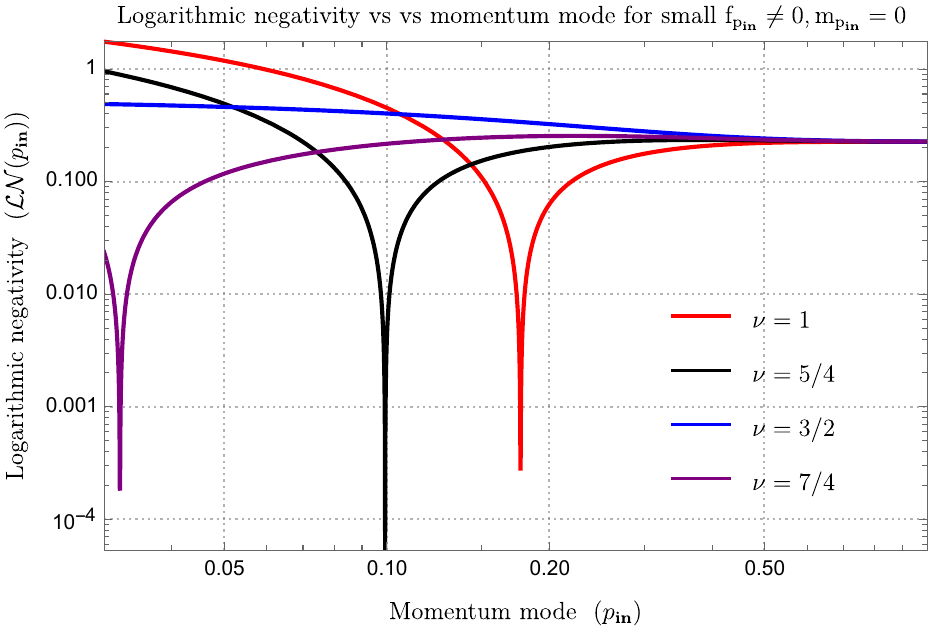}
        \label{LNp2}
       }
    \caption[Optional caption for list of figures]{Logarithmic negativity vs the momentum mode for a specified mass parameter. } 
    \label{LNp}
    \end{figure*}  
    
        The real root from the $(m_{\bf p_{in}},m_{\bf p_{in}}+1)$ block is given by:
        \bea \widetilde{\lambda}_{m_{\bf p_{in}}} &=&\frac{1}{3}\Bigg[\overline{A}_{m_{\bf p_{in}}}+\frac{f(\overline{A}_{m_{\bf p_{in}}},\overline{B}_{m_{\bf p_{in}}},\overline{C}_{m_{\bf p_{in}}})}{\sqrt[3]{2}} -\frac{\sqrt[3]{2} \left(3 \overline{B}_{m_{\bf p_{in}}}-\overline{A}^2_{m_{\bf p_{in}}}\right)}{f(\overline{A}_{m_{\bf p_{in}}},\overline{B}_{m_{\bf p_{in}}},\overline{C}_{m_{\bf p_{in}}})}\Bigg]. \eea
        where we define the newly defined function $f(\overline{A}_{m_{\bf p_{in}}},\overline{B}_{m_{\bf p_{in}}},\overline{C}_{m_{\bf p_{in}}})$ which is defined as:
        \bea f(\overline{A}_{m_{\bf p_{in}}},\overline{B}_{m_{\bf p_{in}}},\overline{C}_{m_{\bf p_{in}}}):&=&\Bigg[2 \overline{A}^3_{m_{\bf p_{in}}}-9 \overline{A}_{m_{\bf p_{in}}} \overline{B}_{m_{\bf p_{in}}}-27 \overline{C}_{m_{\bf p_{in}}}\nonumber\\
 &&+3 \sqrt{3} \Bigg\{18 \overline{A}_{m_{\bf p_{in}}}\overline{ B}_{m_{\bf p_{in}}}\overline{ C}_{m_{\bf p_{in}}}+4 \overline{B}^3_{m_{\bf p_{in}}}+27 \overline{C}^2_{m_{\bf p_{in}}}\nonumber\\
 &&-4 \overline{A}^3_{m_{\bf p_{in}}} \overline{C}_{m_{\bf p_{in}}}-\overline{A}^2_{m_{\bf p_{in}}} \overline{B}^2_{m_{\bf p_{in}}}\Bigg\}^{\displaystyle \frac{1}{2}} \Bigg]^{\displaystyle \frac{1}{3}}.\quad\eea      
      Then the logarithmic negativity is given by:
      \bea {\cal LN}&=&\ln\left(2\sum_{\widetilde{\lambda}_{m_{\bf p_{in}}}<0}\widetilde{\lambda}_{m_{\bf p_{in}}}+1\right)\nonumber\\
      &=&\ln\Bigg(\frac{2}{3}\Bigg[\overline{A}_{m_{\bf p_{in}}}+\frac{f(\overline{A}_{m_{\bf p_{in}}},\overline{B}_{m_{\bf p_{in}}},\overline{C}_{m_{\bf p_{in}}})}{\sqrt[3]{2}} -\frac{\sqrt[3]{2} \left(3 \overline{B}_{m_{\bf p_{in}}}-\overline{A}^2_{m_{\bf p_{in}}}\right)}{f(\overline{A}_{m_{\bf p_{in}}},\overline{B}_{m_{\bf p_{in}}},\overline{C}_{m_{\bf p_{in}}})}\Bigg]+1\Bigg).\quad\quad\quad\eea
 
Here  in figure (\ref{LEN}(a)) and (\ref{LEN}(b)),  we have shown the logarithmic negativity with mass parameter.  Also,  in figure (\ref{LNp}(a)) and (\ref{LNp}(b)),  we have shown the logarithmic negativity with the momentum mode for the given value of mass parameter.  

     \section{Quantum discord in Axiverse}\label{ka9}

     In this section our aim is to compute quantum discord, which is the measure of the quantumness  of the present system under consideration. For this purpose, let us start with the general definition of quantum discord, which is given by:
     \bea {\cal D}(A,B)=S(\rho_A)-S(\rho_{A,B})+S(B|A).\eea
     Now the  equation (\ref{wl}) can be further recast as:
\bea \rho_{A,B}:&=& \frac{\left(1-|\gamma_{p_{\bf in}}|^2\right)}{2\left(1+f_{p_{\bf in}}\right)}\Bigg\{|0_{\bf p_{out}}\rangle_{\bf BD}~{}_{\bf BD}\langle 0|{}_{{\bf L}^{'}}\otimes {\cal M}_{00}+|0_{\bf p_{out}}\rangle_{\bf BD}~{}_{\bf BD}\langle 1|{}_{{\bf L}^{'}}\otimes {\cal M}_{01}\nonumber\\
&&\quad\quad\quad\quad\quad\quad\quad\quad+|1_{\bf p_{out}}\rangle_{\bf BD}~{}_{\bf BD}\langle 0|{}_{{\bf L}^{'}}{\cal M}_{10}+|1_{\bf p_{out}}\rangle_{\bf BD}~{}_{\bf BD}\langle 1|{}_{{\bf L}^{'}}{\cal M}_{11}\Bigg\}\nonumber\\
&&+\frac{f^{2}_{p_{\bf in}}}{2\left(1+f_{p_{\bf in}}\right)}\Bigg\{|0_{\bf p_{out}}\rangle_{\bf BD}~{}_{\bf BD}\langle 0|{}_{{\bf L}^{'}}\otimes \overline{\cal M}_{00}+|0_{\bf p_{out}}\rangle_{\bf BD}~{}_{\bf BD}\langle 1|{}_{{\bf L}^{'}}\otimes \overline{\cal M}_{01}\nonumber\\
&&\quad\quad\quad\quad\quad\quad\quad\quad+|1_{\bf p_{out}}\rangle_{\bf BD}~{}_{\bf BD}\langle 0|{}_{{\bf L}^{'}}\overline{\cal M}_{10}+|1_{\bf p_{out}}\rangle_{\bf BD}~{}_{\bf BD}\langle 1|{}_{{\bf L}^{'}}\overline{\cal M}_{11}\Bigg\}.\eea
where we define:
   \bea {\cal M}_{00}:&=&\sum^{\infty}_{m_{\bf p_{in}}=0}|\gamma_{p_{\bf in}}|^{2m_{\bf p_{in}}}|m_{\bf p_{in}}\rangle_{{\bf L}^{'}}~{}_{{\bf L}^{'}}\langle m_{\bf p_{in}}|,\\
   {\cal M}_{01}:&=&\sum^{\infty}_{m_{\bf p_{in}}=0}|\gamma_{p_{\bf in}}|^{2m_{\bf p_{in}}}\sqrt{m_{\bf p_{in}}+1}\Bigg\{\Delta^{*}_2\gamma_{p_{\bf in}}|m_{\bf p_{in}}+1\rangle_{{\bf L}^{'}}~{}_{{\bf L}^{'}}\langle m_{\bf p_{in}}|\nonumber\\&&\quad\quad\quad\quad\quad\quad\quad\quad\quad\quad\quad\quad\quad\quad\quad+\Delta^{*}_1 |m_{\bf p_{in}}\rangle_{{\bf L}^{'}}~{}_{{\bf L}^{'}}\langle m_{\bf p_{in}}+1|\Bigg\},\\
   \\
   {\cal M}_{10}:&=&\sum^{\infty}_{m_{\bf p_{in}}=0}|\gamma_{p_{\bf in}}|^{2m_{\bf p_{in}}}\sqrt{m_{\bf p_{in}}+1}\Bigg\{\Delta_2\gamma^{*}_{p_{\bf in}}|m_{\bf p_{in}}\rangle_{{\bf L}^{'}}~{}_{{\bf L}^{'}}\langle m_{\bf p_{in}}+1|\nonumber\\&&\quad\quad\quad\quad\quad\quad\quad\quad\quad\quad\quad\quad\quad\quad\quad+\Delta_1 |m_{\bf p_{in}}+1\rangle_{{\bf L}^{'}}~{}_{{\bf L}^{'}}\langle m_{\bf p_{in}}|\Bigg\},\\
   {\cal M}_{11}:&=&\sum^{\infty}_{m_{\bf p_{in}}=0}|\gamma_{p_{\bf in}}|^{2m_{\bf p_{in}}}(m_{\bf p_{in}}+1)\Bigg\{|\Delta_2|^{2}|m_{\bf p_{in}}\rangle_{{\bf L}^{'}}~{}_{{\bf L}^{'}}\langle m_{\bf p_{in}}|\nonumber\\&&\quad\quad\quad\quad\quad\quad\quad\quad\quad\quad\quad\quad\quad\quad\quad+|\Delta_1|^{2}|m_{\bf p_{in}}+1\rangle_{{\bf L}^{'}}~{}_{{\bf L}^{'}}\langle m_{\bf p_{in}}+1|\Bigg\}\nonumber\\
   &&+\sum^{\infty}_{m_{\bf p_{in}}=0}|\gamma_{p_{\bf in}}|^{2m_{\bf p_{in}}}\sqrt{(m_{\bf p_{in}}+1)(m_{\bf p_{in}}+2)}\Bigg\{\Delta^{*}_1 \Delta_2 \gamma^{*}_{p_{\bf in}}|m_{\bf p_{in}}\rangle_{{\bf L}^{'}}~{}_{{\bf L}^{'}}\langle m_{\bf p_{in}}+2|\nonumber\\&&\quad\quad\quad\quad\quad\quad\quad\quad\quad\quad\quad\quad\quad+\Delta_1 \Delta^{*}_2|m_{\bf p_{in}}+2\rangle_{{\bf L}^{'}}~{}_{{\bf L}^{'}}\langle m_{\bf p_{in}}|\Bigg\}.\\
   \overline{\cal M}_{00}:&=&\sum^{\infty}_{m_{\bf p_{in}}=0}\sum^{\infty}_{s=0}|\Gamma_{p_{\bf in},s}|^{2m_{\bf p_{in}}}|s,m_{\bf p_{in}}\rangle_{{\bf L}^{'}}~{}_{{\bf L}^{'}}\langle s,m_{\bf p_{in}}|,\\
   \overline{\cal M}_{01}:&=&\sum^{\infty}_{m_{\bf p_{in}}=0}\sum^{\infty}_{s=0}|\Gamma_{p_{\bf in},s}|^{2m_{\bf p_{in}}}\sqrt{m_{\bf p_{in}}+1}\Bigg\{\Delta^{*}_{4,s}\Gamma_{p_{\bf in},s}|s,(m_{\bf p_{in}}+1)\rangle_{{\bf L}^{'}}~{}_{{\bf L}^{'}}\langle s,m_{\bf p_{in}}|\nonumber\\&&\quad\quad\quad\quad\quad\quad\quad\quad\quad\quad\quad\quad\quad\quad\quad+\Delta^{*}_{3,s} |s,m_{\bf p_{in}}\rangle_{{\bf L}^{'}}~{}_{{\bf L}^{'}}\langle s,(m_{\bf p_{in}}+1)|\Bigg\},\\
   \overline{\cal M}_{10}:&=&\sum^{\infty}_{m_{\bf p_{in}}=0}\sum^{\infty}_{s=0}|\Gamma_{p_{\bf in},s}|^{2m_{\bf p_{in}}}\sqrt{m_{\bf p_{in}}+1}\Bigg\{\Delta_{4,s}\Gamma^{*}_{p_{\bf in},s}|s,m_{\bf p_{in}}\rangle_{{\bf L}^{'}}~{}_{{\bf L}^{'}}\langle s,(m_{\bf p_{in}}+1)|\nonumber\\&&\quad\quad\quad\quad\quad\quad\quad\quad\quad\quad\quad\quad\quad\quad\quad+\Delta_{3,s} |s,(m_{\bf p_{in}}+1)\rangle_{{\bf L}^{'}}~{}_{{\bf L}^{'}}\langle s,m_{\bf p_{in}}|\Bigg\},\\
      \overline{\cal M}_{11}:&=&\sum^{\infty}_{m_{\bf p_{in}}=0}\sum^{\infty}_{s=0}|\Gamma_{p_{\bf in},s}|^{2m_{\bf p_{in}}}(m_{\bf p_{in}}+1)\Bigg\{|\Delta_{4,s}|^{2}|s,m_{\bf p_{in}}\rangle_{{\bf L}^{'}}~{}_{{\bf L}^{'}}\langle s,m_{\bf p_{in}}|\nonumber\\&&\quad\quad\quad\quad\quad\quad\quad\quad\quad\quad\quad\quad\quad\quad\quad+|\Delta_{3,s}|^{2}|s,(m_{\bf p_{in}}+1)\rangle_{{\bf L}^{'}}~{}_{{\bf L}^{'}}\langle s,(m_{\bf p_{in}}+1)|\Bigg\}\nonumber\\
   &&+\sum^{\infty}_{m_{\bf p_{in}}=0}\sum^{\infty}_{s=0}|\Gamma_{p_{\bf in},s}|^{2m_{\bf p_{in}}}\sqrt{(m_{\bf p_{in}}+1)(m_{\bf p_{in}}+2)}\nonumber\\
&&\quad\quad\quad\quad\quad\quad\quad\quad\quad\quad\quad\quad\Bigg\{\Delta^{*}_{3,s} \Delta_{4,s} \Gamma^{*}_{p_{\bf in},s}|s,m_{\bf p_{in}}\rangle_{{\bf L}^{'}}~{}_{{\bf L}^{'}}\langle s,(m_{\bf p_{in}}+2)|\nonumber\\&&\quad\quad\quad\quad\quad\quad\quad\quad\quad\quad\quad\quad\quad+\Delta_{3,s} \Delta^{*}_{4,s}|s,(m_{\bf p_{in}}+2)\rangle_{{\bf L}^{'}}~{}_{{\bf L}^{'}}\langle s,m_{\bf p_{in}}|\Bigg\}.
   \eea
Here it is important to note that the maximally entangled state under consideration can be further factorized into two dimensional Alices's subsystem ($A$) and infinite dimensional Bob's subsystem ($B$). After taking the partial trace over the Bob's subsystem ($B$) one can construct the density matrix of the Alices's subsystem ($A$), which is given by: \bea \rho_A={\rm Tr}_B \rho_{A,B}&=&\frac{1}{2}\frac{\left(1-|\gamma_{p_{\bf in}}|^2\right)}{\left(1+f_{p_{\bf in}}\right)}\Bigg(\frac{1}{\left(1-|\gamma_{p_{\bf in}}|^2\right)}|0\rangle\langle 0|+\frac{\left(|\Delta_1|^2+|\Delta_2|^2\right)}{\left(1-|\gamma_{p_{\bf in}}|^2\right)^2}|1\rangle\langle 1|\Bigg)\nonumber\\
&&+\frac{1}{2}\frac{f^2_{p_{\bf in}}}{\left(1+f_{p_{\bf in}}\right)}\Bigg(\frac{1}{f_{p_{\bf in}}}|0\rangle\langle 0|+\sum^{\infty}_{s=0}\frac{\left(|\Delta_{3,s}|^2+|\Delta_{4,s}|^2\right)}{\left(1-|\Gamma_{p_{\bf in},s}|^2\right)^2}|1\rangle\langle 1|\Bigg).\eea
Here we have the following facts:
\bea &&\left(|\Delta_1|^2+|\Delta_2|^2\right)=\left(1-|\gamma_{p_{\bf in}}|^2\right),\\
&& \left(|\Delta_{3,s}|^2+|\Delta_{4,s}|^2\right)=\left(1-|\Gamma_{p_{\bf in},s}|^2\right),\eea
and
\bea &&\sum^{\infty}_{m_{\bf p_{in}}=0}|\gamma_{p_{\bf in}}|^{2m_{\bf p_{in}}}(m_{\bf p_{in}}+1)=\frac{1}{\left(1-|\gamma_{p_{\bf in}}|^2\right)^2},\\
&&\sum^{\infty}_{m_{\bf p_{in}}=0}|\Gamma_{p_{\bf in},s}|^{2m_{\bf p_{in}}}(m_{\bf p_{in}}+1)=\frac{1}{\left(1-|\Gamma_{p_{\bf in},s}|^2\right)^2}.\eea
Hence, we have the following simplified expression:
\bea \rho_A&=&\frac{1}{2}\frac{1}{\left(1+f_{p_{\bf in}}\right)}\Bigg(|0\rangle\langle 0|+|1\rangle\langle 1|\Bigg)\nonumber\\
&&+\frac{1}{2}\frac{f^2_{p_{\bf in}}}{\left(1+f_{p_{\bf in}}\right)}\Bigg(\frac{1}{f_{p_{\bf in}}}|0\rangle\langle 0|+\sum^{\infty}_{s=0}\frac{1}{\left(1-|\Gamma_{p_{\bf in},s}|^2\right)}|1\rangle\langle 1|\Bigg)\nonumber\\
&=&\frac{1}{2}\frac{1}{\left(1+f_{p_{\bf in}}\right)}\Bigg(|0\rangle\langle 0|+|1\rangle\langle 1|\Bigg)+\frac{1}{2}\frac{f_{p_{\bf in}}}{\left(1+f_{p_{\bf in}}\right)}\Bigg(|0\rangle\langle 0|+|1\rangle\langle 1|\Bigg)\nonumber\\
&=&\frac{1}{2}\Bigg(|0\rangle\langle 0|+|1\rangle\langle 1|\Bigg).\eea
Consequently, we have the following expression for the von Neumann entropy of the subsystem $A$:
\bea S(\rho_A)=-{\rm Tr}\left(\rho_A \log_2 \rho_A\right)=1.\eea
Now, to compute the von Neumann entropy of the system as a whole, one needs to find out the eigenvalues of the total density matrix $\rho_{A,B}$ in numerical fashion.

Now, our job is to compute the quantum version of the conditional entropy $S(B|A)$ for which we restrict our attention to the projective measurements on the subsystem $A$, characterized by a complete set of projectors:
\bea \Pi_{\pm}:=\frac{1}{2}\left(I\pm {\bf x}.{\bf \sigma}\right)=\frac{1}{2}\begin{pmatrix}
1\pm x_3 & \pm(x_1-ix_2)\\
\pm(x_1+ix_2) & 1\mp x_3
\end{pmatrix}.\eea
Here we have:
\bea {\bf x}.{\bf x}:=x^2_1+x^2_2+x^2_3=1.\eea
Also, $I$ is a $2\times 2$ identity matrix, and ${\bf \sigma}$ represents the Pauli spin matrices. The specific choices of the components $x_i$ in the present context is directly related to the choice of the measurement. For this specific reason, we are interested in the particular type of measurement process, which  minimizes the disturbances and fluctuations appearing on the system under consideration. Then, after performing the measurement, the density matrix in the present context of discussion (particularly in the projective measurement basis that we have introduced before)  can be expressed by the following  simplified expression:
\bea \rho_{B,\pm}:=\frac{1}{L_{\pm}}\Pi_{\pm}\rho_{A,B}\Pi_{\pm}.\eea
Here, performing the trace operation on the above-mentioned density matrix, we get the following simplified expression:
\bea {\rm Tr}(\rho_{B,\pm})
&=&\frac{1}{L_{\pm}}{\rm  Tr}\Bigg(\Pi^2_{\pm}\rho_{A,B}\Bigg)\nonumber\\
&=&\frac{1}{L_{\pm}}{\rm  Tr}\Bigg(\Pi_{\pm}\rho_{A,B}\Bigg)\nonumber\\
&=&\frac{1}{4L_{\pm}}\Bigg[\frac{\left(1-|\gamma_{p_{\bf in}}|^2\right)}{\left(1+f_{p_{\bf in}}\right)}\Bigg\{(1\pm x_3) {\cal M}_{00}\pm (x_1+ix_2) {\cal M}_{01}\pm (x_1-ix_2){\cal M}_{10}+(1\mp x_3){\cal M}_{11}\Bigg\}\nonumber\\
&&+\frac{f^{2}_{p_{\bf in}}}{\left(1+f_{p_{\bf in}}\right)}\Bigg\{(1\pm x_3) \overline{\cal M}_{00}\pm (x_1+ix_2) \overline{\cal M}_{01}\pm (x_1-ix_2)\overline{\cal M}_{10}+(1\mp x_3)\overline{\cal M}_{11}\Bigg\}\Bigg].\nonumber\\
&&\eea
Here we have used the cyclic property of the trace and also $\Pi^2_{\pm}=\Pi_{\pm}$. Additionally, it is important to note the definition of the factor $L_{\pm}$, is given by the following expression:
\bea L_{\pm}:&=&{\rm Tr}_{A,B}\Bigg(\Pi^2_{\pm}\rho_{A,B}\Bigg)\nonumber\\
&=&{\rm Tr}_{A,B}\Bigg(\Pi_{\pm}\rho_{A,B}\Bigg)\nonumber\\
&=&\frac{1}{4}\Bigg[\frac{\left(1-|\gamma_{p_{\bf in}}|^2\right)}{\left(1+f_{p_{\bf in}}\right)}\Bigg\{(1\pm x_3) {\rm Tr}{\cal M}_{00}+(1\mp x_3){\rm Tr}{\cal M}_{11}\Bigg\}\nonumber\\
&&+\frac{f^{2}_{p_{\bf in}}}{\left(1+f_{p_{\bf in}}\right)}\Bigg\{(1\pm x_3) {\rm Tr}\overline{\cal M}_{00}+(1\mp x_3){\rm Tr}\overline{\cal M}_{11}\Bigg\}\Bigg].\eea
Here, we use the following results for the further computational purpose:
\bea {\rm Tr}{\cal M}_{00}&=&\sum^{\infty}_{m_{\bf p_{in}}=0}|\gamma_{p_{\bf in}}|^{2m_{\bf p_{in}}}=\frac{1}{\left(1-|\gamma_{p_{\bf in}}|^2\right)},\\
    {\rm Tr}{\cal M}_{11}&=&\left(|\Delta_1|^2+|\Delta_2|^2\right)\sum^{\infty}_{m_{\bf p_{in}}=0}|\gamma_{p_{\bf in}}|^{2m_{\bf p_{in}}}(m_{\bf p_{in}}+1)\nonumber\\
    &=&\frac{\left(|\Delta_1|^2+|\Delta_2|^2\right)}{\left(1-|\gamma_{p_{\bf in}}|^2\right)^2}\nonumber\\
    &=&\frac{1}{\left(1-|\gamma_{p_{\bf in}}|^2\right)},\\
    {\rm Tr}\overline{\cal M}_{00}&=&\sum^{\infty}_{m_{\bf p_{in}}=0}\sum^{\infty}_{s=0}|\Gamma_{p_{\bf in},s}|^{2m_{\bf p_{in}}}=\sum^{\infty}_{s=0}\frac{1}{\left(1-|\Gamma_{p_{\bf in},s}|^2\right)}=f^{-1}_{p_{\bf in}},\\
    {\rm Tr}\overline{\cal M}_{11}&=&\sum^{\infty}_{m_{\bf p_{in}}=0}\sum^{\infty}_{s=0}|\Gamma_{p_{\bf in},s}|^{2m_{\bf p_{in}}}(m_{\bf p_{in}}+1)\left(|\Delta_{3,s}|^2+|\Delta_{4,s}|^2\right)\nonumber\\
    &=&\sum^{\infty}_{s=0}\frac{\left(|\Delta_{3,s}|^2+|\Delta_{4,s}|^2\right)}{\left(1-|\Gamma_{p_{\bf in},s}|^2\right)^2}\nonumber\\
    &=&\sum^{\infty}_{s=0}\frac{1}{\left(1-|\Gamma_{p_{\bf in},s}|^2\right)}\nonumber\\
    &=&f^{-1}_{p_{\bf in}}.\eea
    Using the above-mentioned results, we get:
\bea L_{\pm}:&=&\frac{1}{2}\Bigg[\frac{1}{\left(1+f_{p_{\bf in}}\right)}+\frac{f_{p_{\bf in}}}{\left(1+f_{p_{\bf in}}\right)}\Bigg]=\frac{1}{2}.\eea
Now, we are going to use the following preferred angular parametrization for the further simplification purpose:
\bea x_1&=&\sin\theta \cos\phi,\\
x_2&=&\sin\theta \sin\phi,\\
x_3&=&\cos\theta.\eea
Consequently, we found that the expression for the density matrix in the projected basis for the subsystem $B$ in the angular parametrized form can be expressed as:
\bea \rho_{B,\pm}&=&\frac{1}{2}\Bigg[\frac{\left(1-|\gamma_{p_{\bf in}}|^2\right)}{\left(1+f_{p_{\bf in}}\right)}\Bigg\{(1\pm \cos\theta) {\cal M}_{00}\pm \sin\theta {\cal M}_{01}\pm \sin\theta {\cal M}_{10}+(1\mp \cos\theta){\cal M}_{11}\Bigg\}\nonumber\\
&&+\frac{f^{2}_{p_{\bf in}}}{\left(1+f_{p_{\bf in}}\right)}\Bigg\{(1\pm \cos\theta) \overline{\cal M}_{00}\pm \sin\theta \overline{\cal M}_{01}\pm \sin\theta\overline{\cal M}_{10}+(1\mp \cos\theta)\overline{\cal M}_{11}\Bigg\}\Bigg].\nonumber\\
&&\eea
Here in the final step the $\phi$ dependent phase factors are absorbed appropriately. Finally, the simplified form of the quantum discord can be recast as:
\bea {\cal D}_{\theta}=1+{\rm Tr}\Bigg(\rho_{A,B}\log_2\rho_{A,B}\Bigg)-\frac{1}{2}\Bigg[{\rm Tr}\Bigg(\rho_{B,+}\log_2\rho_{B,+}\Bigg)+{\rm Tr}\Bigg(\rho_{B,-}\log_2\rho_{B,-}\Bigg)\Bigg].\eea
Here, our next prime job is to find out numerically the eigenvalues of $\rho_{A,B}$, $\rho_{B,+}$ and $\rho_{B,-}$ and to find out the specific value of the angular parameter $\theta$ that minimizes the above-mentioned expression for the quantum discord, represented by the symbol ${\cal D}_{\theta}$. Here, for the computational purpose it has to be noted that the convergence of the summation for $\rho_{A,B}$ is not very fast for the values of the mass parameters, $\nu=1/2$ and $\nu=3/2$, particularly in the limit when we take $p_{\bf in}\rightarrow 0$, the factor appearing in the summation $|\gamma_{ p_{\bf in}}|^{2m_{p_{\bf in}}}\rightarrow 1$. For this reason the nuumerical analysis has to be truncated for the small values of the momentum mode. The above argument we have written for the situation where we don't have any source i.e. $f_{p_{\bf in}}=0$. For the situation where we have source, i.e. $f_{p_{\bf in}}\neq 0$, we have the additional convergence criteria needs to be imposed for $p_{\bf in}\rightarrow 0$ along with smooth $s$. In this limiting situation, $|\Gamma_{ p_{\bf in},s}|^{2m_{p_{\bf in}}}\rightarrow 1$.

Further, we analytically compute the expression for the quantum discord in the limiting situation $p_{\bf in}\rightarrow \infty$ for any arbitrary values of the mass parameter $\nu$. In this situation, we have the following two constraints:
\bea &&\lim_{p_{\bf in}\rightarrow \infty}\gamma_{ p_{\bf in}}\rightarrow 0,\\
&&\lim_{p_{\bf in}\rightarrow \infty}\Gamma_{ p_{\bf in},s}\rightarrow 0.\eea
Consequently, in this particular limiting situation the density matrix $\rho_{A,B}$ can be expressed as:
\bea \lim_{p_{\bf in}\rightarrow \infty}\rho_{A,B}&=&\frac{1}{2\left(1+f_{p_{\bf in}}\right)}\sum^{\infty}_{m_{\bf p_{in}}=0}|\gamma_{p_{\bf in}}|^{2m_{\bf p_{in}}}\Bigg[|0,m_{\bf p_{in}}\rangle\langle 0,m_{\bf p_{in}}|\nonumber\\
&&+(m_{\bf p_{in}}+1)\Bigg\{|\Delta_2|^ 2|1,m_{\bf p_{in}}\rangle\langle 1,m_{\bf p_{in}}|+|\Delta_1|^ 2|1,m_{\bf p_{in}}+1\rangle\langle 1,m_{\bf p_{in}}+1|\Bigg\}\nonumber\\
&&+\sqrt{m_{\bf p_{in}}+1}\Bigg\{\Delta^{\star}_1|0,m_{\bf p_{in}}\rangle\langle 1,m_{\bf p_{in}}+1|+\Delta_1|1,m_{\bf p_{in}}+1\rangle\langle 0,m_{\bf p_{in}}|\Bigg\}\Bigg]\nonumber\\
&&+\frac{f^{2}_{p_{\bf in}}}{2\left(1+f_{p_{\bf in}}\right)}\sum^{\infty}_{m_{\bf p_{in}}=0}\sum^{\infty}_{s=0}|\Gamma_{p_{\bf in},s}|^{2m_{\bf p_{in}}}\Bigg[|0,s,m_{\bf p_{in}}\rangle\langle 0,s,m_{\bf p_{in}}|\nonumber\\
&&+(m_{\bf p_{in}}+1)\Bigg\{|\Delta_{4,s}|^ 2|1,s,m_{\bf p_{in}}\rangle\langle 1,s,m_{\bf p_{in}}|+|\Delta_{3,s}|^ 2|1,s,m_{\bf p_{in}}+1\rangle\langle 1,s,m_{\bf p_{in}}+1|\Bigg\}\nonumber\\
&&+\sqrt{m_{\bf p_{in}}+1}\Bigg\{\Delta^{\star}_{3,s}|0,s,m_{\bf p_{in}}\rangle\langle 1,s,m_{\bf p_{in}}+1|+\Delta_{3,s}|1,s,m_{\bf p_{in}}+1\rangle\langle 0,s,m_{\bf p_{in}}|\Bigg\}\Bigg]\nonumber\\
&\sim&\frac{1}{2\left(1+f_{p_{\bf in}}\right)}\Bigg[|0,0\rangle\langle 0,0|+\Bigg\{|\Delta_2|^ 2|1,0\rangle\langle 1,0|+|\Delta_1|^ 2|1,1\rangle\langle 1,1|\Bigg\}\nonumber\\
&&\quad\quad\quad\quad\quad\quad\quad\quad\quad\quad+\Bigg\{\Delta^{\star}_1|0,0\rangle\langle 1,1|+\Delta_1|1,1\rangle\langle 0,0|\Bigg\}\Bigg]\nonumber\\
&&+\frac{f^{2}_{p_{\bf in}}}{2\left(1+f_{p_{\bf in}}\right)}\sum^{\infty}_{s=0}\Bigg[|0,0\rangle_s {}_s\langle 0,0|\nonumber\\
&&+\Bigg\{|\Delta_{4,s}|^ 2|1,0\rangle_s {}_s\langle 1,0|+|\Delta_{3,s}|^ 2|1,1\rangle_s {}_s\langle 1,1|\Bigg\}\nonumber\\
&&+\Bigg\{\Delta^{\star}_{3,s}|0,0\rangle_s {}_s\langle 1,1|+\Delta_{3,s}|1,1\rangle_s {}_s\langle 0,0|\Bigg\}\Bigg]
\nonumber\\
&=&\frac{1}{2\left(1+f_{p_{\bf in}}\right)}\Bigg[\Bigg(|0,0\rangle+\Delta_1|1,1\rangle\Bigg)\Bigg(\langle 0,0|+\Delta^{\star}
_1\langle 1,1|\Bigg)+|\Delta_2|^ 2|1,0\rangle\langle 1,0|\Bigg]\nonumber\\
&&+\frac{f^{2}_{p_{\bf in}}}{2\left(1+f_{p_{\bf in}}\right)}\sum^{\infty}_{s=0}\Bigg[\Bigg(|0,0\rangle_s+\Delta_{3,s}|1,1\rangle_s\Bigg)\Bigg({}_s\langle 0,0|+\Delta^{\star}
_{3,s}{}_s\langle 1,1|\Bigg)+|\Delta_{4,s}|^ 2|1,0\rangle_s {}_s\langle 1,0|\Bigg]\nonumber\\
&&\eea
The two eigenvalues of the above-mentioned density matrix are given by the following expressions:
\bea \lambda^{(A,B)}_1&=&\frac{1}{2\left(1+f_{p_{\bf in}}\right)}+\frac{f^{2}_{p_{\bf in}}}{2\left(1+f_{p_{\bf in}}\right)}\sum^{\infty}_{s=0}1\nonumber\\
&=&\frac{1}{2\left(1+f_{p_{\bf in}}\right)}+\frac{f^{2}_{p_{\bf in}}}{2\left(1+f_{p_{\bf in}}\right)}\left(1+\zeta(0)\right)\nonumber\\
&=&\frac{1}{2\left(1+f_{p_{\bf in}}\right)}\Bigg\{1+\frac{f^{2}_{p_{\bf in}}}{2}\Bigg\},\\
\lambda^{(A,B)}_2&=&\frac{|\Delta_2|^ 2}{2\left(1+f_{p_{\bf in}}\right)}+\frac{f^{2}_{p_{\bf in}}}{2\left(1+f_{p_{\bf in}}\right)}\sum^{\infty}_{s=0}|\Delta_{4,s}|^2.\eea
Here we use:
\bea \sum^{\infty}_{s=0}1=1+\sum^{\infty}_{s=1}1=1+\zeta(0)=1-\frac{1}{2}=\frac{1}{2}\quad\quad {\rm where}\quad\quad \zeta(0)=-\frac{1}{2}.\eea
Similarly, in the same limiting case, the density matrix $\rho_{B,\pm}$ can be recast into the following form:
\bea \lim_{p_{\bf in}\rightarrow \infty}\rho_{B,\pm}&=&\frac{1}{2\left(1+f_{p_{\bf in}}\right)}\sum^{\infty}_{m_{\bf p_{in}}=0}|\gamma_{p_{\bf in}}|^{2m_{\bf p_{in}}}\Bigg[(1\pm \cos\theta) |m_{\bf p_{in}}\rangle\langle m_{\bf p_{in}}|\nonumber\\
&&\pm \sin\theta \sqrt{m_{\bf p_{in}}+1}\Delta^{*}_1 |m_{\bf p_{in}}\rangle\langle m_{\bf p_{in}}+1|\nonumber\\
&&\pm \sin\theta \sqrt{m_{\bf p_{in}}+1}\Delta_1 |m_{\bf p_{in}}+1\rangle\langle m_{\bf p_{in}}|\nonumber\\
&&+(1\mp \cos\theta)(m_{\bf p_{in}}+1)\Bigg\{|\Delta_2|^{2}|m_{\bf p_{in}}\rangle\langle m_{\bf p_{in}}|+|\Delta_1|^{2}|m_{\bf p_{in}}+1\rangle\langle m_{\bf p_{in}}+1|\Bigg\}\Bigg]\nonumber\\
&&+\frac{f^{2}_{p_{\bf in}}}{2\left(1+f_{p_{\bf in}}\right)}\sum^{\infty}_{m_{\bf p_{in}}=0}\sum^{\infty}_{s=0}|\Gamma_{p_{\bf in},s}|^{2m_{\bf p_{in}}}\Bigg[(1\pm \cos\theta) |m_{\bf p_{in}}\rangle_s {}_s\langle m_{\bf p_{in}}|\nonumber\\
&&\pm \sin\theta \sqrt{m_{\bf p_{in}}+1}\Delta^{*}_{3,s} |m_{\bf p_{in}}\rangle_s{}_s\langle m_{\bf p_{in}}+1|\nonumber\\
&&\pm \sin\theta\sqrt{m_{\bf p_{in}}+1}\Delta_{3,s} |m_{\bf p_{in}}+1\rangle_s {}_s\langle  m_{\bf p_{in}}|\nonumber\\
&&+(1\mp \cos\theta)(m_{\bf p_{in}}+1)\Bigg\{|\Delta_{4,s}|^{2}|m_{\bf p_{in}}\rangle_s{}_s\langle m_{\bf p_{in}}|+|\Delta_{3,s}|^{2}|m_{\bf p_{in}}+1\rangle_s{}_s\langle m_{\bf p_{in}}+1|\Bigg\}\Bigg].\nonumber\\
&\sim&\frac{1}{2\left(1+f_{p_{\bf in}}\right)}\Bigg[(1\pm \cos\theta) |0\rangle\langle 0|\pm \sin\theta \Delta^{*}_1 |0\rangle\langle 1|\pm \sin\theta \Delta_1 |1\rangle\langle 0|\nonumber\\
&&+(1\mp \cos\theta)\Bigg\{|\Delta_2|^{2}|0\rangle\langle 0|+|\Delta_1|^{2}|1\rangle\langle 1|\Bigg\}\Bigg]\nonumber\\
&&+\frac{f^{2}_{p_{\bf in}}}{2\left(1+f_{p_{\bf in}}\right)}\sum^{\infty}_{s=0}\Bigg[(1\pm \cos\theta) |0\rangle_s {}_s\langle 0|\pm \sin\theta \Delta^{*}_{3,s} |0\rangle_s{}_s\langle 1|\pm \sin\theta\Delta_{3,s} |1\rangle_s {}_s\langle  0|\nonumber\\
&&+(1\mp \cos\theta)\Bigg\{|\Delta_{4,s}|^{2}|0\rangle_s{}_s\langle 0|+|\Delta_{3,s}|^{2}|1\rangle_s{}_s\langle 1|\Bigg\}\Bigg].\nonumber\\
&&\eea
    \begin{figure*}[htb]
    \centering
    \subfigure[For $f_{p_{\bf in}}=0$.]{
        \includegraphics[width=14.2cm,height=9.1cm] {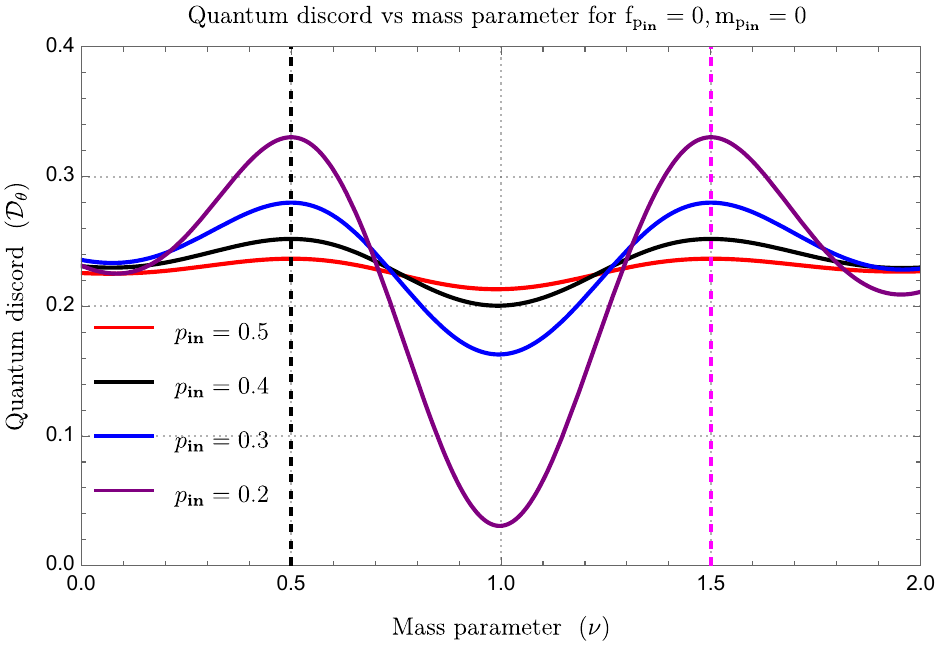}
        \label{Qd1}
    }
    \subfigure[For small $f_{p_{\bf in}}\neq 0$.]{
        \includegraphics[width=14.2cm,height=9.1cm] {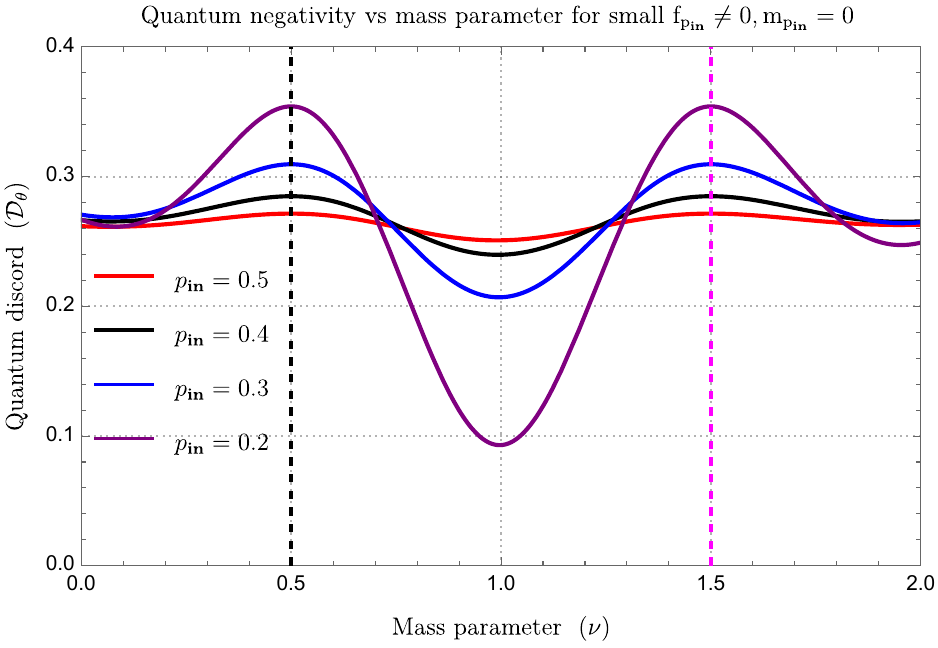}
        \label{Qd2}
       }
    \caption[Optional caption for list of figures]{Quantum discord vs mass parameter for a specified momentum mode. Here we fix $\theta=\pi/2$ at which the quantum discord achieves its maximum value.  } 
    \label{Qda}
    \end{figure*}  
    \begin{figure*}[htb]
    \centering
    \subfigure[For $f_{p_{\bf in}}=0$.]{
        \includegraphics[width=14.2cm,height=9.1cm] {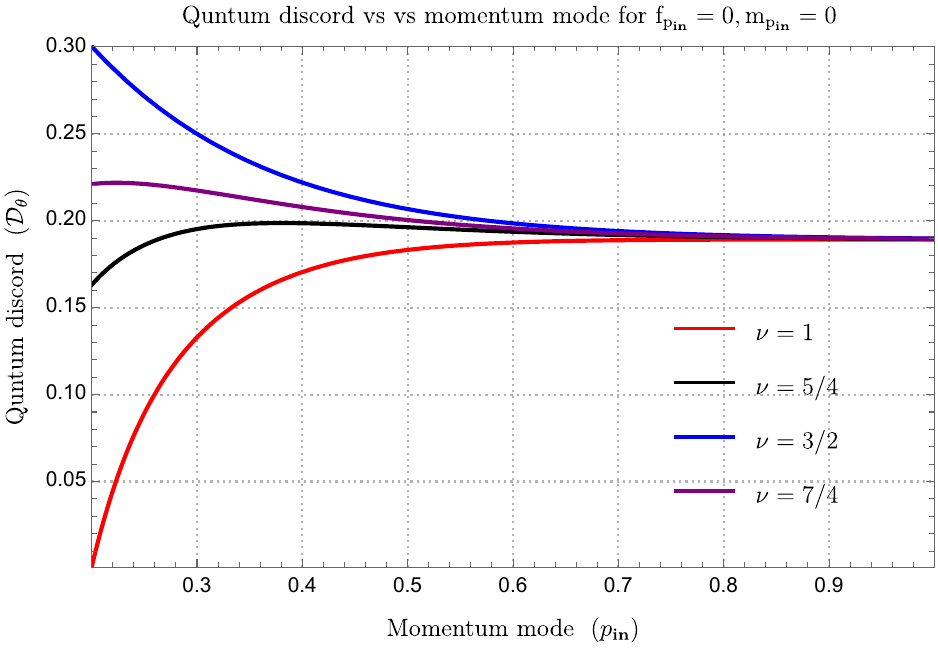}
        \label{Qdd1}
    }
    \subfigure[For small $f_{p_{\bf in}}\neq 0$.]{
        \includegraphics[width=14.2cm,height=9.1cm] {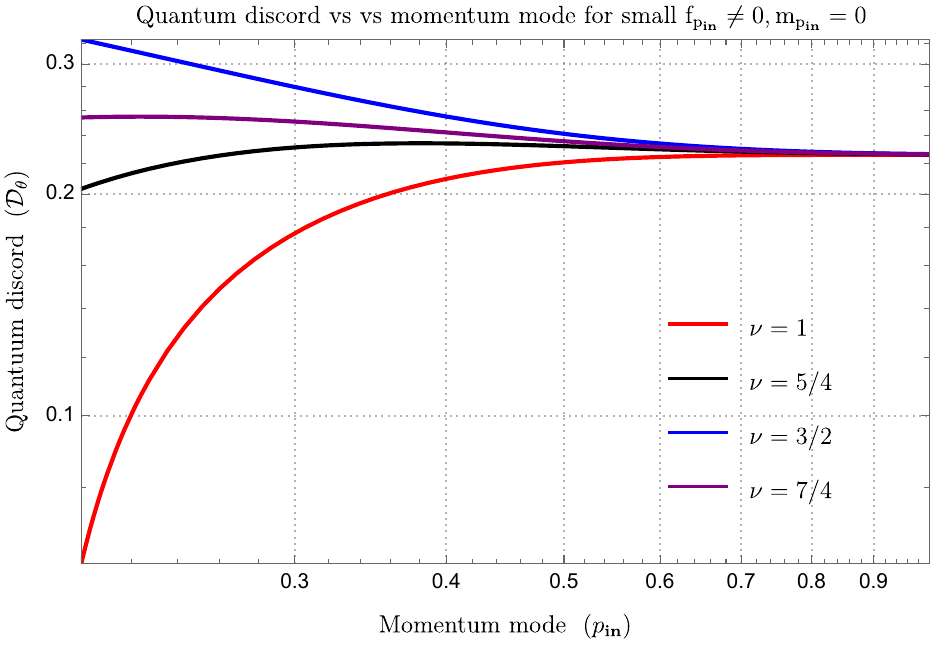}
        \label{Qdd2}
       }
    \caption[Optional caption for list of figures]{Quantum discord vs the momentum mode for a specified mass parameter. Here we fix $\theta=\pi/2$ at which the quantum discord achieves its maximum value.} 
    \label{Qdb}
    \end{figure*}  
Then the eigenvalues of the above-mentioned density matrix can be further computed as:
\bea \lambda_{B,\pm}=\frac{1}{2}\Bigg[\alpha_{\theta}\pm \sqrt{\alpha^2_{\theta}-4\beta_{\theta}}\Bigg],\eea
where we introduce two new symbols, $\alpha_{\theta}$ and $\beta_{\theta}$, which are in the present context is defined as:
\bea \alpha_{\theta}:&=&\Bigg(A_{11}(\theta)+B_{11}(\theta)+A_{00}(\theta)+B_{00}(\theta)\Bigg),\\
\beta_{\theta}:&=&\Bigg(A_{00}(\theta)+B_{00}(\theta)\Bigg)\Bigg(A_{11}(\theta)+B_{11}(\theta)\Bigg)\nonumber\\
&&-\Bigg(A_{01}(\theta)+B_{01}(\theta)\Bigg)\Bigg(A_{10}(\theta)+B_{10}(\theta)\Bigg),\eea
where, the angular parameters, $A_{ij}(\theta)$ and $B_{ij}(\theta)$ for $i,j=0,1$ are defined as:
\bea A_{00}(\theta):&=&\frac{1}{2\left(1+f_{p_{\bf in}}\right)}\Bigg\{\left(1+|\Delta_2|^2\right)\pm \cos\theta \left(1-|\Delta_2|^2\right)\Bigg\},\\ 
A_{01}(\theta):&=&\pm \frac{1}{2\left(1+f_{p_{\bf in}}\right)}\Delta^{\star}_1\sin\theta,\\
A_{10}(\theta):&=&\pm \frac{1}{2\left(1+f_{p_{\bf in}}\right)}\Delta_1\sin\theta,\\
A_{11}(\theta):&=& \frac{1}{2\left(1+f_{p_{\bf in}}\right)}|\Delta_1|^2\left(1\mp\cos\theta\right),\\B_{00}(\theta):&=&\frac{f^2_{p_{\bf in}}}{4\left(1+f_{p_{\bf in}}\right)}\Bigg\{\left(1+2\sum^{\infty}_{s=0}|\Delta_{4,s}|^2\right)\pm \cos\theta \left(1-2\sum^{\infty}_{s=0}|\Delta_{4,s}|^2\right)\Bigg\},\\
B_{01}(\theta):&=&\pm \frac{f^2_{p_{\bf in}}}{2\left(1+f_{p_{\bf in}}\right)}\sum^{\infty}_{s=0}\Delta^{\star}_{3,s}\sin\theta,\\
B_{10}(\theta):&=&\pm \frac{f^2_{p_{\bf in}}}{2\left(1+f_{p_{\bf in}}\right)}\sum^{\infty}_{s=0}\Delta_{3,s}\sin\theta,\\
B_{11}(\theta):&=& \frac{f^2_{p_{\bf in}}}{2\left(1+f_{p_{\bf in}}\right)}\sum^{\infty}_{s=0}|\Delta_{3,s}|^2\left(1\mp\cos\theta\right).\eea
Finally, the analytical expression for quantum discord in the small scale limit can be expressed as:
\bea {\cal D}_{\theta}&=&1+\frac{1}{2\left(1+f_{p_{\bf in}}\right)}\Bigg\{1+\frac{f^{2}_{p_{\bf in}}}{2}\Bigg\}\log_2\Bigg(\frac{1}{2\left(1+f_{p_{\bf in}}\right)}\Bigg\{1+\frac{f^{2}_{p_{\bf in}}}{2}\Bigg\}\Bigg)\nonumber\\
&&+\Bigg(\frac{|\Delta_2|^ 2}{2\left(1+f_{p_{\bf in}}\right)}+\frac{f^{2}_{p_{\bf in}}}{2\left(1+f_{p_{\bf in}}\right)}\sum^{\infty}_{s=0}|\Delta_{4,s}|^2\Bigg)\log_2\Bigg(\frac{|\Delta_2|^ 2}{2\left(1+f_{p_{\bf in}}\right)}+\frac{f^{2}_{p_{\bf in}}}{2\left(1+f_{p_{\bf in}}\right)}\sum^{\infty}_{s=0}|\Delta_{4,s}|^2\Bigg)\nonumber\\
&&-\frac{1}{2}\Bigg[\frac{1}{2}\Bigg(\alpha_{\theta}+ \sqrt{\alpha^2_{\theta}-4\beta_{\theta}}\Bigg)\log_2\Bigg(\frac{1}{2}\Bigg(\alpha_{\theta}+ \sqrt{\alpha^2_{\theta}-4\beta_{\theta}}\Bigg)\Bigg)\nonumber\\
&&+\frac{1}{2}\Bigg(\alpha_{\theta}- \sqrt{\alpha^2_{\theta}-4\beta_{\theta}}\Bigg)\log_2\Bigg(\frac{1}{2}\Bigg(\alpha_{\theta}- \sqrt{\alpha^2_{\theta}-4\beta_{\theta}}\Bigg)\Bigg)\Bigg].\eea
In this limiting situation, at the angular scale $\theta=\pi/2$ the quantum discord takes the following simplified form:
\bea {\cal D}_{\pi/2}&=&1+\frac{1}{2\left(1+f_{p_{\bf in}}\right)}\Bigg\{1+\frac{f^{2}_{p_{\bf in}}}{2}\Bigg\}\log_2\Bigg(\frac{1}{2\left(1+f_{p_{\bf in}}\right)}\Bigg\{1+\frac{f^{2}_{p_{\bf in}}}{2}\Bigg\}\Bigg)\nonumber\\
&&+\Bigg(\frac{|\Delta_2|^ 2}{2\left(1+f_{p_{\bf in}}\right)}+\frac{f^{2}_{p_{\bf in}}}{2\left(1+f_{p_{\bf in}}\right)}\sum^{\infty}_{s=0}|\Delta_{4,s}|^2\Bigg)\log_2\Bigg(\frac{|\Delta_2|^ 2}{2\left(1+f_{p_{\bf in}}\right)}+\frac{f^{2}_{p_{\bf in}}}{2\left(1+f_{p_{\bf in}}\right)}\sum^{\infty}_{s=0}|\Delta_{4,s}|^2\Bigg)\nonumber\\
&&-\frac{1}{2}\Bigg[\frac{1}{2}\Bigg(\alpha_{\pi/2}+ \sqrt{\alpha^2_{\pi/2}-4\beta_{\pi/2}}\Bigg)\log_2\Bigg(\frac{1}{2}\Bigg(\alpha_{\pi/2}+ \sqrt{\alpha^2_{\pi/2}-4\beta_{\pi/2}}\Bigg)\Bigg)\nonumber\\
&&+\frac{1}{2}\Bigg(\alpha_{\pi/2}- \sqrt{\alpha^2_{\pi/2}-4\beta_{\pi/2}}\Bigg)\log_2\Bigg(\frac{1}{2}\Bigg(\alpha_{\pi/2}- \sqrt{\alpha^2_{\pi/2}-4\beta_{\pi/2}}\Bigg)\Bigg)\Bigg].\eea
where, the angular parameters, $A_{ij}(\pi/2)$ and $B_{ij}(\pi/2)$ for $i,j=0,1$ are defined as:
\bea A_{00}(\pi/2):&=&\frac{1}{2\left(1+f_{p_{\bf in}}\right)}\left(1+|\Delta_2|^2\right),\\ 
A_{01}(\pi/2):&=&\pm \frac{1}{2\left(1+f_{p_{\bf in}}\right)}\Delta^{\star}_1,\\
A_{10}(\pi/2):&=&\pm \frac{1}{2\left(1+f_{p_{\bf in}}\right)}\Delta_1,\\
A_{11}(\pi/2):&=& \frac{1}{2\left(1+f_{p_{\bf in}}\right)}|\Delta_1|^2,\\B_{00}(\pi/2):&=&\frac{f^2_{p_{\bf in}}}{4\left(1+f_{p_{\bf in}}\right)}\left(1+2\sum^{\infty}_{s=0}|\Delta_{4,s}|^2\right),\\
B_{01}(\pi/2):&=&\pm \frac{f^2_{p_{\bf in}}}{2\left(1+f_{p_{\bf in}}\right)}\sum^{\infty}_{s=0}\Delta^{\star}_{3,s},\\
B_{10}(\pi/2):&=&\pm \frac{f^2_{p_{\bf in}}}{2\left(1+f_{p_{\bf in}}\right)}\sum^{\infty}_{s=0}\Delta_{3,s},\\
B_{11}(\pi/2):&=& \frac{f^2_{p_{\bf in}}}{2\left(1+f_{p_{\bf in}}\right)}\sum^{\infty}_{s=0}|\Delta_{3,s}|^2.\eea
Here  in figure (\ref{Qda}(a)) and (\ref{Qda}(b)),  we have shown the quantum discord with mass parameter.  Also,  in figure (\ref{Qdb}(a)) and (\ref{Qdb}(b)),  we have shown the quantum discord with the momentum mode for the given value of mass parameter. It has to be noted that, an oscialltory repeated behaviour is observed in the case of figure (\ref{Qdb}(a)) and (\ref{Qdb}(b)). In both of the sets of plots we fix $\theta=\pi/2$ at which the quantum discord achieves its maximum value. Both vanishing $f_{p_{\bf in}}=0$ and a very small value but $f_{p_{\bf in}}\neq 0$ have been taken into consideration in these graphs.  The following is a point-by-point account of the intriguing results and physical interpretation of these plots:
\begin{enumerate}
    \item Figures (\ref{Qda}(a)) and (\ref{Qda}(b)) show that quantum discord almost disappears at $\nu=1$ for $f_{p_{\bf in}}=0$, but it is non-zero but extremely small for small $f_{p_{\bf in}}\neq 0$.  It also suggests that the contribution from the big scales is almost insignificant at the value of the mass parameter $\nu=1$. It's also crucial to remember that this precise result is achieved for a certain momentum mode value, $p_{\bf in}=0.2$.

    \item However, we have discovered that the obtained value of the quantum discord from the current theoretical setup reaches its highest value at $\nu=1/2$ and $\nu=3/2$, which also corresponds to the maximum correlation.  This result holds for the momentum mode, $p_{\bf in}=0.2$.   The small value of the momentum mode $p_{\bf in}$ corresponds to the large scale limit, which is an essential point to note.

    \item In comparison to the example we have examined for the momentum mode $p_{\bf in}=0.2$, we have discovered that the variation with regard to the mass parameter $\nu$ is smaller for the other momentum mode values falling within the window $0.2<p_{\bf in}<0.5$.   We have discovered that if we increase the value of $p_{\bf in}$, the corresponding variation is decreased and we obtain intermediate values of discord, which correspond to the intermediate amount of quantum mechanical entanglement, by comparing all the results obtained for the various momentum modes within the specified range.

    \item  Figures (\ref{Qdb}(a)) and (\ref{Qdb}(b)) show that for large values of the momentum mode $p_{\bf in}>0.5$, the corresponding quantum discord estimated from the stipulated theoretical set up saturates to a constant non-zero, positive, and negligible value.  This implies a consistent level of quantum entanglement for any arbitrary positive real value of the mass parameter $\nu$.  However, in this asymptotic limit, it is impossible to differentiate the individual effect of the mass parameter in the current computation. This also means that the low momentum modes are preferable for the current investigation to identify the individual effects of the mass parameter $\nu$.

    \item We demonstrated that even in the limit where the entanglement negativity disappears, quantum discord persists.
\end{enumerate}

\section{Conclusion}
\label{ka10}

We wrap off our conversation with the following conclusions drawn from the analysis we conducted for this publication:
\begin{itemize}
\item  First, in the context of quantum information theory, we have begun with a fundamental description of the quantum discord for a broad quantum mechanical setup.  The technical information for the associated calculations from a broad quantum mechanical setup has been supplied.  Additionally, we have given a suitable physical explanation for why the corresponding measure is physically important or significant for the computation we wish to carry out for the global de Sitter space open chart.

\item  Furthermore, we have thoroughly discussed the details of the geometrical arrangement of the open chart of the de Sitter space, the platform on which we plan to perform the remaining calculation. We have independently ascertained the structure of the metric in the region between {\bf L} and {\bf R}, which is an essential piece of information for estimating the behavior of scalar modes based on our computations.

\item  Next, we computed the explicit equation for the mode function using the string theory-driven Axiverse, with which we created the Bunch Davies vacuum state and subsequently the expression for the reduced density matrix.

\item Additionally, we have calculated the open chart expressions for the Axiverse model's entanglement negativity, logarithmic negativity, and quantum discord.  In comparison to the Von Neumann measure, which is frequently employed in this situation, we have discovered that the recently investigated measures are more significant.

\item Additionally, we discovered that in the entanglement negativity vs. mass parameter squared graphs, the level of quantum entanglement decreases exponentially for large masses.  However, when we take into account the logarithmic negativity measure in this situation, this decaying behavior is slightly different.

\item Additionally, we discovered that two successive peaks of identical height exist in the logarithmic negativity and entanglement negativity spectra for the massless scenario $\nu=3/2$ and the conformal coupling $\nu=1/2$.  We have discovered a truly fascinating feature.  Furthermore, we have discovered that the small mass parameter of the axion field during the de Sitter expansion in the global coordinates causes oscillation in the spectrum in addition to two noticeable peaks for the mass parameter values indicated. If the mass parameter is really tiny, this oscillation accelerates.  Conversely, if the mass parameter is high, the oscillation period is longer and slower.

\item 

The maximally entangled state, which we used to construct the reduced density matrix by extracting all the information from the initial Bunch Davies vacuum state, is the most crucial component of this specific computation.  This solution has also been utilized to calculate the equation for the reduced density matrix's partial transposed version.  Next, we discovered that the mode corresponding to $m_{\bf p_{in}}=0$ (ground state) corresponds to the negative eigen value spectrum. Using this information, we numerically examined the behavior of quantum discord and logarithmic negativity from the specified theoretical setup. We have discovered that the corresponding eigen values for the other values of $m_{\bf p_{in}}$ are mostly positive, which is undesirable when building an Axiverse.

\item The conformally linked case with $\nu=1/2$ and the massless case with $\nu=3/2$ are two extremely unusual spots in the entanglement spectrum in the Axiverse where the quantum correlation is equal and has a high amplitude.  However, we have discovered that the quantity of quantum correlation calculated from the related picture reaches its smallest value for the mass parameter $\nu=1$.  After conducting our investigation on the Axiverse, we were able to gather this information, which is clearly promising.

\item After introducing two observers—one in an open chart of de Sitter space and the other in a global chart—we calculated the quantum discord produced by each observer identifying a mode. This scenario is comparable to the interaction between an observer in one of the two Rindler wedges in flat space and another in Minkowski space. It is well known that in Rindler space, entanglement disappears as the relative acceleration approaches infinity. In contrast, the observer's relative acceleration in de Sitter space is proportional to the open chart's curvature scale.

\item We demonstrated that even in the limit where the entanglement negativity disappears, quantum discord persists.
\end{itemize}

Here are some intriguing directions for the near future where our analysis can be expanded:
\begin{itemize}

\item By taking into account the Bunch Davies quantum vacuum state, we have limited our research to the calculation of entanglement negativity, logarithmic negativity, and quantum discord.  Our analysis can be immediately extended to a general non-Bunch Davies vacua, like $\alpha$ vacua.  Extending our research for non-Bunch Davies vacua is anticipated to yield many intriguing results since it will reveal quantum correlations and their numerous unidentified applications.

\item Because the global and planar coordinates of de Sitter space are linked by coordinate transformation, it is useful to understand how the current results can be justified within the context of primordial cosmology.   This is a potential that should be thoroughly considered for future work using observation.

\item The direct association between elevated point quantum correlations and various quantum information theoretic measures, alongside the manifestations of quantum entanglement within quantum correlations calculated in the quantum field theory of de Sitter space and primordial cosmological frameworks, presents intriguing avenues for future investigation based on the current framework established in this paper. 

\item Future research could benefit greatly from expanding the current computation to examine quantum mechanical decoherence \cite{Burgess:2022nwu,Martin:2021znx,Martin:2018lin,Martin:2018zbe,Liu:2016aaf} and quantum diffusion \cite{Ezquiaga:2022qpw}, which can explain a number of unknown facts from the current setup in both the global and planer patch of de Sitter space.

\item In cosmological settings, the creation of squeezed quantum mechanical states and their implications are frequently studied \cite{Choudhury:2022btc,Choudhury:2021brg,Adhikari:2021ked,Adhikari:2021pvv,Choudhury:2020hil,Bhargava:2020fhl,Adhikari:2022oxr,Park:2022xyf,Ruan:2021wep,DiGiulio:2021fhf,Bhattacharyya:2020rpy,Li:2021kfq,Grishchuk:1993ds,Albrecht:1992kf}.  If we could create a squeezed quantum state using the current theoretical framework that we are examining in this paper, that would be fantastic. This will assist in determining different quantum information theoretic measurements and their applicability in different situations.  In the event that the building of a squeezed state is not feasible, it is also possible to investigate alternative options beyond the current configuration \cite{Adhikari:2022whf,Adhikari:2021ckk,Choudhury:2022xip,Choudhury:2022dox}.

\item Because we are discussing closed quantum systems, the computation is now limited to a quantum system that is entirely adiabatic in nature.  Studying the open quantum system version of the current setup within the context of de Sitter space quantum field theory would be fantastic \cite{Colas:2022kfu,Colas:2022hlq,Choudhury:2022ati,Banerjee:2021lqu,Choudhury:2020dpf,Banerjee:2020ljo,Akhtar:2019qdn,Choudhury:2018rjl,Choudhury:2018bcf,Chaykov:2022zro,Chaykov:2022pwd,Burgess:2021luo,Kaplanek:2021fnl,Burgess:2015ajz}.

\item In near future we haver a plan to extend our analysis to study the imprints of stringy Axiverse in the context of features of small and large primordial fluctuations \cite{Choudhury:2025hnu,Choudhury:2025vso,Choudhury:2011sq,Choudhury:2012yh,Choudhury:2013zna,Choudhury:2013jya,Choudhury:2013iaa,Choudhury:2014sxa,Choudhury:2014uxa,Choudhury:2014kma,Choudhury:2014sua,Choudhury:2015pqa,Choudhury:2015hvr,Choudhury:2017cos}, including primordial black hole formation \cite{Choudhury:2025kxg,Choudhury:2024kjj,Choudhury:2024aji,Choudhury:2024dzw,Choudhury:2024dei,Choudhury:2024jlz,Choudhury:2024ybk,Choudhury:2024one,Choudhury:2023fjs,Choudhury:2023fwk,Choudhury:2023hfm,Choudhury:2023kdb,Choudhury:2023hvf,Choudhury:2023rks,Choudhury:2023jlt,Choudhury:2023vuj,Choudhury:2013woa,Choudhury:2011jt}, dark matter production and to study the gravitational collapse mechanisms in detail.
\end{itemize}

\newpage



	\subsection*{Acknowledgements}
SC would like to thank the work
friendly environment of The Thanu Padmanabhan Centre For Cosmology and Science Popularization (CCSP), 
Shree Guru Gobind Singh Tricentenary (SGT) University,  Gurugram,  Delhi-NCR for providing tremendous support in research and offer the Assistant Professor (Senior Grade) position.  SC would like to thank The North American Nanohertz Observatory for Gravitational
Waves (NANOGrav) collaboration and the National Academy of Sciences (NASI), Prayagraj, India, for being elected as an associate member and the member of the academy
respectively. SC also thanks
all the members of virtual international
non-profit consortium Quantum Aspects of the Space Time \& Matter (QASTM) for elaborative discussions.  Last but not
least, we would like to acknowledge our debt to the people belonging to the various parts of the world for their
generous and steady support for research in natural sciences.

\clearpage

\addcontentsline{toc}{section}{References}
\bibliographystyle{utphys}
\bibliography{references2}

\end{document}